\documentclass[a4paper,11pt]{article}

\usepackage{fullpage}%
\usepackage{amsmath}
\usepackage[utf8]{inputenc}
\usepackage[usenames, dvipsnames]{color} 
\usepackage{comment} 
\usepackage{graphicx}
\usepackage{caption}
\usepackage{subcaption}
\usepackage{mathtools}
\usepackage[toc,page]{appendix}
\usepackage[english]{babel}
\usepackage{enumerate}

\usepackage{comment} 

\usepackage{amsthm}
\usepackage{amssymb}
\usepackage{amsmath}

\usepackage{algorithm}
\usepackage[noend]{algpseudocode}

\usepackage{stmaryrd}

\usepackage{amsmath, amssymb, amsfonts, amsthm}
\usepackage{algorithm}
\usepackage[noend]{algpseudocode}

\usepackage{tablefootnote}
\usepackage{pgf}
\usepackage{ulem}
\usepackage{smallsec}
\usepackage{hyperref}
\usepackage{colored-comments}

\usepackage{array}
\usepackage{multirow}

\usepackage{tikz}
\usetikzlibrary{arrows,shapes,snakes,automata,backgrounds,positioning}

\tikzstyle{player}=[state,draw,rounded rectangle,align=center]
\tikzstyle{widget}=[draw=red,rectangle, rounded rectangle=10pt,dashed,minimum size=6mm,fill=yellow]
\tikzset{every loop/.style={looseness=7}}

\tikzstyle{player1}=[state,draw,rounded rectangle,align=center]
\tikzstyle{player2}=[state,draw,rectangle,align=center]

\definecolor{armygreen}{rgb}{0.29, 0.33, 0.13}
\definecolor{ao(english)}{rgb}{0.0, 0.5, 0.0}
\definecolor{myblue}{rgb}{0.3, 0.4, 0.6}
\definecolor{pakistangreen}{rgb}{0.2, 0.8, 0.2}
\definecolor{hardness}{rgb}{0.16, 0.5, 0.0}
\definecolor{pseudo}{rgb}{1.0, 0.5, 0.0} 
\definecolor{bscc}{rgb}{0.24, 0.7, 0.44}
\definecolor{mikadoyellow}{rgb}{1.0, 0.77, 0.05}
\definecolor{cblue}{rgb}{0.19, 0.73, 0.56}

\definecolor{bluemdp}{rgb}{0.25, 0.29, 0.3}
\definecolor{redmdp}{rgb}{0.6, 0.0, 0.0}

\DeclareMathOperator*{\argmax}{arg\,max}

\newtheorem{theorem}{Theorem}
\newtheorem{corollary}{Corollary}[theorem]
\newtheorem{lemma}[theorem]{Lemma}
\newtheorem{remark}[theorem]{Remark}
\newtheorem{proposition}[theorem]{Proposition}

\newcommand{\initState}{\ensuremath{s_{{\sf init}}} }

\newcommand{\graph}{\ensuremath{\mathcal{G}} }

\newcommand{\looping}{\ensuremath{{\sf loop}} }

\newcommand{\final}{\ensuremath{{\sf final}} }

\newcommand{\MDP}{\ensuremath{\Gamma}}
\newcommand{\maxEndCompMDP}{\ensuremath{\Gamma^{MEC}}}
\newcommand{\MDPtoMC}[2]{\ensuremath{{#1}^{[{#2}]}}}

\newcommand{\nat}{\ensuremath{\mathbb{N}} }

\newcommand{\rat}{\ensuremath{\mathbb{Q}} }
\newcommand{\reals}{\ensuremath{\mathbb{R}} }

\newcommand{\play}{\ensuremath{\pi} }

\newcommand{\last}[1]{\ensuremath{\mathsf{Last}(#1)} }

\newcommand{\mpay}{\ensuremath{\mathsf{MP}} }

\newcommand{\sizeMin}{\ensuremath{l_{max}} }

\newcommand{\goodWindowMPObj}{\ensuremath{\mathsf{GW}(\lambda, \sizeMin)} }
\newcommand{\directFixedWindowMPObj}{\ensuremath{\mathsf{DirFixWMP}(\lambda, \sizeMin)} }
\newcommand{\directBoundedWindowMPObj}{\ensuremath{\mathsf{DirBWMP}(\lambda)} }
\newcommand{\fixedWindowMPObj}{\ensuremath{\mathsf{FixWMP}(\lambda, \sizeMin)} }
\newcommand{\boundedWindowMPObj}{\ensuremath{\mathsf{BWMP}(\lambda)} }

\newcommand{\wmp}{\ensuremath{\mathsf{WMP}} }
\newcommand{\wtp}{\ensuremath{\mathsf{WTP}} }

\newcommand{\fnBoundedWindow}{\ensuremath{\mathsf{f_{BWMP}}} }
\newcommand{\fnFixedWindow}{\ensuremath{\mathsf{f_{FixWMP}^{\sizeMin}}} }

\newcommand{\fnDirFixedWindow}{\ensuremath{\mathsf{f_{DirFixWMP}^{\sizeMin}}} }

\newcommand{\goodWindowMCObj}{\ensuremath{\mathsf{GWC}(\lambda, \sizeMin)} }
\newcommand{\directFixedWindowMCObj}{\ensuremath{\mathsf{DirFixWMC}(\lambda, \sizeMin)} }
\newcommand{\directBoundedWindowMCObj}{\ensuremath{\mathsf{DirBWMC}(\lambda)} }
\newcommand{\fixedWindowMCObj}{\ensuremath{\mathsf{FixWMC}(\lambda, \sizeMin)} }
\newcommand{\boundedWindowMCObj}{\ensuremath{\mathsf{BWMC}(\lambda)} }

\newcommand{\wmc}{\ensuremath{\mathsf{WMC}} }
\newcommand{\wtc}{\ensuremath{\mathsf{WTC}} }

\newcommand{\playSuffix}[1]{\ensuremath{\play(#1,\infty)} }

\newcommand{\dists}{\ensuremath{\mathcal{D}} }

\newcommand{\expect}{\ensuremath{\mathbb{E}} }

\newcommand{\poly}{\ensuremath{\mathsf{poly}} }

\newcommand{\fpaths}{\ensuremath{\mathsf{Fpaths}}}

\newcommand{\zug}[1]{\langle #1  \rangle}
\newcommand{\stam}[1]{}

\newcommand{\realpos}{\mbox{I$\!$R}_{\ge 0}}

\title{Expected Window Mean-Payoff}

\author{Benjamin Bordais\\ENS Rennes \and Shibashis Guha\\Université libre de Bruxelles \and Jean-François Raskin\\Université libre de Bruxelles}

\begin{document}

\maketitle

	\begin{abstract}
In the window mean-payoff objective, given an infinite path, instead of considering a long run average, we consider the minimum payoff that can be ensured at every position of the path over a finite window that slides over the entire path.
In \cite{CDRR15}, the problem to decide if in a two-player game, Player $1$ has a strategy to ensure a window mean-payoff of at least $0$, has been studied.

In this work, we consider a function that given a path returns the supremum value of the window mean-payoff that can be ensured over the path and we show how to compute its expected value in Markov chains and Markov decision processes. We consider two variants of the function: \emph{Fixed window mean-payoff} in which a fixed window length $\sizeMin$ is provided; and \emph{Bounded window mean-payoff} in which we compute the maximum possible value of the window mean-payoff over all possible window lengths.
Further, for both variants, we consider (i) a \emph{direct} version of the problem where for each path, the payoff that can be ensured from its very beginning and (ii) a \emph{non-direct} version that is the prefix independent counterpart of the direct version of the problem.

In the case of Markov chains, we show that computing the expected value of the fixed window mean-payoff is polynomial in the size of the Markov chain and in the window size $\sizeMin$, that is, when $\sizeMin$ is given in unary, while the bounded window mean-payoff problem is polynomial in the size of the Markov chain.
In the case of MDPs, computing an optimal strategy that maximizes the expected value of the non-direct fixed window mean-payoff is polynomial in the size of the MDP and the window size $\sizeMin$, while the direct version of the problem is polynomial in the size of the MDP and exponential in $\sizeMin$.
As a lower bound, for the non-direct fixed objective, we show that the problem is at least as hard as two-player window mean-payoff games, while for the direct version, we show that the problem is {\sc PSpace}-hard even when $\sizeMin$ is given in unary implying that one cannot have an algorithm that is polynomial in $\sizeMin$ unless P={\sc PSpace}.
For bounded window mean-payoff problem, we show that computing the optimal strategy to maximize the expected value of the bounded window mean-payoff is in NP $\cap$ coNP and is at least as hard as two-player mean-payoff games.
Finally, we show that the optimal value of the expected bounded window mean-payoff in an MDP equals the supremum of the fixed window mean-payoff over all $\sizeMin$ and all strategies.
\end{abstract}
	
	\section{Introduction} \label{sec:intro}

Markov Decision processes (MDPs) are a classical model for decision-making inside stochastic environments~\cite{Put94,BK08}.
In this model, a stochastic model of the environment is formalized and we aim at finding strategies that maximize the expected performance of the system with that stochastic environment. This performance in turn is formalized by a function that maps each infinite path in the MDP to a value. One classical such function is the mean-payoff function that maps an infinite path to the limit of the means of the payoffs obtained on its prefixes. While this measure is classical, alternatives to the mean-payoff measure have been studied in the literature, e.g. one of the most studied alternative notion is the notion of discounted sum~\cite{Put94}. The main drawback of the mean-payoff value is that it does not guarantee local stability of the values along the path: if the limit mean-value of an infinite path converges to value $v$, it may be the case that for arbitrarily long infixes of the infinite path, the mean-payoff of the infix is largely away from this value. There have been several recent contributions~\cite{CDRR13,BCFK17,CDRR15,BFKN16} in the literature to deal with this possible fluctuations from the mean-payoff value along a path. In this paper, we study the notion of window mean-payoff that was introduced in~\cite{CDRR13,CDRR15} for two player games, in the context of MDPs, and we provide algorithms and prove computational complexities for the expected value of window mean-payoff objectives.

As introduced in~\cite{CDRR15}, in a window mean-payoff objective instead of the limit of the mean-payoffs along the whole sequence of payoffs, we consider payoffs over a local finite length window sliding along the infinite sequence: the objective asks that the mean-payoff must always reach a given threshold within the window length $\sizeMin$. This objective is clearly a strengthening of the mean-payoff objective: for all lengths $\sizeMin$, and all infinite sequences $\pi$, payoffs that satisfy the window mean-payoff objective for threshold $\lambda$ imply that $\pi$ has a mean-payoff value of at least $\lambda$. It can also be shown that this additional stability property can always be met at the cost of a small degradation of mean-payoff performances in two-player games: whenever there exists a strategy that forces a mean-payoff value $v$ against any behavior of the adversary then for every $\epsilon > 0$, there is a window length $\sizeMin$ and a strategy that ensure that the window mean-payoff objective for threshold $v-\epsilon$ is eventually satisfied for windows of length $\sizeMin$ (see Proposition~1 in~\cite{HPR18}).

In this paper, we study how to maximize the expected value of the window mean-payoff function ${\sf f^{\sizeMin}_{WMP}}$ in MDPs. The value of an infinite sequence of integer values $\pi : \mathbb{N} \rightarrow \mathbb{Z}$ for this function is defined as follows: 

\begin{equation}
{\sf f^{\sizeMin}_{WMP}}(\pi)=\sup \{  \lambda \in \mathbb{R} \mid \forall i \in \mathbb{N} : \lambda=\max_{1 \leq j \leq i} \frac{1}{j} \sum_{k=0}^{j-1} \pi(i+k) \geq \lambda \}
\end{equation}

\noindent
i.e., it returns the supremum of all window mean-payoff thresholds that are enforced by the sequence of payoffs $\pi$ over the window length $\sizeMin$. As in~\cite{HPR18}, we study natural variants of this measure: $(i)$ when the length of the window is fixed or it is left unspecified but needs to be bounded, and $(ii)$ when the window property needs to be true from the beginning of the path, or a prefix independent variant which asks the window property to eventually hold from some position in the path (leading to a prefix independent variant).

\paragraph{Main contributions}
Our results are as follows. 
First, for the prefix independent version of the measure ${\sf f^{\sizeMin}_{WMP}}$ and for a fixed window length $\sizeMin$, we provide an algorithm to compute the best expected value of ${\sf f^{\sizeMin}_{WMP}}$ with a time complexity that is polynomial in the size of the MDP $M$ and in $\sizeMin$ (Theorem~\ref{complexity_mdp_fixed}). It is worth to note that, since the main motivation for introducing window mean-payoff objective is to ensure strong stability over reasonable period of time, it is very natural to assume that $\sizeMin$ is bounded polynomially by the size of the the MDP $M$. This in turn implies that our algorithm is fully polynomial for the most interesting cases. We also note that this complexity matches the complexity of computing the value of the function ${\sf f^{\sizeMin}_{WMP}}$ for two-player games~\cite{CDRR15}, and we provide a relative hardness result: the problem of deciding the existence of a winning strategy in a window mean-payoff game can be reduced to the problem of deciding if the maximal expected mean-payoff value of a MDP for ${\sf f^{\sizeMin}_{WMP}}$ is larger than or equal to a given threshold $\lambda$ (Theorem~\ref{fix_mdp_hardness}).

 Second, we consider the case where the length $\sizeMin$ in the measure ${\sf f^{\sizeMin}_{WMP}}$ is not fixed but only required to be bounded. In that case, we provide an algorithm which is in~{\sc NP}$\cap${\sc coNP} (Theorem~~\ref{complexity_mdp_bounded}). In addition, we show that providing a polynomial time solution to our problem would also provide a polynomial time solution to the value problem in mean-payoff games (Theorem~\ref{bounded_mdp_hardness}), this is a long-standing open problem in the area~\cite{ZP96}.

 Third, we consider the case where the good window property needs to be imposed from the start of the path. In that case, surprisingly, the problem of computing if there is a strategy to obtain an expected value above a threshold $\lambda$ is provably harder than for two-player games unless {\sc P=PSpace}. Indeed, while the threshold problem for the worst-case value can be solved in time polynomial in the size of the game and in $\sizeMin$, we show that for the expected value in a MDP, the problem is {\sc PSpace-Hard} even if $\sizeMin$ is given in unary (Theorem~\ref{direct_hardness}). To solve the problem, we provide an algorithm that executes in time that is polynomial in the size of the MDP, polynomial in the largest payoff appearing in the MDP, and exponential in the length $\sizeMin$ (Theorem~\ref{complexity_mdp_direct_fixed}).

 Finally, while our main results concentrate on MDP, we also systematically provide results for the special case of Markov chains.

\paragraph{Related Works} As already mentioned, the window mean-payoff objective was introduced in~\cite{CDRR13} for two-player games. We show in this paper that the complexity of computing maximal expected value for the window mean-payoff value function is closely related to the computation of the worst-case value of a game inside end-components of the MDP (see Theorem~\ref{val_MEC} and~\ref{val_MEC_bounded}) for the prefix independant version of our objective. For the non-prefix independent version, surprinsingly, computing the expected value for MDP seems computationally more demanding than computing the mean-payoff value for games (unless {\sc P=PSpace}.).  The window mean-payoff objectives were also considered in games with imperfect information in~\cite{HPR18}, and in combinaison with omega-regular constraints in~\cite{BHR16}.

Stability issues of the mean-payoff measure have been studied in several contributions. In~\cite{BCFK17}, the authors study MDP where the objective is to optimize the expected mean-payoff performance and stability. They propose alternative definitions to the classical notions of statistical variance. The notion of stability that is offered by window mean-payoff objective and that has been studied in this paper is stronger than the one proposed in~\cite{BCFK17}. The techniques needed to solve the two problems are very different too as they mainly rely on solving sets of quadratic constraints.

In~\cite{BFKN16}, window-stability objectives have been introduced. Those objectives are inspired from the window mean-payoff objective of~\cite{BCFK17} but they are different in that they do not enjoy the so called inductive window property because of the stricter stability constraints that those objectives impose. The authors have considered the window-stability objectives in the context of games (2 players) and graphs (1 player) but they did not consider the case of MDP ($1\frac{1}{2}$ players).

MDP with classical mean-payoff objectives have been extensively studied both for the probabilistic threshold and the expectation payoff problem, see e.g.~\cite{Put94}. Combination of both type of constraints have been considered in~\cite{BBCFK14}.

\paragraph{Structure of the paper}
Section \ref{sec:prelims} introduces the  definitions and the formal concepts used in this paper.
In Section \ref{sec:MC}, we study the expected window mean-payoff problems in weighted Markov chains, while in Section \ref{sec:MDP}, we study the problems for weighted Markov decision processes.
We give algorithms to solve the problems, as well as we give hardness results for the problems we study.
In Section \ref{sec:relate}, we show that in an MDP, the value of the bounded window mean-payoff problem equals the supremum of the fixed window mean-payoff problem over all window lengths and over all strategies.
Section \ref{sec:WMC} defines window mean-cost instead of window mean-payoff where the objective is to minimize the cost as opposed to maximizing payoff.
Finally, we conclude in Section \ref{sec:conclusion} with a summary of the complexity and the hardness results that we obtain for both Markov chains and MDPs.
Some details related to some of the proofs and an additional algorithms for the direct fixed window mean-payoff problem in Markov chains have been moved to the appendix.

	\section{Preliminaries} \label{sec:prelims}
		\subsection{Weighted Markov Chains}
		%$M = \zug{V, E, Prob}$, 
Given a finite set $A$, a (rational) \textit{probability distribution} over $A$ is a function
$Pr \colon A \rightarrow [0, 1] \cap \rat$ such that $\sum_{a\in A} Pr(a) = 1$. 
We denote the set of probability distributions over $A$ by $\dists(A)$. 
%The \textit{support} of the probability distribution $\dist$ on $A$ is $\supp(\dist) = \left\lbrace a \in A \;\vert\; \dist(a) > 0\right\rbrace$. 
%A distribution is called \emph{Dirac} if $|\supp(\dist)| = 1$.

A finite weighted \emph{Markov chain} (MC, for short) is a tuple $\mathcal{M} = \zug{S,E,s_{init},w,\mathbb{P}}$, where $S$ is the finite set of states, $s_{init} \in S$ is the initial state, $E \subseteq S \times S$ is the set of edges, the function $w:E\mapsto \rat$ defines the \emph{weights} of the edges, and $\mathbb{P}: S \rightarrow \dists(E)$ is a function that assigns a probability distribution  -- on the set $E(s)$ of outgoing edges from $s$ -- to all states $s \in S$.
A Markov chain is infinite if the set $S$ is not finite.
Unless stated otherwise, we always consider a Markov chain to be finite.

Given a state $s \in S$, we define the \emph{support} of $s$ to be $\mathsf{Supp}(s) = \{s' \in S \mid (s,s') \in E\}$, that is, every edge has a non-zero probability. Then, for a state $s \in S$, we define the set of infinite paths in $\mathcal{M}$ starting from $s$ as $Paths^{\mathcal{M}}(s) = \lbrace \pi = s_0 s_1 \ldots \in S^\omega \mid s_0 = s, \forall\, n \in \nat,\, s_{n+1}\in\mathsf{Supp}(s_n) \rbrace$, and set of infinite paths in the whole Markov chain is denoted $Paths^{\mathcal{M}}$ and is equal to $\bigcup_{s \in S} Paths^{\mathcal{M}}(s)$. Then, the set of finite paths from a state $s$ is denoted $FPaths^{\mathcal{M}}(s)$ and corresponds to the set of all prefixes of all the paths in $Paths^{\mathcal{M}}(s)$. Similarly, the set of finite paths in the whole Markov chain $FPaths^{\mathcal{M}}$ is equal to $\bigcup_{s \in S} FPaths^{\mathcal{M}}(s)$. We denote by $\last{\rho}$ the last vertex $s_n$ of the finite path $\rho = s_0 \dots s_n$. Finally, the set paths of length $l \geq 1$ starting from $s$ can be defined as $FPaths^{\mathcal{M}}_{l}(s) = \lbrace \pi \in FPaths^{\mathcal{M}}(s) \ \mid \ |\pi| = l \rbrace$, where $|\pi|$ denotes the number of edges in $\pi$. The set of finite paths of length $l$ in $\mathcal{M}$, that is $FPaths^{\mathcal{M}}_{\sizeMin}$, is equal to $\bigcup \limits_{s \in S} FPaths^{\mathcal{M}}_{\sizeMin}(s)$.

If we consider a weighted Markov chain $\mathcal{M} = \langle S,E,s_{init},w,\mathbb{P} \rangle$ and a path $\rho \in FPaths^{\mathcal{M}}$, we denote by $Cyl(\rho)$ the cylinder set generated by $\rho$ and it is defined as $Cyl(\rho) = \lbrace \pi \in Paths^{\mathcal{M}} \mid \rho \text{ is a prefix of } \pi \rbrace$. The interesting property of these cylinder sets is that, for any path $\rho \in FPaths^{\mathcal{M}}$, we have: $Pr(Cyl(\rho)) = \mathbb{P}(\rho)$.

The bottom strongly connected components (BSCCs for short) in a Markov chain $\mathcal{M}$ are the strongly connected components from which it is impossible to exit. That is, a BSCC $\mathcal{B}$ ensures that $\forall s \in \mathcal{B},\; Supp(s) \subseteq \mathcal{B}$.  We denote by $BSCC(\mathcal{M})$ the set of BSCCs of the Markov chain $\mathcal{M}$.
A nice property of these components is that every infinite path $\pi \in Paths^{\mathcal{M}}$ eventually ends up in one of the BSCCs almost surely. Formally, we have the proposition:
\begin{proposition}
	\label{eventually_bscc}
	For all state $s \in S$, we have: $Pr(\pi \in Paths^{\mathcal{M}}(s) \mid \exists \mathcal{B} \in BSCC(\mathcal{M}),\; \pi \models \lozenge \square \mathcal{B}) = 1$.
\end{proposition} 

To compute the different expected values we are interested in, we will need some computations on Markov chains. 
More specifically, given a weighted Markov chain $\mathcal{M} = \langle S, E,s_{init},w,\mathbb{P} \rangle$, we would like to compute $Pr_s(\lbrace \pi \mid \pi \models \lozenge B \rbrace)$ or $Pr_s(\lbrace \pi \mid \pi \models C \mathcal{U} B \rbrace)$ from a state $s \in S$, where $B$ and $C$ are two sets of states, $\lozenge B$ denotes eventually reaching some state in $B$ and $C U B$ denotes until some state in $B$ is reached, only states in $C$ are visited. 
In the following, $lbrace \pi \mid \pi \models \lozenge B \rbrace$ and $\lbrace \pi \mid \pi \models C \mathcal{U} B \rbrace$ are denoted $\lozenge B$ and $B \mathcal{U} C$ respectively. The corresponding probabilities can be computed by solving a linear equation system (see \cite{BK08}). By using the Gauss-Jordan method, we have an algorithm that computes these probabilities in time $O(|S|^3)$. We denote by $Mat(|S|)$ the time to computing these probabilities.

\subsection{Weighted Markov Decision Processes}
\label{prelim_mdp}
A finite weighted \emph{Markov decision process} (MDP, for short) is a tuple $\Gamma = \langle S,E,Act,s_{init},w,\mathbb{P} \rangle$, where $S$ is the set of states, $s_{init} \in S$ is the initial state of this Markov decision process, $Act$ is the set of actions, and $E \subseteq S \times Act \times S$ is set of edges. The function $w:E\mapsto \rat$ defines the \emph{weights} of the edges, and $\mathbb{P}: S \times Act \rightarrow \dists(E)$ is a partial function that assigns a probability distribution -- on the set $E(s,a)$ of outgoing edges from $s$ -- to all states $s \in S$ if action $a \in Act$ is taken in $s$. Given a state $s \in S$, we denote by $Act(s)$ the set of actions $\lbrace a \in Act \mid \exists s' \in S,\; \mathbb{P}(s,a,s') > 0 \rbrace$. Then, given an action $a \in Act(s)$, we define $\mathsf{Post}(s,a) = \lbrace s' \in S \mid \mathbb{P}(s,a,s') > 0 \rbrace$. We assume that, for all $s \in S$, $Act(s) \neq \emptyset$.

A strategy in $\Gamma$ is a function $\sigma: S^{+} \longrightarrow Act$ such that $\sigma(s_0 \ldots s_n) \in Act(s_n)$, for all $s_0 \ldots s_n \in S^{+}$. A finite-memory strategy can be seen as tuple $\langle M,act,\delta,start \rangle$ where:
\begin{itemize}
	\item $M$ is a set of modes;
	\item $act: M \times S \rightarrow Act$ selects an action such that, for all $m \in M$ and $s \in S$, $act(m,s) \in Act(s)$;
	\item $\delta: M \times S \rightarrow M$ is a mode update function;
	\item $start: S \rightarrow M$ selects the initial mode for any state $s \in S$.
\end{itemize}
The amount of memory used by such a strategy is defined to be $|M|$. 
Once we fix a strategy in an MDP, we obtain an MC.
If the strategy is not finite memory, we may obtain an infinite Markov chain.
For $\sigma \in strat(\Gamma)$, we denote by $\MDPtoMC{\Gamma}{\sigma}$ the Markov chain that we obtain by applying strategy $\sigma$ to $\Gamma$. 
A strategy is said to be \emph{memoryless} if $|M| = 1$, that is, the choice of action only depends on the current state where the choice is made. 
Formally, a strategy is said to be memoryless if for all finite paths $\rho_1$ and $\rho_2$ in $FPaths^{\MDPtoMC{\Gamma}{\sigma}}$ such that $Last(\rho_1) = Last(\rho_2)$, we have $\sigma(\rho_1) = \sigma(\rho_2)$.
We denote by $strat(\Gamma)$ the set of all strategies, and by $strat_0(\Gamma)$ the set of all memoryless strategies. Note that this set is finite.

We now define maximal end components in an MDP.
To do this, we first define the notion of sub-MDP and the directed graph induced by it.		
For an MDP $\Gamma = \langle S,E,Act,s_{init},w,\mathbb{P} \rangle$, a pair $(T,A)$ where $\emptyset \neq T \subseteq S$ and $A: T \longrightarrow 2^{Act}$ such that:
\begin{itemize}
	\item $\emptyset \neq A(s) \subseteq Act(s)$ for all state s $\in S$;
	\item for $s \in T,\; a \in A(s)$, we have $Post(s,a) \subseteq T$;
\end{itemize}
is called a sub-MDP of $\Gamma$. A sub-MDP $(T,A)$ induces a directed graph $\graph_{(T,A)}$ whose vertex are $T \cup \lbrace (s,a) \in T \times A(s) \rbrace$ and whose edges are $\lbrace (s,(s,a)) \mid s \in T, a \in A(s) \rbrace \cup \lbrace ((s,a),t) \mid t \in Post(s,a) \rbrace$. 
Then, an \emph{end-component} is a sub-MDP whose induced graph is strongly connected. Finally, a \emph{maximal end component} (MEC, for short) is an end component that is included in no other end component. We denote by $MEC(\Gamma)$ the set of all maximal end components of $\Gamma$. Computing the set $MEC(\Gamma)$ can be done in time $O(|S| \cdot (|S| + |E|))$ (see \cite{BK08}). Any infinite path will eventually end up in one maximal end component almost surely, whatever strategy is considered. 
%Maximal end components in MDPs are what stand for BSCCs in MCs.
That is, we have the following proposition:
\begin{proposition}
	\label{eventually_mec}
	For all strategy $\sigma \in \textsf{strat}(\Gamma)$, for all state $s \in T$, we have: $Pr(\pi \in Paths^{\MDPtoMC{\Gamma}{\sigma}}(s) \mid \exists M \in MEC(\Gamma),\; \pi \models \lozenge \square M) = 1$.
\end{proposition} 

In several cases, we will need to compute the (optimal) expected value of the mean-payoff in an MDP $\Gamma = \langle S,E,Act,s_{init},w,\mathbb{P} \rangle$. This can be done in polynomial time (see \cite{Put94}). In the following, we will denote the complexity of computing the optimal expected value in $\Gamma$ by $Avg(|S|,|Act|)$.

The size of the MDP $\Gamma$, that is denoted $|\Gamma|$, corresponds to the value $|S| + |E| + \log(W)$ where $W = \max_{e \in E} w(e)$.
\subsection{Weighted Two-Player Games}
\label{subsec_prelim_game}
We introduce here the notion of a finite weighted two-player game. 
%We do not study this structure, but we will use some algorithms that already exist (especially from\cite{CDRR15}) to solve the problems we are interested in inside Markov chains and Markov decision processes.

A finite two-player game is a weighted graph $G = \langle S_1,S_2,s_{init},E,w \rangle$ where the set of vertices $S = S_1 \uplus S_2$ is partitioned into the vertices belonging to Player 1, that is $S_1$, and the vertices belonging to Player 2, that is $S_2$, and $s_{init} \in S_1$ is the initial vertex.
The set of edges $E \subseteq S \times S$ is such that for all $s \in S$, there exists $s' \in S$ such that $(s,s') \in E$. The weight function $w$ is such that $w: E \rightarrow \rat$.

The strategies in a two-player game are analogous to the strategies in an MDP, except that Player $1$ chooses an edge from the states in $S_1$, while Player 2 chooses an edge from the states in $S_2$
%		, and they both chooses directly the next edge to take 
(instead of choosing an action, that leads to a probabilistic distribution over the edges as in an MDP). 
We denote by $strat_1(G) = \lbrace \sigma: S^* S_1 \mapsto E \rbrace$ and $strat_2(G) = \lbrace \sigma: S^* S_2 \mapsto E \rbrace$ the set of all strategies of Player 1 and 2 respectively available in the game $G$.

Given an MDP $\Gamma = \langle S,E,Act,s_{init},w,\mathbb{P} \rangle$, we denote by $G_{\Gamma}$ the two-player game $\langle S_1,S_2,s_{init}, E',w' \rangle$ \emph{resulting from} $\Gamma$ where:
\begin{itemize}
	\item $S_1 = S$;
	\item $S_2 = S \times Act$;
	\item $E' = \underbrace{\lbrace (s,(s,a)) \in S_1 \times S_2 \mid s \in S,\; a \in Act(s) \rbrace}_{\text{denoted } E_1} \cup \underbrace{\lbrace ((s,a),s') \in S_2 \times S_1,\; \mathbb{P}(s,a,s') > 0 \mid \rbrace}_{\text{denoted } E_2}$;
	\item 
	$\left\{
	\begin{array}{l c}
	w'(e) = 0 & \text{ if } e \in E_1\\
	w'(e) = w(e) & \text{ if } e \in E_2
	\end{array}
	\right.
	$
\end{itemize}

It has to be noted that any strategy $\sigma_1$ of Player 1 in $G_\Gamma$ can be seen as a strategy in the MDP $\Gamma$. Reciprocally, any strategy $\sigma$ in the MDP $\Gamma$ can be seen as a strategy for Player 1 in $G_\Gamma$. Therefore, we have $strat(\Gamma) = strat_1(G_\Gamma)$.

In a two-player game, there is no randomness. Therefore, once the deterministic strategies are fixed, there is a unique path that can occurs n the game, from the initial state $s$. For a game $G$, given two strategies $\sigma_1 \in strat_1(G)$ and $\sigma_2 \in strat_2(G)$, we denote by $\pi_{(G,s,\sigma_1,\sigma_2)}$ the path that occur in the two-player game $G$ under strategies $\sigma_1$ and $\sigma_2$ from state $s$. Then, if we consider a function $f$ that associate a value to any infinite path of a two-player game, we denote by $V_s^{f}(G,\sigma_1,\sigma_2)$ the value $f(\pi_{(G,s,\sigma_1,\sigma_2)})$.

%		For a strategy $\sigma_1 \in strat_1(G_\Gamma)$ for some MDP $\Gamma$, we define the set of paths that could occur in $G$ from $s$ if strategy $\sigma_1$ is player by Player 1: $Paths^{G,\sigma_1} = \lbrace \pi_{(G,s,\sigma_1,\sigma_2)} \mid \sigma_2 \in strat_2(G) \rbrace$. 
In the two-player game $G_\Gamma$ resulting from an MDP $\Gamma$, the set of paths from a state $s$ that may occur when Player $1$ chooses a strategy $\sigma_1 \in G_\Gamma$ from $s$ is defined as $Paths^{G,\sigma_1}(s) = \lbrace \pi_{(G,s,\sigma_1,\sigma_2)} \mid \sigma_2 \in strat_2(G) \rbrace$. Then, we may consider the function $p_s^{\sigma_1}: Paths^{\MDPtoMC{\Gamma}{\sigma_1}}(s) \longrightarrow Paths^{G_\Gamma,\sigma_1}(s)$ such that for $\pi = s_0 s_1\ldots \in Paths^{\MDPtoMC{\Gamma}{\sigma_1}}(s)$ where $s_0=s$, we have $p_s^{\sigma_1}(\pi) = t_0 t_1\ldots \in Paths^{G_\Gamma,\sigma_1}(s)$ with 
\begin{itemize}
	\item $t_0 = s_0 \in S_1$
	\item $\forall i \geq 1,\;$
	\begin{itemize}
		\item $t_{2 \cdot i - 1} = (s_{i-1},\sigma_1(s_0 \ldots s_{i-1})) \in S_2$
		\item $t_{2 \cdot i} = (s_i) \in S_1$
	\end{itemize}
\end{itemize}	
That is, $p_s^{\sigma_1}$ associate to a path $\pi \in Paths^{\MDPtoMC{\Gamma}{\sigma_1}}(s)$ its counterpart in the game $G_\Gamma$ where Player 1 opts for strategy $\sigma_1$.
%For all strategy $\sigma_1 \in strat_1(G_\Gamma)$, $g^{\sigma_1}$ is a bijection.

\subsection{Further Notations}
For a Markov chain $\mathcal{M} = \zug{S,E,s_{init},w,\mathbb{P}}$, a path $\pi = s_0 s_1 \ldots \in Paths^{\mathcal{M}}$ and $i,l \in \nat$, by $\pi(i \ldots (i+l))$ we refer to the sequence of $l+1$ states $s_i s_{i+1} \ldots s_{i+l}$ (that is also a sequence of $l$ edges) and by $\mathbb{P}(\pi(i \ldots (i+l)))$ we refer to $\prod\limits_{j=0}^{l-1} \mathbb{P}(s_i,s_{i+1})$. 
Consider some measurable function $f: Paths^{\mathcal{M}}(s_{init}) \rightarrow \reals$ associating a value to each infinite path starting from $s_{init}$. 
For an interval $I \subset \reals$, we denote by $f^{-1}(\mathcal{M},s_{init},I)$ the set $\lbrace \pi \in Paths^{\mathcal{M}}(s_{init}) \mid f(\pi) \in I \rbrace$, and for $r \in \reals,\; f^{-1}(\mathcal{M},s_{init},r)$ refers to $f^{-1}(\mathcal{M},s_{init},[r,r])$. 
Since the set of paths $Paths^\mathcal{M}(s_{init})$ forms a probability space, and $f$ is a random variable, we denote by $\expect^{\mathcal{M}}_{s_{init}}(f)$ the \textit{expected value} of $f$ over the set of paths starting from
$s_{init}$.

Finally, for $k \in \mathbb{N}$, we denote respectively by $[k]_0$ and $[k]$ the set of natural numbers $\lbrace 0, \ldots, k \rbrace$ and $\lbrace 1, \ldots, k \rbrace$ respectively.

\subsection{Decision Problems}
Consider a set $E$, a payoff function $w: E \mapsto \rat$, and $\rho = e_0 \ldots e_{l-1} \in E^l$. We first define the function $WTP$ (that stands for \textit{Window Total-Payoff}) such that: 
\begin{displaymath}
	{\it WTP}(\rho,l) = \max_{k \in [l]} \sum_{i = 1}^{k} w(e_i)
\end{displaymath} 
The value $WTP(\rho)$ is the maximum total payoff one can ensure over a window of length $k \in [l]$ starting from %$s_0$. 
the initial vertex of the sequence of edges.

Similarly, we define $WMP$ (that stands for \textit{Window Mean-Payoff}) such that:
\begin{displaymath}
	{\it WMP}(\rho,l) = \max_{k \in [l]} \frac{1}{k} \sum_{i = 1}^{k} w(e_i)
\end{displaymath} 
The value $WMP(\rho)$ is the maximum mean-payoff one can ensure over a window of length $k \in [l]$. For a given infinite path $\pi$, a threshold $\lambda \in \rat$, a position $i \in \nat$ and $l \in [\sizeMin]$, we say that the window $\pi(i \ldots (i+l))$ is \emph{closed} if $WMP(\pi(i \ldots (i+l))) \geq \lambda$. Otherwise, the window is \emph{open}.
We note that $WMP(\rho,l') \ge WMP(\rho,l)$ for $l' \ge l$.

Given a Markov chain $\mathcal{M}$ with an initial state $s_{init}$ and a rational threshold $\lambda \in \rat$, we define the following objectives.
\begin{itemize}
	\item Given $\sizeMin \in \nat$, the \textit{good window} objective
	
	\begin{align}	
	\goodWindowMPObj = \Big\lbrace\play \in Paths^{\mathcal{M}}(s_{init}) \;\vert\; \lambda \leq WMP(\pi(0 \ldots \sizeMin),l)\Big\rbrace\label{eq:goodWindowObj},
	\end{align}
	
	requires that there exists a window starting in the first position of $\pi$ and of size at most $\sizeMin$ over which the mean-payoff is bounded below by the threshold $\lambda$.
	Again, $\mathsf{GW}(\lambda, \sizeMin') \supseteq \goodWindowMPObj$ for $\sizeMin' \ge \sizeMin$.
	
	\item Given $\sizeMin \in \nat$, the \textit{direct fixed window mean-payoff} objective
	
	\begin{align}	
	\directFixedWindowMPObj =  \Big\lbrace\play \in Paths^{\mathcal{M}}(s_{init}) \;\vert\; \forall\, j \geq 0,\; \playSuffix{j} \in \goodWindowMPObj \Big\rbrace\label{eq:directFixedWindowObj}
	\end{align}
	requires that good windows of size at most $\sizeMin$ exist in all positions along the path.
	
	\item The \textit{direct bounded window mean-payoff} objective
	
	\begin{align}	
	\directBoundedWindowMPObj =  \Big\lbrace\play \in Paths^{\mathcal{M}}(s_{init}) \;\vert\; \exists \sizeMin>0,\; \play \in \directFixedWindowMPObj \Big\rbrace\label{eq:directBoundeddWindowObj}
	\end{align}
	requires that there exists a bound $\sizeMin$ such that the path satisfies the direct fixed objective for the length $\sizeMin$.
	
	\item Given $\sizeMin \in \nat$, the \textit{fixed window mean-payoff}  objective
	\begin{align}
	\fixedWindowMPObj =  \Big\lbrace\play \in Paths^{\mathcal{M}}(s_{init}) \;\vert\; \exists\, i \in \mathbb{N},\; \playSuffix{i} \in \directFixedWindowMPObj \Big\rbrace\label{eq:fixedWindowObj}
	\end{align}
	is the \textit{prefix-independent} version of the direct fixed window objective: it requires for the existence of a suffix of the path satisfying it.
	
	\item The \textit{bounded window mean-payoff} objective
	
	\begin{align}	
	\boundedWindowMPObj =  \Big\lbrace\play \in Paths^{\mathcal{M}}(s_{init}) \;\vert\; \exists \sizeMin>0,\; \play \in \fixedWindowMPObj \Big\rbrace\label{eq:boundedWindowObj}
	\end{align}
	is the prefix-independent version of the direct bounded window objective.
\end{itemize}

\subsection{Functions of Interest}
\label{functions of interest}
For each of those objective, we associate a value to every infinite path. We define the following functions, respectively for the \textit{fixed}, \textit{direct fixed}, \textit{bounded} and \textit{direct bounded} \textit{window mean-payoff} problem:
\begin{center}
	$
	\begin{array}{l c l}
	f^{\sizeMin}_{FixWMP}(\pi) & = & \sup \{\lambda \in \rat \mid \pi \in FixWMP(\lambda,\sizeMin)\}\\
	f^{\sizeMin}_{DirFixWMP}(\pi) & = & \sup \{\lambda  \in \rat \mid \pi \in DirFixWMP(\lambda,\sizeMin)\}\\
	f_{DirBWMP}(\pi) & = & \sup \{\lambda \in \rat \mid \pi \in DirBWMP(\lambda)\}\\
	f_{BWMP}(\pi) & = & \sup \{\lambda \in \rat \mid \pi \in BWMP(\lambda)\}\\
	\end{array}
	$
\end{center}
It is now possible to study the complexity of finding the expected value of these functions in a Markov chain and in a Markov decision process.
In the following, in the MCs and MDPs w.l.o.g. we consider only non-negative integer weights.
Note that if the weights belong to $\rat$, then one can multiply them with the LCM of their denominators to obtain integer weights.
Among the resultant set of integer weights, if the minimum integer weight $\kappa$ is negative, then we add -$\kappa$ to the weight of each edge so as to obtain weights that are natural numbers.

\subsection{A First Result}
The first observation is that the functions $f_{BWMP}$ and $f_{DirBWMP}$ always produce the same result. That is stated in the following theorem.
\begin{lemma}
	\label{th_first}
	Let $\mathcal{M} = \zug{S,E,s_{init},w,\mathbb{P}}$ be a Markov chain and let $\pi = s_0 \ldots \in Paths^\mathcal{M}$. Then:
	\begin{displaymath}
		f_{DirBWMP}(\pi) = 	f_{BWMP}(\pi)
	\end{displaymath}
\end{lemma}
%shibashis: It is difficult to undersand deficit.
%This comes from the fact that, for the direct bounded window mean-payoff function, the deficit that could occur by considering the payoffs from the very beginning of the paths is caught up since we consider arbitrarily large window lengths. 
It is easy to see that $f_{DirBWMP}(\pi) \leq f_{BWMP}(\pi)$.
Now for every $\varepsilon \in \realpos$, a window mean-payoff value of $f_{BWMP}(\pi)- \varepsilon$ can be ensured from the beginning of the path $\pi$ by considering appropriately large window length.
Since $f_{DirBWMP}(\pi)$ has been defined as the supremum of the window mean-payoff values that can be ensured along the path $\pi$ with arbitrarily large window lengths, the result follows.
The detailed proof is given in Appendix \ref{lem:dirBWMPequalsBWMP}.

	%\section{Expected value and complexity in a weighted Markov chain}
\section{Expected Window Mean-payoff in Markov Chains}
\label{sec:MC}
In this section, we study the fixed, direct fixed and bounded window mean-payoff problems on weighted Markov chains.
We show that while the bounded window mean-payoff problem can be solved in polynomial time, the algorithm for the fixed window mean-payoff problem is polynomial in the value of the window length, that is, it is polynomial when the window length is given in unary, and pseudopolynomial if it is given in binary. Then, the algorithm for the direct fixed window mean-payoff problem is polynomial in the value of the window length and the weights appearing in the Markov chains. Therefore, it is pseudopolynomial when they are given in binary.
\subsection{\textit{Fixed Window Mean-Payoff}}
We are interested in the expected value of $f^{\sizeMin}_{FixWMP}$ for a given $\sizeMin$ in a Markov chain. More specifically, consider a Markov chain $\mathcal{M}$ and an initial state $s_{init}$. Then, we study
\begin{displaymath}
\mathbb{E}_{s_{init}}^{\mathcal{M}}(f^{\sizeMin}_{FixWMP}) = \int_{r \in \mathbb{Q}} r \cdot Pr((f^{\sizeMin}_{FixWMP})^{-1}(\mathcal{M},s_{init},r))
\end{displaymath} 

In order to compute this value, first consider the bottom strongly connected components (BSCCs for short) of the Markov chain $\mathcal{M}$. In the preliminaries, we mentioned that any infinite path will eventually end up in one of them almost surely. Since the \textit{fixed window mean-payoff} is prefix independent, the value of a path $\pi$, that is $f^{\sizeMin}_{FixWMP}(\pi)$, is only determined by the BSCC in which it ends up. Moreover, we have the following result:

\begin{proposition}
	\label{BSCC_inf_often}
	Let $\mathcal{B}$ be a BSCC. Let $\rho = s_0 \ldots s_n \in FPaths^{\mathcal{B}}$. Then, every path in $\mathcal{B}$ will almost surely visit the sequence of states $\rho$ infinitely often. Formally:
	\begin{displaymath}
		Pr(\{ \pi \in Paths^{\mathcal{B}} \mid \pi \in \square \lozenge \rho\}) = 1
	\end{displaymath}	
\end{proposition}

\begin{proof}
	Consider an infinite path $\pi \in Paths^{\mathcal{B}}$. Then, because we are in a BSCC, $\pi$ will almost surely go infinitely often through $s_0$. Moreover, in $s_0$, there is a positive probability of taking the path $\rho$, that is, visiting the states in $\rho$ in sequence. It follows that $\pi$ will almost surely reach the sequence of states $\rho$ infinitely often. 
\end{proof}

In particular, in a BSCC $\mathcal{B} \in BSCC(\mathcal{M})$, the sequence of $\sizeMin + 1$ states $s_0 \ldots s_{\sizeMin} \in \mathcal{B}$ that minimizes $\wmp$ is visited infinitely often by any infinite path almost surely. Hence, the following theorem:
\begin{theorem} \label{thm:fixMC}
	The expected value of $f^{\sizeMin}_{FixWMP}$ over paths that are entirely contained in a BSCC does not depend on the starting state of the paths considered. Moreover, if we define the expected value in a BSCC $\mathcal{B}$ as the expected value starting from any state, then we have:
	\begin{displaymath}
	\mathbb{E}^{\mathcal{B}}(f^{\sizeMin}_{FixWMP}) = \underbrace{\min_{s \in \mathcal{B}} \min_{\pi \in FPaths^{\mathcal{M}}_{\sizeMin}(s)} \wmp(\pi(0 \ldots \sizeMin))}_{\text{denoted }m_{\mathcal{B}}}
	\end{displaymath} 
\end{theorem}

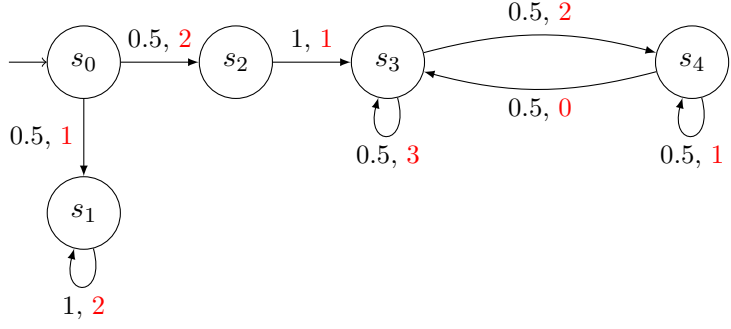
\begin{figure}
	\begin{center}
		\vspace{-15pt}
\centering
\scalebox{1}{
	\begin{tikzpicture}
		\node[player,initial,initial text={}] (sinit) at (0,0) {$s_0$} ;
		\node[player] (s1) at (0,-2) {$s_1$} ;
		\node[player] (s2) at (2,0) {$s_2$} ;
		\node[player] (s3) at (4,0) {$s_3$} ;
		\node[player] (s4) at (8,0) {$s_4$} ;
		
 \path[-latex]  (sinit) edge node[left] {\small 0.5, \textcolor{red}{1}} (s1)
				(sinit) edge node[above] {\small 0.5, \textcolor{red}{2}} (s2)
				(s1) edge[loop below] node[below] {\small 1, \textcolor{red}{2}} (s1)
				(s2) edge node[above] {\small 1, \textcolor{red}{1}} (s3)
				
				(s3) edge[bend left=15] node[above] {\small 0.5, \textcolor{red}{2}} (s4)
				(s3) edge[loop below] node[below] {\small 0.5, \textcolor{red}{3}} (s3)
				
				(s4) edge[bend left=15] node[below] {\small 0.5, \textcolor{red}{0}} (s3)
				(s4) edge[loop below] node[below] {\small 0.5, \textcolor{red}{1}} (s4)
		;
	\end{tikzpicture}
}
	\end{center}
	\caption{An example of a weighted Markov chain} 
	\label{eq_weighted_markov}
\end{figure}

Before proving this theorem, let us look at the example in Figure~\ref{eq_weighted_markov} of a weighted Markov chain to illustrate the definition of $m_{\mathcal{B}}$ for some BSCC $\mathcal{B}$. Such an example can be seen on Figure~\ref{eq_weighted_markov}. The probabilities are written next to each edge in black, the weights are in red. Let us consider the case where $\sizeMin = 2$. In this Markov chain, there are two BSCCs: $\mathcal{B}_1 = \{ s_1 \}$ and $\mathcal{B}_2 = \{ s_3, s_4 \}$. The value of $m_{\mathcal{B}_1}$ is obvious since there is only one path of length $\sizeMin = 2$ in $\mathcal{B}_1$: $\pi = s_1 s_1 s_1$ with $\wmp(\pi) = 2$. Let us now consider $m_{\mathcal{B}_2}$. In the BSCC $\mathcal{B}_2$, there are eight paths of length $\sizeMin = 2$.
\begin{itemize}
	\item From $s_3$: 
	\begin{itemize}
		\item $\pi_{3_1} = s_3 s_3 s_3$, with $\wmp(\pi_{3_1}) = 3$;
		\item $\pi_{3_2} = s_3 s_3 s_4$, with $\wmp(\pi_{3_2}) = 3$;
		\item $\pi_{3_3} = s_3 s_4 s_3$, with $\wmp(\pi_{3_3}) = 2$;
		\item $\pi_{3_2} = s_3 s_4 s_4$, with $\wmp(\pi_{3_2}) = 2$;.
	\end{itemize}
	\item From $s_4$: 
	\begin{itemize}
		\item $\pi_{4_1} = s_4 s_3 s_3$, with $\wmp(\pi_{4_1}) = 1.5$;
		\item $\pi_{4_2} = s_4 s_3 s_4$, with $\wmp(\pi_{4_2}) = 1$;
		\item $\pi_{4_3} = s_4 s_4 s_3$, with $\wmp(\pi_{4_3}) = 1$;
		\item $\pi_{4_4} = s_4 s_4 s_4$, with $\wmp(\pi_{4_4}) = 1$.
	\end{itemize}
\end{itemize}
The minimum over these paths is achieved for paths $\pi_{4_2}$, $\pi_{4_3}$ and $\pi_{4_4}$. Therefore, $m_{\mathcal{B}_2} = \wmp(\pi_{4_3}) = 1$. Moreover, since both BSCCs have the same probability ($0.5$) of being reached, by applying Formula~\ref{expectedvalue}, we obtain that the expected value of the fixed window mean-payoff in this Markov chain $\mathcal{M}$ is equal to $\mathbb{E}^{\mathcal{M}}(f^{\sizeMin}_{FixWMP}) = 0.5 \cdot 2 + 0.5 \cdot 1 = 1.5$.

\begin{proof}
	Let $s \in \mathcal{B}$. We prove:
	\begin{itemize}
		\item[1.] For all $p < m_{\mathcal{B}}$, we have $Pr((f^{\sizeMin}_{FixWMP})^{-1}(\mathcal{M},s_{init},p)) = 0$;
		\item[2.] For all $q > m_{\mathcal{B}}$, we have $Pr((f^{\sizeMin}_{FixWMP})^{-1}(\mathcal{M},s_{init},q)) = 0$.
	\end{itemize}
	\begin{itemize}
		\item[1.] Consider a path $\pi \in Paths^{\mathcal{M}}(s)$. We prove that $\pi \in FixWMP(m_{\mathcal{B}},\sizeMin)$. Let $i \in \mathbb{N}$ and let $m_i = \wmp(\pi(i \ldots (i+\sizeMin)))$. Then, by definition of $m_{\mathcal{B}}$, we have for all $i \in \mathbb{N}$, $m_i \geq m_{\mathcal{B}}$. Therefore, we have that $\pi \in FixWMP(m_{\mathcal{B}},\sizeMin)$. This implies that $f^{\sizeMin}_{FixWMP}(\pi) \geq m_{\mathcal{B}}$. Hence, for all $p < m_{\mathcal{B}}$, we have $(f^{\sizeMin}_{FixWMP})^{-1}(\mathcal{M},s,p) = \emptyset$ and therefore, $Pr((f^{\sizeMin}_{FixWMP})^{-1}(\mathcal{M},s,p)) = 0$. 
		
		\item[2.] Let us prove the second point. Let us denote by $\rho_{m_{\mathcal{B}}}$ the sequence of $\sizeMin + 1$ states such that $m_{\mathcal{B}} = \wmp(\rho_{m_\mathcal{B}})$. Because $\mathcal{B}$ is a BSCC and according to Proposition~\ref{BSCC_inf_often}, we know that $Pr(s \models \square \lozenge \rho_{m_\mathcal{B}}) = 1$, which is equivalent to say that the set $P_{\mathcal{B}} = \{\pi \in \textit{Path}(s) \mid \pi \models \square \lozenge \rho_{m_\mathcal{B}}\}$ of paths that reach infinitely often the sequence of states $\pi_{m_{\mathcal{B}}}$ is such that $Pr(P_\mathcal{B}) = 1$. Let $\pi \in P_{\mathcal{B}}$. Then $f^{\sizeMin}_{FixWMP}(\pi) \leq m_{\mathcal{B}}$. Hence, for all $q > m_{\mathcal{B}}$, we have $(f^{\sizeMin}_{FixWMP})^{-1}(\mathcal{M},s,q) \subseteq (P_{\mathcal{B}})^{C}$ and therefore, $Pr((f^{\sizeMin}_{FixWMP})^{-1}(\mathcal{M},s,q)) \leq Pr((P_{\mathcal{B}})^{C}) = 1 - Pr(P_{\mathcal{B}}) = 0$.  
	\end{itemize}
	
	Concluding from 1 and 2, we thus have, for all $s \in \mathcal{B},\; \sum_{r \in \mathbb{Q}} r \cdot Pr((f^{\sizeMin}_{FixWMP})^{-1}(\mathcal{M},s,r)) = m_{\mathcal{B}}$. Therefore, the expected value in a BSCC $\mathcal{B}$ does not depend on the starting state and is always equal to $m_{\mathcal{B}}$. This concludes the proof.
\end{proof}

To sum up, the fixed window mean-payoff of a path only depends on where it ends up. Moreover, almost surely, a path end up in a BSCC. Furthermore, the value of any path that ends up in a BSCC $\mathcal{B}$ is, almost surely, equal to $m_{\mathcal{B}}$. Thus the expected value of $f^{\sizeMin}_{FixWMP}$ in the Markov chain $\mathcal{M}$ from the initial state $s_{init}$ is given by:
\begin{align}
\mathbb{E}_{s_{init}}^{\mathcal{M}}(f^{\sizeMin}_{FixWMP}) = \sum_{\mathcal{B} \in BSCC(\mathcal{M})} Pr(\lozenge \mathcal{B}) \cdot m_{\mathcal{B}}\label{expectedvalue}
\end{align} 

\begin{figure}
	\begin{center}
		\vspace{5pt}
\centering
\scalebox{1}{
	\begin{tikzpicture}
		\node[player] (sinit) at (0,0) {$s_0$} ;
		\node[player] (s1) at (2,0) {$s_1$} ;
		\node[player] (s2) at (4,0) {$s_2$} ;
		\node[player] (s3) at (6,0) {$s_2$} ;
		
		\path[-latex]  (sinit) edge node[above] {\small 1, \textcolor{red}{2}} (s1)
				(s1) edge node[above] {\small 1, \textcolor{red}{5}} (s2)
				(s2) edge node[above] {\small 1, \textcolor{red}{4}} (s3)
				(s3) edge[bend left] node[below] {\small 1, \textcolor{red}{2}} (sinit)
		;
	\end{tikzpicture}
}
	\end{center}
	\caption{An example of a BSCC $\mathcal{B}$ where we want to compute $m_{\mathcal{B}}$, for $\sizeMin = 3$} 
	\label{ex_2_5_4}
\end{figure}
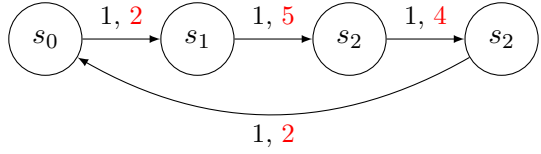

Let us focus on computing $m_{\mathcal{B}}$ for a BSCC $\mathcal{B}$. Ideally, we would compute this value using dynamic programming. More specifically, for $s \in \mathcal{B}$ and $l \in [\sizeMin]$, we would express $\min \limits_{\pi \in FPaths^{\mathcal{B}}_{l}(s)} \wmp(\pi(0 \ldots l))$ as a function of $\min \limits_{\pi \in FPaths^{\mathcal{B}}_{l-1}(s')} \wmp(\pi(0 \ldots l-1))$ for $s' \in Supp(s)$. However, it seemed not possible in some cases. For instance, consider the BSCC of Figure~\ref{ex_2_5_4} with $\sizeMin = 3$. It is obvious that, from state $s_1$, the best window mean-payoff is $5$ by only considering the first edge. And from state $s_0$ the best window mean-payoff is equal to $(2+5+4)/3 = 11/3$ by considering three edges. Therefore, from state $s_1$, the edge $(s_2,s_3)$ is not interesting since $w(s_2,s_3) = 4 < 5 = w(s_1,s_2)$. But, from state $s_0$, $(s_2,s_3)$ is interesting since $w(s_2,s_3) = 4 > 3.5 = (w(s_0,s_1) + w(s_1,s_2))/2$. The problem comes from the fact that whether the edge $(s_2,s_3)$ is interesting or not depends on the current mean of the path.

\begin{figure}
	\begin{center}
		\vspace{5pt}
\scalebox{1}{
	\begin{tikzpicture}
		\node[player] (sinit) at (0,0) {$s_0$} ;
		\node[player] (s1) at (2,0) {$s_1$} ;
		\node[player] (s2) at (4,0) {$s_2$} ;
		\node[player] (s3) at (6,0) {$s_2$} ;
		
		\path[-latex]  (sinit) edge node[above] {\small 1, \textcolor{red}{-1}} (s1)
			(s1) edge node[above] {\small 1, \textcolor{red}{2}} (s2)
			(s2) edge node[above] {\small 1, \textcolor{red}{1}} (s3)
			(s3) edge[bend left] node[below] {\small 1, \textcolor{red}{-1}} (sinit)
		;
	\end{tikzpicture}
}
	\end{center}
	\caption{An example of a BSCC $\mathcal{B}$ where we want to decide if $m_{\mathcal{B}} \geq 0$, for $\sizeMin = 3$} 
	\label{ex_threshold}
\end{figure}
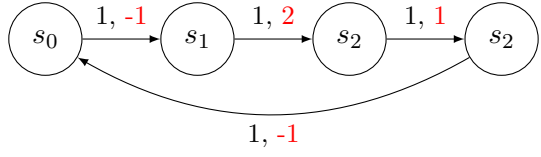

Now consider the decision problem where we have to decide if $m_{\mathcal{B}} \geq \lambda$ for some $\lambda \in \rat$. This is equivalent to decide whether or not $m_{\mathcal{B}} \geq 0$ in the BSCC where every weight have been reduced by $\lambda$. Now, an interesting property is that, for every $\pi \in FPaths^{\mathcal{B}}_{\sizeMin}$, $\wmp(\pi(0 \ldots \sizeMin)) \geq 0 \Leftrightarrow \wtp(\pi(0 \ldots \sizeMin)) \geq 0$. Therefore, we can look at the total payoff, instead of looking at the mean-payoff. In that case, whether an edge is interesting to consider or not does not depend on the current sum of the path, it only depends on the value of the edge (if it is positive or negative). For example, let us consider the case where $\lambda = 3$ in the previous BSCC. The BSCC obtained by reducing every weight by 3 can be seen in Figure~\ref{ex_threshold}. Now, the edge $(s_2,s_3)$ is interesting from $s_0$ and $s_1$ since $w(s_2,s_3) = 1 > 0$. 

The following lemma generalize this result:

\begin{lemma}
	\label{rec_total_payoff}
	Let $\mathcal{B}$ be a BSCC. 
	
	For $s \in \mathcal{B}$ and $l \in [\sizeMin]$, we define $TP_l(s) = \min \limits_{\pi \in FPaths^{\mathcal{M}}_{l}(s)} \wtp(\pi(0 \ldots l))$. Then
	\begin{displaymath}
		\forall s \in \mathcal{B},\; \forall l \in [\sizeMin],\; TP_l(s) = \min \limits_{s' \in Supp(s)} \max (w(s,s'),\; w(s,s') + TP_{l-1}(s'))
	\end{displaymath}
	with $TP_0(s) = 0$ for all $s \in \mathcal{B}$.
\end{lemma}

\begin{proof}
	To prove this lemma, we just use the fact that a path of length $l$ consists of an edge followed by a path of length $l-1$. More specifically, let $s \in \mathcal{B}$ and $l \in [\sizeMin]$, then:
	\begin{align*}
	TP_l(s) & = \min \limits_{\pi \in FPaths^{\mathcal{M}}_{l}(s)} \wtp(\pi(0 \ldots l)) \\
	& = \min \limits_{s' \in Supp(s)} \min \limits_{\pi \in FPaths^{\mathcal{M}}_{l-1}(s')} \wtp(s \; \pi(0 \ldots l-1)) \\
	& = \min \limits_{s' \in Supp(s)} \min \limits_{\pi = s_0 \ldots s_{|\pi|} \in FPaths^{\mathcal{M}}_{l-1}(s')} \max(w(s,s'),\; \max_{k \in [1..l-1]} w(s,s') + \sum_{i = 0}^{k-1} w(s_i,s_{i+1})) \\
	& = \min \limits_{s' \in Supp(s)} \max(w(s,s'),\; w(s,s') + \min \limits_{\pi = s_0 \ldots s_{|\pi|} \in FPaths^{\mathcal{M}}_{l-1}(s')} \max_{k \in [1..l-1]}  \sum_{i = 0}^{k-1} w(s_i,s_{i+1}))  \\
	& = \min \limits_{s' \in Supp(s)} \max(w(s,s'),\; w(s,s') + TP_{l-1}(s'))  
	\end{align*}
\end{proof}

Algorithm~\ref{goodwin}\footnote{This algorithm is an adaptation of the GoodWin Algorithm from \cite{CDRR15} where every state belongs to Player 2.} computes and returns the set of states $S_{pos} = \lbrace s \in S \mid TP_{\sizeMin}(s) \geq 0 \rbrace$ by using Lemma~\ref{rec_total_payoff}. More precisely, at iteration $i \in [\sizeMin]$ of the second \textsf{for} loop, $C(s) = TP_i(s)$.

The number of operations in that algorithm is bounded by $|E| \cdot \sizeMin$ since inside the second \textsf{for} loop (line 4-5), every edge is visited exactly once. If $W$ be the maximal weight that appears in the Markov chain, then each operation requires a time linear in $\log(W)$. Thus, the complexity of this algorithm is in $O(|E| \cdot \sizeMin \cdot \log(W))$.

\begin{algorithm}
	\caption{NonNegWindowBSCC($\mathcal{B},\sizeMin$)}
	\label{goodwin}
	\begin{algorithmic}[1]
		\Require{$\mathcal{B} = (S, E, w)$ and $\sizeMin \in \mathbb{N}_0$}
		\Ensure{$S_{pos}$ is the set of states that ensure a non-negative window total payoff for any sequence states of $\sizeMin$ starting in them}
		\ForAll {$s \in S$}
		\State {$C(s) := 0$}
		\EndFor
		\ForAll {$i \in [\sizeMin]$}
		\ForAll {$s \in S$}
		\State {$C(s) := \min \limits_{(s,s') \in E} {max \{w((s,s')), w((s,s')) + C(s')\}}$}
		\EndFor
		\EndFor
		\State 
		\Return {$S_{pos} := \{s \in S \mid C(s) \geq 0\}$}
	\end{algorithmic}
\end{algorithm}

The previous algorithm allows us to know if $m_{\mathbb{B}} \geq \lambda$, for some $\lambda \in \rat$. Algorithm~\ref{exp_val_bscc} uses a binary search between a lower and an upper bound, respectively called $L_B$ and $U_B$ and initially set to 0 and to $W+1$, $W$ being the maximum weight in the Markov chain. This algorithm ensures that, at the end of the binary search, $L_B \leq m_{\mathcal{B}} < U_B$ with $U_B - L_B = \epsilon \leq \frac{1}{\sizeMin^{2}}$ (this is justified in the following). This can be ensured by calling successively Algorithm~\ref{goodwin} where the weights have been reduced by a given $\lambda$. If all states ensure a non negative total payoff, then $\lambda \leq m_{\mathcal{B}}$, otherwise $\lambda > m_{\mathcal{B}}$. Once the binary search is over, we have to find the exact value of $m_{\mathcal{B}}$ between $L_B$ and $U_B$. We know that there exists $(p,q) \in \mathbb{N} \times [\sizeMin]$ such that $m_{\mathcal{B}} = \frac{p}{q}$. Moreover, for any $((k,l),(k',l')) \in (\mathbb{N} \times [\sizeMin])^2$ such that if $\frac{k}{l} \neq \frac{k'}{l'}$ then $\big| \frac{k}{l} - \frac{k'}{l'} \big| = \big| \frac{kl' - k'l}{ll'}\big| > \frac{1}{\sizeMin^{2}} \geq \epsilon$. This implies that if a pair $(p,q) \in (\mathbb{N} \times [\sizeMin])^2$ verifying $L_B \leq \frac{p}{q} < U_B$ is found, then necessarily $m_{\mathcal{B}} = \frac{p}{q}$. That is computed in the last \textsf{for} loop (line 15-18).

\begin{algorithm}
	\caption{ExpValBSCC($\mathcal{B},\sizeMin$)}
	\label{exp_val_bscc}
	\begin{algorithmic}[1]
		\Require{$\mathcal{B}= \langle S,E,w \rangle$ is a BSCC from a weighted Markov Chain and $\sizeMin \in \mathbb{N}_0$}
		\Ensure{$\mu$ is equal to $\mathbb{E}^{\mathcal{B}}(f^{\sizeMin}_{FixWMP})$}
		\State {$W := \max \limits_{e \in E} |w(e)|$}
		\State {$L_B := 0,\; U_B := W+1$}
		\State {$\epsilon := L_B + U_B$}
		\While {$\epsilon > \frac{1}{\sizeMin^{2}} $}
		\State {$\lambda := \frac{L_B + U_B}{2}$}
		\State {$\epsilon := \frac{\epsilon}{2}$}
		\State {$S' := NonNegWindowBSCC(\mathcal{B} = \langle S,E,w-\lambda \rangle)$}
		\If {$(S == S')$}
		\State {$L_B := \lambda$}
		\Else
		\State {$U_B := \lambda$}
		\EndIf 
		\EndWhile
		\For {$l \in [\sizeMin] $}
		\State {$l_B = l \times L_B; u_B = l \times U_B$}
		\If {$(\lceil l_B \rceil + 1 == \lceil u_B \rceil)$}
		\State {$\mu := \frac{\lfloor u_B \rfloor}{l}$}
		\State
		\Return {$\mu$}
		\EndIf
		\EndFor
	\end{algorithmic}
\end{algorithm}

Let us now discuss the complexity of the Algorithm~\ref{exp_val_bscc}. We consider the number $n_{while}$ of times that we enter the \textsf{while} loop. Initially, $\epsilon = W$, and at each step $\epsilon$ is divided by 2. This stops whenever $\epsilon \leq \frac{1}{\sizeMin^2}$. Therefore, $n_{while} \leq \lceil \log(W \cdot \sizeMin^2) \rceil$. At each step, the most expensive (in terms of time of computation) operation is the call of Algorithm~\ref{goodwin} of complexity $O(|E| \cdot \sizeMin \cdot \log(W))$. Finally, the last \textsf{for} loop is taking $O(\sizeMin \cdot \log(W))$ time. Therefore, the complexity of Algorithm~\ref{exp_val_bscc} is in $O(|E| \cdot \sizeMin \cdot \log(W) \cdot \log(W \cdot \sizeMin))$.
	
\begin{algorithm}
	\caption{FixWMP($\mathcal{M},s_{init},\sizeMin$)}
	\label{exp_markov}
	\begin{algorithmic}[1]
		\Require{$\mathcal{M}= \langle S,E,s_{init},\mathbb{P},w \rangle$ is a weighted Markov Chain, $s_{init} \in S$ and $\sizeMin \in \mathbb{N}_0$}
		\Ensure{$E$ is equal to $\mathbb{E}_s^{\mathcal{M}}(f^{\sizeMin}_{FixWMP})$}
		\State {$BSCC\_Vect := BSCC(\mathcal{M})$}
		\State {$E := 0$}
		\For {$\mathcal{B} \in BSCC\_Vect$}
		\State {$Prob_{\mathcal{B}} := Pr_{s_{init}}(\lozenge \mathcal{B}) $}
		\State {$E_\mathcal{B} := ExpValBSCC(\mathcal{B},\sizeMin)$}
		\State {$E \mathrel{+}= Prob_{\mathcal{B}} \cdot E_\mathcal{B}$}
		\EndFor
		\Return {$E$}
	\end{algorithmic}
\end{algorithm}

Algorithm~\ref{exp_markov} only consists in applying Formula~\ref{expectedvalue} from the initial state $s_{init}$. In this algorithm, we first need to find the BSCCs of the Markov chain. One can use the Tarjan algorithm\cite{Tarjan72} to extract the strongly connected components, and then check which of them are BSCCs. This would require $O(|S| + |E|)$ steps. Inside the \textsf{for} loop, computing the probability of reaching a BSCC is done in time $O(Mat(|S|))$ and computing the expected value of the BSCC consists in calling Algorithm~\ref{exp_val_bscc} of complexity $O(|E| \cdot \sizeMin \cdot \log(W) \cdot \log(W \times \sizeMin))$. The number of BSCCs in a Markov chain is bounded by $|S|$. Therefore, the complexity of Algorithm~\ref{exp_markov} is in $O(|S| \cdot (Mat(|S|) + |E| \cdot \sizeMin \cdot \log(W) \cdot \log(W \times \sizeMin)))$.

	\subsection{Bounded Window Mean-Payoff}
	\label{BWMP_MC}
	We are interested in the expected value of $f_{BWMP}$ in a Markov chain. More specifically, consider a Markov chain $\mathcal{M} = \zug{S,E,s_{init},w,\mathbb{P}}$ and an initial state $s_{init}$. Then, we study
	\begin{displaymath}
	\mathbb{E}_{s_{init}}^{\mathcal{M}}(f_{BWMP}) = \int_{r \in \mathbb{Q}} r \cdot Pr((f_{BWMP})^{-1}(\mathcal{M},s_{init},r))
	\end{displaymath} 
	
	Since the \textit{bounded window mean-payoff} is \textit{prefix independent}, the value of a path only depends on the BSCC in which it ends up. Moreover, like in the \textit{fixed window mean-payoff} case, the expected value of $f_{BWMP}$ in a BSCC does not depend on the starting state. We first introduce the function $\mpay: FPaths^\mathcal{M} \longrightarrow \rat$ such that:
	\begin{displaymath}
		\forall \rho = s_0 \ldots s_{|\rho|} \in FPaths^\mathcal{M},\; \mpay(\rho) = \frac{1}{|\rho|} \sum\limits_{i = 0}^{|\rho|-1} w(s_i,s_{i+1})
	\end{displaymath}
	
	Then, we have the following result:
	\begin{theorem} \label{thm:MC_BWMP}
		The expected value of $f_{BWMP}$ in a BSCC $\mathcal{B}$ does not depend on the starting state of the paths considered. Moreover, if we denote that expected value by $\mathbb{E}^{\mathcal{B}}(f_{BWMP})$, then we have:
		\begin{displaymath}
		\lim\limits_{l \rightarrow \infty} \mathbb{E}^{\mathcal{B}}(f^l_{FixWMP}) = \mathbb{E}^{\mathcal{B}}(f_{BWMP}) = \underbrace{\min_{\rho \in ElemCycle(\mathcal{B})} \mpay(\rho)}_{\text{noted }c_{\mathcal{B}}}
		\end{displaymath} 
		where $ElemCycle(\mathcal{B})$ denotes the (finite) set of elementary cycles in $\mathcal{B}$, that is the set of cycles composed by states seen only once, except for one which is seen twice.
	\end{theorem}
	
	\begin{proof}
		From Theorem~\ref{thm:fixMC} we have that, for all $l \geq 1$, the expected value of $f^l_{FixWMP}$ inside a BSCC $\mathcal{B}$ does not depend on the starting state and is denoted $\mathbb{E}^{\mathcal{B}}(f^{l}_{FixWMP})$.
		
		First, we prove that the limit of the series $(\mathbb{E}^{\mathcal{B}}(f^l_{FixWMP}))_{l \geq 1}$ exists:
		\begin{itemize}
			\item $\forall l \in \nat$, $\mathbb{E}^{\mathcal{B}}(f^l_{FixWMP}) \leq \mathbb{E}^{\mathcal{B}}(f^{l+1}_{FixWMP})$ since $\forall \pi \in Paths^{\mathbb{B}}$, $f^l_{FixWMP}(\pi) \leq f^{l+1}_{FixWMP}(\pi)$;
			\item $\forall l \in \nat$, $\mathbb{E}^{\mathcal{B}}(f^l_{FixWMP}) \leq W$, where $W$ is the maximum over all weights that appear in the Markov chain.
		\end{itemize}
		Therefore, according to the monotone convergence theorem, the series $(\mathbb{E}^{\mathcal{B}}(f^l_{FixWMP}))_{l \geq 1}$ is converging.
	
		Now, let $s \in \mathcal{B}$. We prove that:
		\begin{itemize}
			\item[1.] $\lim\limits_{l \rightarrow \infty} \mathbb{E}^{\mathcal{B}}(f^l_{FixWMP}) \leq \mathbb{E}^{\mathcal{B}}_s(f_{BWMP})$;
			\item[2.] $\mathbb{E}^{\mathcal{B}}_s(f_{BWMP}) \leq c_{\mathcal{B}}$; 
			\item[3.] $c_{\mathcal{B}} \leq \lim\limits_{l \rightarrow \infty} \mathbb{E}^{\mathcal{B}}(f^l_{FixWMP})$.
		\end{itemize}
		\begin{itemize}
			\item[1.] $\forall l \in \nat$, $\forall \pi \in Paths^{\mathbb{B}}(s)$, $f^l_{FixWMP}(\pi) \leq f_{BWMP}(\pi)$ by definition of $f_{BWMP}$. Therefore, $\lim\limits_{l \rightarrow \infty} \underbrace{\mathbb{E}^{\mathcal{B}}(f^l_{FixWMP})}_{= \mathbb{E}^{\mathcal{B}}_s(f^l_{FixWMP})} \leq \mathbb{E}^{\mathcal{B}}_s(f_{BWMP})$.
			
			\item[2.] Let $r > c_{\mathcal{B}}$. Let $\rho_{c_{\mathcal{B}}} = s_0 \ldots s_n \in ElemCycle(\mathcal{B})$ be an elementary cycle (we have $s_n = s_0$) such that $\mpay(\rho_{c_{\mathcal{B}}}) = c_{\mathcal{B}} < r$. Consider the smallest index $i$ in
			$\argmax\limits_{i \in [n-1]_0} \sum\limits_{k = 0}^{i-1} (w(s_k,s_{k+1}) - c_{\mathcal{B}})$. The state $s_{i}$ is called a high point of the cycle (see, for instance, \cite{CDRR15} in the proof of Lemma 10). We prove that the mean-payoff from $s_i$ to $s_j$ along the cycle $\rho_{c_{\mathcal{B}}}$ is at most $c_{\mathcal{B}}$ for all $j \in [n-1]_0$. Recall that $\rho_{c_{\mathcal{B}}}$ is a cycle, therefore $s_k = s_{k \mod n}$, for all $k \geq 0$:
			\begin{itemize}
				\item If $i < j$, then $\sum\limits_{k = 0}^{j-1} (w(s_k,s_{k+1}) - c_{\mathcal{B}}) = \sum\limits_{k = 0}^{i-1} (w(s_k,s_{k+1}) - c_{\mathcal{B}}) + \sum\limits_{k = i}^{j-1} (w(s_k,s_{k+1}) - c_{\mathcal{B}}) \leq \sum\limits_{k = 0}^{i-1} (w(s_k,s_{k+1}) - c_{\mathcal{B}})$, by definition of $i$. Therefore, $\sum\limits_{k = i}^{j-1} (w(s_k,s_{k+1}) - c_{\mathcal{B}}) \leq 0$;
				\item It $j \leq i$, then $0 = \sum\limits_{k = 0}^{n-1} (w(s_k,s_{k+1}) - c_{\mathcal{B}}) = \underbrace{\sum\limits_{k = j}^{i-1} (w(s_k,s_{k+1}) - c_{\mathcal{B}})}_{\geq 0, \text{ by definition of }i} + \sum\limits_{k = i}^{j-1} (w(s_{k},s_{k+1}) - c_{\mathcal{B}}) \geq \sum\limits_{k = i}^{j-1} (w(s_{k},s_{k+1}) - c_{\mathcal{B}})$. That is, $\sum\limits_{k = i}^{j-1} (w(s_{k},s_{k+1}) - c_{\mathcal{B}}) \leq 0$.
			\end{itemize}
			This is true for all $j \in [n-1]_0$. Then, we denote by $\rho_{c_{\mathcal{B}}}^{n}(s_i) = s_i \ldots s_{i + n \cdot |\rho_{c_{\mathcal{B}}}|}$ the sequence of states that starts from $s_i$ and goes $n$ times through the cycle $\rho_{c_{\mathcal{B}}}$. In that case, for every $n \in \nat$, we have:
			\begin{displaymath}
				\forall l \leq n \cdot |\pi_{c_{\mathcal{B}}}|, \wmp(\rho_{c_{\mathcal{B}}}^{n}(0 \ldots l)) \leq c_{\mathcal{B}} < r 
			\end{displaymath}
			Therefore, it implies that for all $l \geq 1$, these exists $\pi \in FPaths^{\mathcal{B}}_l(s)$ such that $\forall l_0 \leq l,\; \wmp(\pi(0 \ldots l_0)) < r$. Therefore:
			\begin{displaymath}
			\forall l \geq 1,\; Pr(\underbrace{(f^{l}_{FixWMP})^{-1}(\mathcal{B},s,r)}_{\text{ denoted } D_l}) = 0
			\end{displaymath}
			
			In fact:
			\begin{itemize}
				\item $D_1 \subset D_2 \subset D_3 \subset \ldots$;
				\item $\forall l \geq 1,\; Pr(D_{l}) = 0$.
			
			Therefore:
			\end{itemize}
			\begin{displaymath}
			Pr((f_{BWMP})^{-1}(\mathcal{B},s,r)) = Pr(\bigcup \limits_{1 \leq l} D_l) = \lim \limits_{l \rightarrow \infty} \underbrace{Pr(D_l)}_{=0} = 0
			\end{displaymath}
			
			Since this is true for any $r > c_{\mathcal{B}}$, we have $\mathbb{E}^{\mathcal{B}}_s(f_{BWMP}) \leq c_{\mathcal{B}}$.
			
			\item[3.] Let $\lambda < c_{\mathcal{B}}$. We prove that there exists $l \in \nat$, such that $\forall \pi \in Paths^{\mathcal{B}},\; f^l_{FixWMP}(\pi) \geq \lambda$ (which implies that $\mathbb{E}^{\mathcal{B}}(f^l_{FixWMP}) \geq \lambda$). We consider the weighted Markov chain restricted to $\mathcal{B}$ where every weight has been subtracted $\lambda$. In that case, for all $\pi \in Paths^{\mathcal{B}}$, $f^l_{FixWMP}(\pi) \geq \lambda$ in the original Markov chain is equivalent to $f^l_{FixWMP}(\pi) \geq 0$ in the new Markov chain. We work in that new Markov chain. Let $d = c_{\mathcal{B}} - \lambda > 0$ and let $l = \lceil \frac{(|\mathcal{B}| \cdot W) \cdot (|\mathcal{B}| - 1) }{d} + (|\mathcal{B}| - 1) \rceil \in \nat$ where $W$ is the maximum value of a weight that appears in the new Markov chain. 
			
			Let $\pi \in Paths^{\mathcal{B}}$. We prove that $f^l_{FixWMP}(\pi) \geq 0$. Let $i \in \nat$ and let $\pi^l_i = \pi(i \ldots (i+l))$. We consider a decomposition of $\pi^l_i$ into elementary cycle: $\pi^l_i = \mathcal{A}, \mathcal{C}_1, \ldots, \mathcal{C}_n$ constructed as follows: the states of $\pi^l_i$ are pushed in a stack, and as soon as a state already seen is pushed in that stack, an elementary cycle, that we remove from the stack, is added to the decomposition\footnote{This decomposition is exactly the same as the one used in \cite{CDRR15} to prove Lemma 2}. In that decomposition, $\mathcal{A}$ corresponds to the acyclic part of $\pi_i^l$. It has to be noted that the cycles in that decomposition may appear more than once. Consider the function $Sum$ that associates to every finite sequences of edges the sum of the weights that appear in that sequence. Because $\mathcal{A}$ constitutes the acyclic part, we have $|\mathcal{A}| \leq |\mathcal{B}| - 1$. Therefore, $Sum(\mathcal{A}) \geq - (|\mathcal{B}| - 1) \cdot W$. Moreover, the number $n$ of cycles has to be greater than $n \geq \frac{l - (|\mathcal{B}| - 1)}{|\mathcal{B}|} \geq \frac{(|\mathcal{B}| - 1) \cdot W}{d}$, since the length of cycle is at most $|\mathcal{B}|$. Therefore:
			\begin{displaymath}
				Sum(\mathcal{C}_1, \ldots, \mathcal{C}_n) = \sum \limits
				_{i = 1}^{n} \underbrace{Sum(\mathcal{C}_i)}_{\geq d} \geq n \cdot d \geq (|\mathcal{B}| - 1) \cdot W
			\end{displaymath}
			It follows that $Sum(\pi^l_i) \geq 0$. In fact, $\forall i \in \nat,\; Sum(\pi(i \ldots (i+l))) \geq 0$, that is: $\forall i \in \nat,\; \wmp(\pi(i \ldots (i+l))) \geq 0$. Thus, $f^l_{FixWMP}(\pi) \geq 0$ in the new Markov chain. 
			
			Therefore, $\pi \in FixWMP(\lambda,l)$ in the original Markov chain, and $f^l_{FixWMP}(\pi) \geq \lambda$ in the original Markov chain. Since this is true for all $\pi \in Paths^{\mathcal{B}}$, we have $\mathbb{E}^{\mathcal{B}}(f^l_{FixWMP}) \geq \lambda$. In fact, $ \forall \lambda < c_{\mathcal{B}},\; \exists l \in \nat,\; \mathbb{E}^{\mathcal{B}}(f^l_{FixWMP}) \geq \lambda$. That is, $c_{\mathcal{B}} \leq \lim\limits_{l \rightarrow \infty} \mathbb{E}^{\mathcal{B}}(f^l_{FixWMP})$.
		\end{itemize}
		
		Thus, the expected value in the BSCC $\mathcal{B}$ starting from any state is equal to the minimum mean-payoff of an elementary cycle in $\mathcal{B}$.
	\end{proof}
	
	Hence, in the same way than in the \textit{fixed window mean-payoff} case, we have:
	\begin{align}
	\mathbb{E}_{s_{init}}^{\mathcal{M}}(f_{BWMP}) = \sum_{\mathcal{B} \in BSCC(\mathcal{M})} Pr(\lozenge \mathcal{B}) \cdot \mathbb{E}^{\mathcal{B}}(f_{BWMP})\label{boundedmarkov}
	\end{align} 
	
	\begin{remark}
		From that formula and Theorem~\ref{thm:MC_BWMP}, it is possible to show that $\lim\limits_{l \rightarrow \infty} \mathbb{E}^{\mathcal{M}}(f^l_{FixWMP}) =  \mathbb{E}^{\mathcal{M}}(f_{BWMP})$.
	\end{remark}
	
	We recall that we are considering a weighted Markov chain $\mathcal{M} = \langle S,E,s_{init},w,\mathbb{P} \rangle$ from an initial state $s_{init} \in S$, with $W$ the maximum weight appearing in the Markov chain.
	
	Finding the expected value in a BSCC $\mathcal{B}$ in equivalent to finding the lowest mean-payoff that an elementary cycle can achieve. There exists an algorithm (from Karp, see \cite{Karp78}) that runs in time $O(|S'| \cdot |E'| \cdot \log(W))$ that solves this problem, where $S'$ and $E'$ are the set of vertices and edges of the BSCC $\mathcal{B}$.
	
	Algorithm~\ref{bounded} only consists in applying Formula~\ref{boundedmarkov} from the initial state $s_{init}$. This algorithm is very similar to the one used in the fixed case. In the two algorithms, the only difference lies in Line 5. Computing the BSCCs requires $O(|S| + |E|)$ steps. Inside the \textsf{for} loop, computing the probability of reaching a BSCC is done in time $O(Mat(|S|))$ and computing the expected value of the BSCC is of complexity $O(|S| \cdot |E| \cdot \log(W))$. The number of BSCCs in a Markov chain is bounded by $|S|$. Therefore, the complexity of Algorithm~\ref{bounded} is in $O(|S| \cdot (Mat(|S|) + (|S| \cdot |E| \cdot \log(W)))$.
	
	\begin{algorithm}
		\caption{BWMP($\mathcal{M},s_{init}$)}
		\label{bounded}
		\begin{algorithmic}[1]
			\Require{$\mathcal{M}=\langle S,E,s_{init},\mathbb{P},w \rangle$ is a weighted Markov Chain and $s_{init} \in S$}
			\Ensure{$E$ is equal to $\mathbb{E}_s^{\mathcal{M}}(f_{BWMP})$}
			\State {$BSCC\_Vect := BSCC(\mathcal{M})$}
			\State {$E := 0$}
			\For {$\mathcal{B} \in BSCC\_Vect$}
			\State {$Prob_{\mathcal{B}} := Pr_{s_{init}}(\lozenge \mathcal{B}) $}
			\State {$E_\mathcal{B} := KarpMeanCycle(\mathcal{B},\sizeMin)$}
			\State {$E = E + Prob_{\mathcal{B}} \cdot E_\mathcal{B}$}
			\EndFor
			\Return {$E$}
		\end{algorithmic}
	\end{algorithm}

		\subsection{\textit{Direct Fixed Window Mean-Payoff}}
	We are interested in the expected value of $f^{\sizeMin}_{DirFixWMP}$ for a given $\sizeMin$ in a Markov chain. More specifically, consider a Markov chain $\mathcal{M}$ and an initial state $s_{init}$. Then, we study
	\begin{displaymath}
	\mathbb{E}_{s_{init}}^{\mathcal{M}}(f^{\sizeMin}_{DirFixWMP}) = \int_{r \in \mathbb{Q}} r \cdot Pr((f^{\sizeMin}_{DrFixWMP})^{-1}(\mathcal{M},s_{init},r))
	\end{displaymath} 
	
	The \textit{direct fixed winvdow mean-payoff} is not prefix independent, and therefore we cannot only look at the BSCCs. We present here two algorithms to compute that expected value. The first one is polynomial in the size of the Markov chain, in $\sizeMin$, and in the maximum weight $W$ that appear in the Markov chain. Thus, the algorithm is pseudopolynomial since $\sizeMin$ and $W$ are given in binary. However, it is fixed-parameter tractable when $\sizeMin$ and $W$ are considered as parameters. The second one is exponential in $\sizeMin$, however it is logarithmic in the maximum weight that appears in the Markov chain. Both algorithms consist in building a new Markov chain and then computing probabilities to reach some states in those new Markov chains. 
	
	For a given Markov chain $\mathcal{M} = \langle S,E,s_{init},w,\mathbb{P} \rangle$, an initial state $s_{init}$ and a threshold $\lambda = \frac{a}{b} \in \rat$, we would like to compute the probability that the direct fixed window mean-payoff of a path starting in $s_{init}$ is at least $\lambda$, that is $Pr((f^{\sizeMin}_{DirFixWMP})^{-1}(\mathcal{M},s_{init},[\lambda,\infty[))$.
	
	We construct a new Markov chain $\mathcal{M}_{\sizeMin}^{\lambda}$ such that the probability that the direct fixed window mean-payoff of a path is at least $\lambda$ in the original Markov chain $\mathcal{M}$ is equal to the probability of reaching a specific state ($trap$) in $\mathcal{M}_{\sizeMin}^{\lambda}$.
	
	Let us first define a new weight function $\bar{w}$ such that:
	\begin{displaymath}
	\forall e \in E,\; \bar{w}(e) = w(e) \cdot b - a
	\end{displaymath}
	and the maximum over the new weights:
	\begin{displaymath}
	W = \max \limits_{e \in E} \; \bar{w}(e)
	\end{displaymath}
	
	We now define $\mathcal{M}_{\sizeMin}^{\lambda} = \langle S',E',w',s'_{init},\mathbb{P}' \rangle$ where:
	
	\begin{itemize}
		\item $S' = (S \times [\sizeMin-1]_0 \times [W \cdot (\sizeMin-1)]_0)
		 \cup (S \times \lbrace trap \rbrace) \cup \lbrace trap \rbrace$;
		\item $E' \subseteq S' \times S'$;
		\item $w'(e = (s,x),(s',y)) = \bar{w}(s,s')$ if $e \in E'$ and $(s,x),(s',y) \in S'$;
		\item $s'_{init} = (s_{init},0,0)$;
		\item $\mathbb{P}'((s,d,w),\; (s',trap)) = p$ if $\mathbb{P}(e = s \rightarrow s') = p$, $d = \sizeMin - 1$ and $w - \bar{w}(e) > 0$;
		\item $\mathbb{P}'((s,d,w),\; (s',d+1,w - \bar{w}(e))) = p$ if $\mathbb{P}(e = s \rightarrow s') = p$, $d < \sizeMin - 1$ and $w - \bar{w}(e) > 0$;
		\item $\mathbb{P}'((s,d,w),\; (s',0,0)) = p$ if $\mathbb{P}(e = s \rightarrow s') = p$ and $w - \bar{w}(e) \leq 0$;
		\item $\mathbb{P}'((s,trap),\; trap) = 1$;
		\item $\mathbb{P}'(trap,\; trap) = 1$.
	\end{itemize}
	
	If the probability $\mathbb{P}'(t,t')$ was not defined above for two states $t,t' \in S'$, then $\mathbb{P}'(t,t') = 0$.
	
	The idea of this construction is that if a path $\pi \in Paths^{\mathcal{M}^{\lambda}_{\sizeMin}}$ visits a state of the form $(s,d,w)$, then it corresponds to a path $\bar{\pi} \in Paths^{\mathcal{M}}$ that visits state $s$. Moreover, the last window that is not closed (for threshold $\lambda$) when $\bar{\pi}$ reaches the state $s$ starts $d < \sizeMin$ states before, and $w$ is the deficit amount, that is, an additional weight of value $w$ needs to be "seen" on the edges in order to close the window. If $\pi$ is in state $trap$, then there is a window that could not be closed in $\bar{\pi}$ within $\sizeMin$ steps. In that case $f^{\sizeMin}_{DirFixWMP}(\bar{\pi}) < \lambda$. The states of the form $(s,trap)$ for $s \in S$ are only here for the sake of the proof of the following theorem. When $\mathcal{M}_{\sizeMin}^{\lambda}$ is constructed in Algorithm~\ref{val_window}, these states are not taken into account, and the edges arriving in them are directly forwarded to the state $trap$. The interest of these states is to avoid the possibility of having two edges coming from the same state arriving in the state $trap$. 
	
	The interest of this new Markov chain can be seen through the following theorem:
	\begin{theorem}
		\label{th_window}
		Let $\mathcal{M}$ be a Markov chain and $\lambda \in \rat$ be a threshold. Then:
		\begin{displaymath}
			Pr((f^{\sizeMin}_{DirFixWMP})^{-1}(\mathcal{M},s_{init},[\lambda,\infty[)) = Pr(\pi \in \mathcal{M}_{\sizeMin}^{\lambda} \mid \pi \models \lnot \lozenge \lbrace trap \rbrace)
		\end{displaymath}
	\end{theorem}
	
	\begin{figure}
		\begin{center}
			\vspace{-15pt}
\scalebox{1}{
	\begin{tikzpicture}
		\node[player,initial,initial text={}] (sinit) at (0,0) {$s_0,0,0$} ;
		\node[player] (s111) at (-2,-2) {$s_1,1,1$} ;
		\node[player] (s10) at (-2,-4) {$s_1,0,0$} ;
		\node[player] (s20) at (2,-2) {$s_2,0,0$} ;
		\node[player] (s30) at (4,-4) {$s_3,0,0$} ;
		\node[player] (s311) at (2,-4) {$s_3,1,1$} ;
		\node[player] (s313) at (4,-6) {$s_3,1,3$} ;
		\node[player] (s3trap) at (1,-8) {$s_3,trap$} ;
		\node[player] (s40) at (2,-6) {$s_4,0,0$} ;
		\node[player] (s411) at (0,-6) {$s_4,1,1$} ;
		\node[player] (s4trap) at (3,-8) {$s_4,trap$} ;
		\node[player] (trap) at (2,-10) {$trap$} ;
		
		\path[-latex]  
				(sinit) edge node[left] {\textcolor{red}{$-1$}} (s111)
				(sinit) edge node[above] {\textcolor{red}{$1$}} (s20)
				
				(s111) edge node[left] {\textcolor{red}{$1$}} (s10)
				(s10) edge[loop below] node[below] {\textcolor{red}{$1$}} (s10)
				(s20) edge node[left] {\textcolor{red}{$-1$}} (s311)
				(s311) edge node[above] {\textcolor{red}{$3$}} (s30)
				(s311) edge node[left] {\textcolor{red}{$1$}} (s40)
				(s30) edge[loop above] node[above] {\textcolor{red}{$3$}} (s0)
				(s30) edge node[above] {\textcolor{red}{$1$}} (s40)
				(s40) edge node[above] {\textcolor{red}{$-3$}} (s313)
				(s40) edge node[above] {\textcolor{red}{$-1$}} (s411)
				(s313) edge node[left] {\textcolor{red}{$3$}} (s30)
				(s313) edge node[left] {\textcolor{red}{$1$}} (s4trap)
				(s411) edge node[left] {\textcolor{red}{$-3$}} (s3trap)
				(s411) edge node[right] {\textcolor{red}{$-1$}} (s4trap)
			
				(s3trap) edge node {} (trap)
				(s4trap) edge node {} (trap)
				
				(trap) edge[loop left] node {} (trap)
		;
	\end{tikzpicture}
}
		\end{center}
		\caption{The Markov chain $\mathcal{M}_{\sizeMin}^{\lambda}$ obtained from the weighted Markov chain of Figure~\ref{eq_weighted_markov}, with $\sizeMin = 2$  and $\lambda = 1.5$. The weights have been modified ($w' = 2 \cdot w - 3$) and appear on each edge. If a state has one outgoing edge, the probability of taking it is 1, if it has two outgoing edges, the probability of taking each one of them is $0.5$.} 
		\label{eq_window_markov}
	\end{figure}
	An illustration of this construction can be seen in Figure~\ref{eq_window_markov}. In that example, the probability of eventually reaching the state $trap$ is equal to 0.5 (that is the probability going to the state $s_1$ from the initial state). Therefore, the probability that the window mean-payoff is above $1.5$ for a infinite path chosen randomly in the original Markov chain is also equal to 0.5.
	
	In order to prove this result, we first consider the inductive property of windows (see \cite{CDRR15}). 
	
	\textbf{Inductive property of windows.}
	Let $\rho = s_0 \ldots s_{|\rho|} \in FPaths^{\mathcal{M}}$. Assume that there are $j \leq j' < |\rho|$ such that the window opened at $s_j$ is still open at $s_{j'}$ and it is closed at $s_{j'+1}$ (with respect to $\lambda$). Then, any window opened between $s_j$ to $s_{j'}$ (included) are closed in $s_{j'+1}$ (with respect to $\lambda$).
	
	We can now see and prove the lemma that will be used in the proof of the theorem. We first denote by
	\begin{align*}
		FP^{\mathcal{M}}_{< \lambda}(s_{init}) = & \lbrace \pi \in FPaths^{\mathcal{M}}(s_{init}) \mid |\pi| \geq \sizeMin \; \wedge \\
		& \forall i < |\pi| - \sizeMin,\; \wmp(\pi(i \ldots (i + \sizeMin))) \geq \lambda \; \wedge \\
		& \wmp(\pi((|\pi| - \sizeMin) \ldots (|\pi|))) < \lambda \rbrace
	\end{align*} 
	the set of finite paths in $\mathcal{M}$ where every window that was opened more than $\sizeMin$ states before the last state of the path is closed, with respect to $\lambda$, and where the window that started $\sizeMin$ states before the last state could not be closed.
	
	We also denote by
	\begin{align*}
		FP^{\mathcal{M}^{\lambda}_{\sizeMin}}_{trap}(s_{start}) = \lbrace \pi = s_0 \ldots s_{|\pi|} \in FPaths^{\mathcal{M}^{\lambda}_{\sizeMin}}(s_{start}) \mid \exists s \in S,\; s_{|\pi|} = (s,trap) \rbrace 
	\end{align*} 
	the set of finite paths of $\mathcal{M}^{\lambda}_{\sizeMin}$ that just entered a state $(s,trap)$ for some $s \in S$. We claim that:
	
	\begin{lemma}
		\label{bijec_window}
		There exists a bijection $h: FP_{< \lambda}^{\mathcal{M}}(s_{init}) \longrightarrow FP^{\mathcal{M}^{\lambda}_{\sizeMin}}_{trap}(s'_{init})$ such that:  

		\begin{displaymath}
			\forall \rho \in FP^{\mathcal{M}}_{< \lambda}(s_{init}),\; \mathbb{P}(\rho) = \mathbb{P}'(h(\rho))
		\end{displaymath}
	\end{lemma}
	
	\begin{proof}
		Let $h: FP^{\mathcal{M}}_{< \lambda}(s_{init}) \rightarrow FP^{\mathcal{M}^{\lambda}_{\sizeMin}}_{trap}(s'_{init})$ be such that for $\rho = s_0 s_1 \ldots s_{|\rho|} \in FP^{\mathcal{M}}_{< \lambda}(s_{init})$, where $s_0 = s_{init}$, we have $h(\rho) = t_0 t_1 \ldots t_{|\rho|} \in FP^{\mathcal{M}^{\lambda}_{\sizeMin}}_{trap}(s'_{init})$ with 
		\begin{itemize}
			\item $t_0 = s'_{init} = (s_{init},0,0) \in S'$
			\item $\forall i \in [|\rho|-1]$
			$$
			t_i = \left\{
			\begin{array}{ll}
				(s_i,d+1,v-w'(s_{i-1},s_i)) & \mbox{if } v-w'(s_{i-1},s_i) > 0 \\
				(s_i,0,0) & \mbox{otherwise}
			\end{array}
			\right.
			$$
			where $t_{i-1} = (s_{i-1},d,v)$.
			\item $t_{|\rho|} = (s_{|\rho|},trap)$
		\end{itemize}
		
		We first prove that $h(\rho)$ is well defined. More specifically, we prove that, for any $\rho \in FP^{\mathcal{M}}_{< \lambda}(s_{init})$, $h(\rho) \in FP^{\mathcal{M}^{\lambda}_{\sizeMin}}_{trap}(s_{start})$. 
		
		Let $\rho = (s_{init} = s_0) \ldots s_{|\rho|} \in FP^{\mathcal{M}}_{< \lambda}(s_{init})$. We prove the following inductive property, defined for $l \in [|\rho|-1]_0$:
		\begin{center}
			$H(l): t_0 \ldots t_l \in FPaths^{\mathcal{M}^{\lambda}_{\sizeMin}}(s_{start})$, and $t_l = (s_l,d,v)$ where
			\begin{itemize}
				\item[(a)] $d \leq \sizeMin - 1$ is equal to the size of the largest window that is still opened in $s_l$, that is, among all windows that are still opened in $s_l$, the window that was opened the earliest (if no window is still opened, $d = 0$);
				\item[(b)] $v = - \sum \limits_{k = 1}^{d} w'(s_{l-k},s_{l-k+1}) \geq 0$. We note that if $d = 0$ then $v = 0$.
			\end{itemize}
		\end{center}
		
		$H(0)$ holds since $t_0 = s_{start} = (s_{init},0,0) \in FPaths^{\mathcal{M}^{\lambda}_{\sizeMin}}(s_{start})$. Let us now assume that $H(k-1)$ hold for $k \leq l \in [|\rho|-1]$. Let $t_{l-1} = (s_{l-1},d,v)$.
		
		We first prove that $t_0 \ldots t_l \in FPaths^{\mathcal{M}^{\lambda}_{\sizeMin}}(s_{start})$.  We assume towards a contradiction that $d = \sizeMin - 1$ and  $v - w'(s_{i-1},s_i) > 0$. Because $d = \sizeMin - 1$, this means that the window starting in $s_{l-1 - (\sizeMin - 1)} = s_{l - \sizeMin}$ is still open in $s_{l-1}$. Moreover, $v - w'(s_{i-1},s_i) = - \sum \limits_{k = 1}^{\sizeMin} w'(s_{l-\sizeMin-1+k},s_{l-\sizeMin+k}) > 0$. That is, $\sum \limits_{k = 1}^{\sizeMin} w(s_{l-\sizeMin-1+k},s_{l-\sizeMin+k}) < \sizeMin \cdot \frac{a}{b} = \sizeMin \cdot \lambda$ by definition of $w'$. Therefore, the window opened in $s_{l-\sizeMin}$ is sill not closed in $s_l$. Hence the contradiction since $\rho \in FP^{\mathcal{M}}_{< \lambda}(s_{init})$, which implies that any window that was opened more than $\sizeMin$ states before the last state of $\rho$ is closed ($s_{l-\sizeMin}$ occurs more than $\sizeMin$ states before the last state of $\rho$ since $l \leq |\rho|-1$). We can conclude that $d < \sizeMin - 1 \lor w - w'(s_{i-1},s_i) \leq 0$ holds, and therefore $\mathbb{P}'(t_{l-1},t_l) = \mathbb{P}(s_{l-1},s_{l}) > 0$. In fact, $t_0 \ldots t_l \in FPaths^{\mathcal{M}^{\lambda}_{\sizeMin}}(s_{start})$. 
		
		Let us now consider conditions (a) and (b). If $v - w'(s_{i-1},s_i) > 0$, it is obvious that (a) and (b) hold for $t_l$. Let us now assume that $v - w'(s_{i-1},s_i) \leq 0$ (which implies that the window opened in $s_{l-1-d}$ is closed in $s_l$). In that case, $t_l = (s_l,0,0)$. We prove that any window opened before $s_l$ is closed. Because the window opened in $s_{l-1-d}$ is the earliest window to be still opened open at $s_{l-1}$, this implies that any window starting before $s_{l-1-d}$ is closed. Moreover, the inductive property of windows gives us that any window opened between $s_{l-d-1}$ and $s_{l-1}$ (included) is closed. Therefore any window opened before $s_l$ is closed. It follows that conditions (a) and (b) also hold in that case.
		
		In fact, $H(l)$ holds for all $l \in [|\rho|-1]_0$.
		
		We now have to prove that $\mathbb{P'}(t_{|\rho|-1},t_{|\rho|}) = \mathbb{P}(s_{|\rho| - 1},s_{|\rho|}) > 0$. That is, if we denote $t_{|\rho| - 1} = (s_{|\rho| - 1},d,v)$, we have to prove that:
		\begin{displaymath}
			d = \sizeMin - 1 \land v - w'(s_{|\rho|-1},s_{|\rho|}) > 0
		\end{displaymath}
		Because $\rho \in FP^{\mathcal{M}}_{< \lambda}(s_{init})$, the window opened in $s_{|\rho| - \sizeMin}$ is not closed in $s_{|\rho|}$. Therefore, according to $H(l)$ for $l \in [|\rho|-1]_0$, the second component of the states $t_l$ was never reset to 0 after $t_{|\rho| - \sizeMin}$. In fact, $d = l_{maw} - 1$. Moreover, according to $H(|\rho|-1)$, $v - w'(s_{|\rho|-1},s_{|\rho|}) = - \underbrace{\sum \limits_{k = 1}^{\sizeMin} w'(s_{l-k},s_{l-k+1})}_{< 0 \text{ since } \sum \limits_{k = 1}^{\sizeMin} w(s_{l-k},s_{l-k+1}) < \lambda} > 0$. We can conclude that $h(\rho) = t_0 t_1 \ldots t_{|\rho|} \in FP^{\mathcal{M}^{\lambda}_{\sizeMin}}_{trap}(s'_{init})$.
		
		Proving that $h$ is a bijection is straightforward from its definition, and the definition of $\mathcal{M}_{\sizeMin}^{\lambda}$. The fact that $\forall \rho \in FP^{\mathcal{M}}_{< \lambda}(s_{init}),\; \mathbb{P}(\rho) = \mathbb{P}'(h(\rho))$ comes from the definition of $\mathcal{M}_{\sizeMin}^{\lambda}$, where the probabilities are the same as in the original Markov chain $\mathcal{M}$.
	\end{proof}
	
	We can now proceed to the proof of Theorem~\ref{th_window}.
	
	\begin{proof}
		We have that:
		\begin{displaymath}
			Pr(\lbrace \pi \in \mathcal{M}_{\sizeMin}^{\lambda} \mid \pi \models \lozenge \lbrace trap \rbrace \rbrace) = Pr(\lbrace \pi \in \mathcal{M}_{\sizeMin}^{\lambda} \mid \pi \models \lozenge \lbrace (s,trap) \mid s \in S \rbrace \rbrace)
		\end{displaymath}
		by definition of the Markov chain $\mathcal{M}^{\lambda}_{\sizeMin}$. Moreover, by definition of the sets $FP^{\mathcal{M}}_{< \lambda}(s_{init})$ and $FP^{\mathcal{M}^{\lambda}_{\sizeMin}}_{trap}(s_{start})$, we have:
		\begin{align*}
			(f^{\sizeMin}_{DirFixWMP})^{-1}(\mathcal{M},s_{init},]-\infty,\lambda[) & = \biguplus \limits_{\pi \in FP^{\mathcal{M}}_{< \lambda}(s_{init})} Cyl(\pi)\\
			\lbrace \pi \in \mathcal{M}_{\sizeMin}^{\lambda} \mid \pi \models \lozenge \lbrace (s,trap) \mid s \in S \rbrace \rbrace & = \biguplus \limits_{\pi \in FP^{\mathcal{M}^{\lambda}_{\sizeMin}}_{trap}(s'_{init})} Cyl(\pi)
		\end{align*}
		Let us denote by $h$ the function defined in Lemma~\ref{bijec_window}.
		Then, we have:
		\begin{center}
			$
			\begin{array}{l l}
			Pr((f^{\sizeMin}_{DirFixWMP})^{-1}(\mathcal{M},s_{init},]-\infty,\lambda[)) & = \sum \limits_{\pi \in FP^{\mathcal{M}}_{< \lambda}(s_{init})} Pr(Cyl(\pi))\\
			= \sum \limits_{\pi \in FP^{\mathcal{M}}_{< \lambda}(s_{init})} \mathbb{P}(\pi)
			& = \sum \limits_{\pi \in FP^{\mathcal{M}}_{< \lambda}(s_{init})} \mathbb{P}'(h(\pi))\\
			= \sum \limits_{\pi \in FP^{\mathcal{M}}_{< \lambda}(s_{init})} Pr(Cyl(h(\pi)))
			& = Pr(\biguplus \limits_{\pi \in FP^{\mathcal{M}}_{< \lambda}(s_{init})} Cyl(h(\pi)))\\
			= Pr(\biguplus \limits_{\pi \in FP^{\mathcal{M}^{\lambda}_{\sizeMin}}_{trap}(s'_{init})} Cyl(\pi))
			& = Pr(\lbrace \pi \in \mathcal{M}_{\sizeMin}^{\lambda} \mid \pi \models \lozenge \lbrace (s,trap) \mid s \in S \rbrace \rbrace)
			\end{array}
			$
		\end{center}
		It follows that:
		\begin{align*}
			Pr((f^{\sizeMin}_{DirFixWMP})^{-1}(\mathcal{M},s_{init},[\lambda,\infty[)) & = 1 - Pr((f^{\sizeMin}_{DirFixWMP})^{-1}(\mathcal{M},s_{init},]-\infty,\lambda[)) \\
			& = 1 - Pr(\lbrace \pi \in \mathcal{M}_{\sizeMin}^{\lambda} \mid \pi \models \lozenge \lbrace (s,trap) \mid s \in S \rbrace \rbrace) \\
			& = 1 - Pr(\lbrace \pi \in \mathcal{M}_{\sizeMin}^{\lambda} \mid \pi \models \lozenge \lbrace trap \rbrace \rbrace) \\ 
			& = Pr(\lbrace \pi \in \mathcal{M}_{\sizeMin}^{\lambda} \mid \pi \models \lnot \lozenge \lbrace trap \rbrace \rbrace)
		\end{align*}
	\end{proof}
	
	This theorem enables us to compute the probability that the value of a path is above a given threshold. However, to design an algorithm that computes the probabilities of various values of $\lambda$ (and then to compute the expected value), we need an additional lemma that specifies the set of possible values of $\lambda$. We observe that for $\pi \in Paths^{\mathcal{M}}(s_{init})$, $\lbrace \wmp(\pi(i \ldots (i+\sizeMin))) \mid i \in \mathbb{N} \rbrace \subset \lbrace \wmp(\pi(0 \ldots \sizeMin)) \mid \pi \in FPaths^{\mathcal{M}}_{\sizeMin} \rbrace \subseteq \lbrace \frac{p}{q} \mid q \in [\sizeMin], p \in [W \cdot q]_0 \rbrace$ which is finite. Hence, the following lemma:
	
	\begin{lemma}
		\label{possible_values}
		Let the set $PossibleVal = \lbrace \lambda \in \reals \mid Pr((f^{\sizeMin}_{DirFixWMP})^{-1}(\mathcal{M},s_{init},\lambda) > 0 \rbrace$.
		
		Then, 
		\begin{displaymath}
		PossibleVal \subseteq \lbrace \frac{p}{q} \mid q \in [\sizeMin], p \in [W \cdot q]_0 \rbrace
		\end{displaymath}
		where $W = \max \limits_{e \in E} w(e)$.
	\end{lemma}

	By definition of the set $PossibleVal$, we have:	
	\begin{align}
		\mathbb{E}_{s_{init}}^{\mathcal{M}}(f^{\sizeMin}_{DirFixWMP}) = \sum_{\lambda \in PossibleVal} \lambda \cdot Pr((f^{\sizeMin}_{DrFixWMP})^{-1}(\mathcal{M},s_{init},\lambda))\label{new_exp}
	\end{align} 
	This lemma ensures that the only values of interest are of the form $\frac{p}{q}$, where $p \in [W \cdot \sizeMin]_0$ and $q \in [\sizeMin]$. That is what is used in Algorithm~\ref{val_window}.
	
	Algorithm~\ref{val_window} computes the expected value of $f^{\sizeMin}_{DirFixWMP}$ in a Markov chain. At the end of the two $\mathsf{for}$ loops (line 9), for all $\lambda \in PossibleVal$, we have $Val(\lambda) = Pr((f^{\sizeMin}_{DirFixWMP})^{-1}(\mathcal{M},s_{init},[\lambda,\infty[))$. Then, the value $Pr((f^{\sizeMin}_{DirFixWMP})^{-1}(\mathcal{M},s_{init},\lambda))$ is computed (line 12) using the fact that if we have $PossivleVal = \lbrace \lambda_1, \lambda_2, \ldots, \lambda_n \rbrace$ with $\lambda_1 > \lambda_2 \ldots > \lambda_n$ then, for all $i \in \lbrace 2, \ldots, n \rbrace$:
	\begin{align*}
		Pr((f^{\sizeMin}_{DirFixWMP})^{-1}(\mathcal{M},s_{init},\lambda_i)) & = Pr((f^{\sizeMin}_{DirFixWMP})^{-1}(\mathcal{M},s_{init},[\lambda_i,\infty[)) \\ & - Pr((f^{\sizeMin}_{DirFixWMP})^{-1}(\mathcal{M},s_{init},[\lambda_{i-1},\infty[)) 
	\end{align*}
	and
	\begin{displaymath}
		Pr((f^{\sizeMin}_{DirFixWMP})^{-1}(\mathcal{M},s_{init},\lambda_1)) = Pr((f^{\sizeMin}_{DirFixWMP})^{-1}(\mathcal{M},s_{init},[\lambda_1,\infty[))
	\end{displaymath}
	Finally, the expected value of $f^{\sizeMin}_{DirFixWMP}$ is computed by using Formula~\ref{new_exp}.
	
	\begin{algorithm}
		\caption{DirFixWMP1($\mathcal{M},s_{init},\sizeMin$)}
		\label{val_window}
		\begin{algorithmic}[1]
			\Require{$\mathcal{M}=\langle S,E,s_{init},\mathbb{P},w \rangle$ is a weighted Markov Chain, $s_{init} \in S$ and $\sizeMin \in \mathbb{N}_0$}
			\Ensure{$E$ is equal to $\mathbb{E}_{s_{init}}^{\mathcal{M}}(f^{\sizeMin}_{DirFixWMP})$}
			\State {$W = \max \limits_{e \in E} |w(e)|$}
			\State {$Val := AssocTable(\textsf{float},\textsf{float}$) \textit{$\;\;$: first elements are sorted in descending order}}
			\For {$q \in [\sizeMin]$}
			\For {$p \in [W \cdot q]_0$}
			\State {$\lambda = \frac{p}{q}$} 			
			\State {$\mathcal{M}^{\lambda}_{\sizeMin} := \langle S',\mathbb{P}',s_{start} \rangle$}
			\State {$Prob_{\geq \lambda} := Pr^{\mathcal{M}^{\lambda}_{\sizeMin}}_{s_{start}}(\lozenge \lbrace trap \rbrace)$}
			\State {$Val(\lambda) := Prob_{\geq \lambda}$}
			\EndFor
			\EndFor
			\State {$E := 0$}
			\State {$Prob_{> \lambda} = 0$}
			\For {$\lambda \in Val$}
			\State {$Prob_{= \lambda} := Val(\lambda) - Prob_{> \lambda}$}
			\State {$E := E + \lambda \cdot Prob_{= \lambda}$}
			\State {$Prob
				_{> \lambda} := Val(\lambda)$}
			\EndFor
			\Return {$E$}
		\end{algorithmic}
	\end{algorithm}
	
	Let us now consider the complexity of Algorithm~\ref{val_window}. We enter $(W + 1) + 2 \cdot (W + 1) + \ldots + \sizeMin \cdot (W + 1) = O(W \cdot \sizeMin^2)$ times the lines 5-8. For $\lambda = \frac{p}{q}$, the size of the Markov chain $\mathcal{M}_{\sizeMin}^{\lambda}$ that is constructed line 6 is of size $O(|S| \cdot \sizeMin \cdot (W \cdot q - p) \cdot \sizeMin) = O(|S| \cdot \sizeMin \cdot W \cdot \sizeMin \cdot \sizeMin)$. The factor $W \cdot q - p$ comes from the fact that we consider a new weight function $w' = q \cdot w - p$.	Then, computing the probability to reach the state $trap$ is in $O(Mat(|S| \cdot \sizeMin^{3} \cdot W))$. Therefore, the complexity of the algorithm is in $O(W \cdot \sizeMin^{2} \cdot Mat(|S| \cdot \sizeMin^{3} \cdot W))$.

We also have a different algorithm to compute $\mathbb{E}_{s_{init}}^{\mathcal{M}}(f^{\sizeMin}_{DirFixWMP})$ based on another construction of Markov chains. That new algorithm is exponential in the window size $\sizeMin$ but is logarithmic in $W$, the maximal weight appearing in the Markov chain. It may be faster than the previous algorithm in the cases where the weights in the Markov chain are very high. The details are given in Appendix~\ref{dirfix_2}.

	%\section{Expected value and complexity in a weighted Markov decision process}
\section{Window Mean-Payoff in Weighted Markov Decision processes}
\label{sec:MDP}
	In this section, given an MDP, we study the problems of finding the optimal expected values of the functions defined in Section~\ref{functions of interest} over all the strategies available in the MDP. 

%\track{Add summary of the results.}
We start with the prefix independent version of the fixed window objective function and show an algorithm that computes the expected value of the objective function in time that is polynomial in the the size of the MDP and the window size.
Further, an optimal strategy requires a finite memory that too is polynomial in the size of the MDP and the window size.
Next we show that the bounded window mean-payoff problem can be solved in time $NP \cap coNP$ and a memoryless optimal strategy exists.
Finally, we study the direct fixed window objective function and find its expected value using an algorithm that runs in time that is exponential in the window length.
An optimal strategy also requires a finite memory that is bounded above by $W^{\sizeMin} \cdot \sizeMin^2$, where %$S$ is the set of states in the MDP, 
$W$ is its maximum weight and $\sizeMin$ is the window length.
We show a {\sc PSpace-hardness} for the problem even when $\sizeMin$is given in unary thus excluding the possibility of having an algorithm that is pseudopolynomial unless {\sc P=PSpace}.
For the prefix independent version of the fixed window objective function, we show that the problem of computing an expected value is at least as hard as computing the value in a two-player window mean-payoff game with the fixed window objective.
For the optimal expected value of the bounded window mean-payoff objective function, we show that the problem of computing the value is at least as hard as solving the value problem in traditional mean-payoff games.

	In the following, we will refer to the optimal expected value as the expected value in the MDP for the corresponding function.
	
	\subsection{\textit{Fixed Window Mean-Payoff}}
	Consider an MDP $\Gamma = \langle S,E,Act,s_{init},w,\mathbb{P} \rangle$. Then, let
	\begin{displaymath}
	 \mathbb{E}_{s_{init}}^{\Gamma}(f^{\sizeMin}_{FixWMP}) =  \sup\limits_{\sigma \in strat(\Gamma)} \mathbb{E}_{s_{init}}^{\MDPtoMC{\Gamma}{\sigma}}(f^{\sizeMin}_{FixWMP})
	\end{displaymath} 
	
	We first design an algorithm that solves this problem in time $O(log(W \cdot \sizeMin^2) \cdot (|S| \cdot |Act|) \cdot (|S| + |E|) \cdot \sizeMin \cdot log(W))$ where $W = \max\limits_{e \in E} w(e)$. Then, we show that the problem of deciding the two-player game for the direct fixed window mean objective can be reduced to the expected fixed window mean-payoff in an MDP in polynomial time.
	
	\subsubsection{Algorithm}
	Since the \textit{fixed window mean-payoff} is prefix independent, we first look at the maximal end components (recall that, for every strategy $\sigma$, each path will almost surely end up in an MEC, by Proposition~\ref{eventually_mec}). Consider an MEC $M = (T,A) \in MEC(\Gamma)$ of $\Gamma$ and a state $s \in T$. Since the subgraph induced by $M$ is strongly connected (since it is an MEC), for every pair of states $s, s' \in T$, there exists a reaching strategy $\sigma_{s,s'} \in strat(\Gamma)$ such that every path starting from $s$ reaches $s'$ almost surely, in the Markov chain $\MDPtoMC{\Gamma}{\sigma_{s,s'}}$. Therefore, if we denote by $v \in T$ a state that maximizes the expected value $\mathbb{E}_{s}^{\Gamma}(f^{\sizeMin}_{FixWMP})$ among all states $s \in T$, that is $\mathbb{E}_{v}^{\Gamma}(f^{\sizeMin}_{FixWMP}) = \displaystyle{\max_{s \in T}}(\mathbb{E}_{s}^{\Gamma}(f^{\sizeMin}_{FixWMP}))$, then $v$ can be reached almost surely from every state $s \in T$.
	Since $f^{\sizeMin}_{FixWMP}$ is prefix independent, we have: 
	\begin{displaymath}
		\forall s \in T,\; \mathbb{E}_{s}^{\Gamma}(f^{\sizeMin}_{FixWMP}) = \mathbb{E}_{v}^{\Gamma}(f^{\sizeMin}_{FixWMP})
	\end{displaymath}
	This is true in every MEC of $\Gamma$. For each $M \in MEC(\Gamma)$, we denote by $\lambda^{\sizeMin}_M$ the (optimal) expected value of $f^{\sizeMin}_{FixWMP}$ from every state of $M$, computed among the paths that stay in $M$, that is $\lambda^{\sizeMin}_M = \mathbb{E}_{v}^{M}(f^{\sizeMin}_{FixWMP})$. In fact, the value $\lambda^{\sizeMin}_M$ can be expressed using a two-player game. Formally, we have the following theorem:
	\begin{theorem}
		\label{val_MEC}
		Let $M = (T,A)$ be an MEC of $\Gamma$. Then we have:
%		\begin{displaymath}
		\begin{equation}  \label{eqn:gameFixWMP}
		\lambda^{\sizeMin}_M = \max\limits_{s \in T} \underbrace{2 \cdot \sup\limits_{\sigma_1 \in strat_1(G_M)} \inf\limits_{\sigma_2 \in strat_2(G_M)} \underbrace{V_s^{f^{2 \cdot \sizeMin}_{DirFixWMP}}(G_M,\sigma_1,\sigma_2)}_{\text{noted }f(s,\sigma_1,\sigma_2)}}_{\text{noted }g(s)}
		\end{equation} 
%		\end{displaymath} 
	\end{theorem}
	
	Before proving this theorem, we make a few observations.
	
	The value $g(s)$ denotes the outcome (multiplied by $2$) of the two-player game resulting from $M$ where the function of interest is the direct fixed window mean-payoff function for a window length $2 \sizeMin$. From \cite{CDRR15}, we know that deterministic finite memory strategies are enough to solve the two player game for the direct fixed window mean-payoff. Moreover, the memory needed to solve that game is bounded above by
	%a function linear in 
	$|S| \cdot l$ with $l = 2 \cdot \sizeMin$ and $S$ denotes the total number of states in the game\footnote{Since there are finitely many strategies with a fixed memory size, the supremum and infimum become maximum and minimum respectively. Also since the two-player game with direct fixed window objective is determined, the $\sup$ and the $\inf$ can be interchanged.}. 
	%Therefore, the supremum and infimum that appear in the expression of $g(s)$ are in fact maximum and minimum, which implies that their order can be swapped. 
	
	When we construct a two-player game from an MDP, an edge of weight $w$ is split into two edges (hence, doubling the number of edges seen on a path) of weight $0$ and $w$ (which divides the mean-payoff of a path by $2$). 
	Therefore, if the fixed window mean-payoff of a path $\pi$ in the MDP equals $2 \cdot \lambda$ for a window length $\sizeMin$, then the fixed window mean-payoff over the corresponding path in the two-player game equals $\lambda$, for a window length of $2 \cdot \sizeMin$.
%	is equal to $\lambda$ for a specific length equals to $2 \cdot l$ in the two-player game, then the fixed window mean-payoff of the corresponding path in the MDP will be equal to $2 \cdot \lambda$, for a length equal to $l$. 
	Formally, for a strategy $\sigma_1 \in strat_1(M)$, and $\pi \in Paths_v^{\MDPtoMC{M}{\sigma_1}}$, we have $2 \cdot f^{2 \cdot l}_{DirFixWMP}(p_v^{\sigma_1}(\pi)) \leq 2 \cdot f^{2 \cdot l}_{FixWMP}(p_v^{\sigma_1}(\pi)) = f^{l}_{FixWMP}(\pi)$. 
The inequality holds since the value of the direct fixed window mean-payoff over a path $\pi$ is at most the the value of fixed window mean-payoff over $\pi$.	
	That is why $g(s)$ equals $2$ times the outcome of the game in which the length considered is $2 \cdot \sizeMin$.
	
	\begin{proof}
		We proceed in two steps:
		\begin{itemize}
			\item[(1)] First we show that $\lambda^{\sizeMin}_M \geq \max\limits_{s \in T} g(s)$. Let $\sigma_1,\sigma_2$ be two optimal strategies for Player $1$ and Player $2$ respectively from a state $v \in T$ that ensures $g(v) = \max\limits_{s \in T} g(s)$ ($v$ is called an optimal state). Let $\pi \in Paths_v^{\MDPtoMC{M}{\sigma_1}}$. Let $\sigma^{\pi}_2 \in strat_2(G_M)$ be a (deterministic) strategy for Player 2 such that the outcome of the game $G_M$ played with strategies $\sigma_1$ and $\sigma^{\pi}_2$ follows exactly the sequence of states 
%			and actions 
			visited by $\pi$ in $\MDPtoMC{M}{\sigma_1}$. Formally, $\sigma^{\pi}_2$ ensures that $\pi_{(G_M,v,\sigma_1,\sigma^{\pi}_2)} = p_v^{\sigma_1}(\pi)$. By definition of $\sigma_2$, we have $g(v) = 2 \cdot f(v,\sigma_1,\sigma_2) \leq 2 \cdot f(v,\sigma_1,\sigma^{\pi}_2) = 2 \cdot f^{2 \cdot \sizeMin}_{DirFixWMP}(\pi_{(G_M,v,\sigma_1,\sigma^{\pi}_2)}) = 2 \cdot f^{2 \cdot \sizeMin}_{DirFixWMP}(p_v^{\sigma_1}(\pi)) \leq f^{\sizeMin}_{FixWMP}(\pi)$, where the function $f$ is defined in Equation \ref{eqn:gameFixWMP} (we justified the last inequality in the paragraph preceding this proof). This is true for every path $\pi \in Paths_v^{\MDPtoMC{M}{\sigma_1}}$. Therefore, $\mathbb{E}_v^{M}(f^{\sizeMin}_{FixWMP}) \geq g(v) = \max\limits_{s \in T} g(s)$. In fact, $\lambda^{\sizeMin}_M \geq \max\limits_{s \in T} g(s)$.
			\item[(2)] Now, we show that $\lambda^{\sizeMin}_M \leq \max\limits_{s \in T} g(s)$. Let $\sigma_1 \in strat(M)$ be a strategy in the MDP and let $\pi' \in Paths^{\MDPtoMC{M}{\sigma_1}}_t$ for some $t \in T$. For every state $s$ occurring in $\pi'$, there exists a strategy $\sigma_s \in strat_2(G_M)$ such that $2 \cdot f^{2 \cdot \sizeMin}_{DirFixWMP}(\pi_{(G_M,s,\sigma_1,\sigma_s)}) \leq g(s)$ (it may be lower than $g(s)$ since $\sigma_1$ may not be an optimal strategy for Player 1). Therefore, in the MDP $M$, there exists a finite sequence of states, that starts in $s$, and that contains a sequence of $\sizeMin$ states whose window mean-payoff is lower than or equal to $g(s) \leq \max\limits_{s' \in T} g(s')$. Thus, the probability of seeing such a sequence of states from every occurrence of $s$ is strictly positive. This is true for every state $s$ occurring in $\pi'$. Hence, infinitely often, there is a non-zero probability of seeing a sequence of states for which the window mean-payoff is lower than or equal to $\max\limits_{s \in T} g(s)$.
			Thus each path almost surely has a fixed window mean-payoff of at most $\max\limits_{s \in T} g(s)$. Formally, $Pr((f^{\sizeMin}_{FixWMP})^{-1}(\MDPtoMC{M}{\sigma_1},t,[0,\max\limits_{s \in T} g(s)])) = 1$, which implies that $\mathbb{E}_t^{\MDPtoMC{M}{\sigma_1}}(f^{\sizeMin}_{FixWMP}) \leq \max\limits_{s \in T} g(s)$ This is true for every strategy $\sigma_1 \in strat(M)$. That is, $\forall \sigma \in strat(M),\; \mathbb{E}_t^{\MDPtoMC{M}{\sigma}}(f^{\sizeMin}_{FixWMP}) \leq \max\limits_{s \in T} g(s)$. Therefore, $\mathbb{E}_t^{M}(f^{\sizeMin}_{FixWMP}) \leq \max\limits_{s \in T} g(s)$. Thus, $\lambda^{\sizeMin}_M \leq \max\limits_{s \in T} g(s)$.
		\end{itemize}
		The theorem follows. This (especially the first paragraph) also proves that an optimal strategy for Player 1 in the two-player game $G_M$ is also an optimal strategy in the MDP $M$. Therefore, there exists an optimal strategy for which the expected value of the fixed window mean-payoff is equal to $\lambda^{\sizeMin}_M$ and whose memory size bounded above by
%		a function linear in 
		$|T| \cdot \sizeMin$.
	\end{proof}
	
	\begin{remark}
		\label{remark_fixed_enough}
		This proof also shows that the theorem holds even if the objective in the two-player game is the fixed window mean-payoff objective. This comes from the fact that the inequality $g(v) \leq f^{\sizeMin}_{FixWMP}(\pi)$ (in the first paragraph) also holds if we consider that new objective. We stated Theorem~\ref{val_MEC} with the direct fixed window mean-payoff objective because the algorithm presented in \cite{CDRR15} to solve the two-player for that objective is faster that the algorithm presented in \cite{CDRR15} to solve the fixed window mean-payoff objective.
	\end{remark}
	
	We now compute the expected value in the whole MDP. We denote by $\maxEndCompMDP = \langle S,E,Act,s_{init},w',\mathbb{P} \rangle$ the MDP where for every $M \in MEC(\Gamma)$, for every edge $e = (s,a,s') \in E$ that appears in $M$, the weight on the edge $e$ is replaced by $\lambda^{\sizeMin}_M$. Formally, for all edges $(s,a,s') \in E$:
	\begin{itemize}
		\item $w'(s,a,s') = \lambda^{\sizeMin}_M$ if $M = (T,A) \in MEC(\Gamma)$ and $a \in A$;
		\item $w'(s,a,s') = w(s,a,s')$ otherwise.
	\end{itemize}
	For every $M \in MEC(\Gamma)$, we denote by $M_\lambda \in MEC(\maxEndCompMDP)$ the corresponding maximal end component in $\maxEndCompMDP$ in which all the weights have been replaced by $\lambda_M^{\sizeMin}$.
	
	For a Markov chain $\mathcal{M} = \langle S,E,s_{init},w,\mathbb{P} \rangle$ and for $\pi = s_0 \ldots \in Paths^{\mathcal{M}}$, we define the function $f_{Mean}: Paths^{\mathcal{M}} \longrightarrow \mathbb{R}$ such that $f_{Mean}(\pi) = \liminf\limits_{n \rightarrow \infty} \frac{1}{n} \sum\limits_{k = 0}^{n-1} w(s_k,s_{k+1})$. This corresponds to the \emph{mean-payoff} of the path $\pi$. Now, we have the following theorem, which states that the expected value of the fixed window mean-payoff in $\Gamma$, that is $\mathbb{E}_{s_{init}}^{\Gamma}(f^{\sizeMin}_{FixWMP})$, is equal to the expected value of the mean-payoff in $\maxEndCompMDP$, that is $\mathbb{E}_{s_{init}}^{\maxEndCompMDP}(f_{Mean})$:
	\begin{theorem}
		\label{window_average}
		For any Markov decision process $\Gamma$, and any initial state $s_{init}$, we have:
		\begin{displaymath}
			\mathbb{E}_{s_{init}}^{\Gamma}(f^{\sizeMin}_{FixWMP}) = \mathbb{E}_{s_{init}}^{\maxEndCompMDP}(f_{Mean})
		\end{displaymath}
	\end{theorem}
	
	\begin{proof}
		Consider an MDP $\Gamma$. By definition, $\Gamma$ and $\maxEndCompMDP$ only differ in the weights inside the MECs. Therefore, $strat(\Gamma) = strat(\maxEndCompMDP)$, and for each $\sigma \in strat(\Gamma)$, we have $Paths^{\MDPtoMC{\Gamma}{\sigma}}(s_{init}) = Paths^{\MDPtoMC{\maxEndCompMDP}{\sigma}}(s_{init})$. 
		
		We also proceed in two steps:
		\begin{itemize}
			\item[(1)] We first prove that $\mathbb{E}_{s_{init}}^{\Gamma}(f^{\sizeMin}_{FixWMP}) \leq \mathbb{E}_{s_{init}}^{\maxEndCompMDP}(f_{Mean})$. Let $\sigma \in strat(\Gamma)$. Let $M \in MEC(\Gamma)$ be a maximal end component. Then, we have:
			\begin{itemize}
				\item\begin{enumerate}				
					\item Since $\sigma$ is any arbitrary strategy, the expected value of $f^{\sizeMin}_{FixWMP}$ in $\MDPtoMC{M}{\sigma}$ ensures $\mathbb{E}^{\MDPtoMC{M}{\sigma}}(f^{\sizeMin}_{FixWMP}) \leq \lambda_M^{\sizeMin}$, by definition of $\lambda_M^{\sizeMin}$;
					\item the construction of $\MDPtoMC{M_\lambda}{\sigma}$ in $\maxEndCompMDP$ ensures that the expected value of $f_{Mean}$ in $\MDPtoMC{M_\lambda}{\sigma}$ equals $\lambda_M^{\sizeMin}$ that is, $\mathbb{E}^{\MDPtoMC{M_\lambda}{\sigma}}(f_{Mean}) = \lambda_M^{\sizeMin}$ since every weight appearing in $\MDPtoMC{M_\lambda}{\sigma}$ is equal to $\lambda^{\sizeMin}_M$.
				\end{enumerate}
				Therefore, $\mathbb{E}^{\MDPtoMC{M}{\sigma}}(f^{\sizeMin}_{FixWMP}) \leq \mathbb{E}^{\MDPtoMC{M_\lambda}{\sigma}}(f_{Mean})$;
				\item The probability of eventually staying forever in $M$ is equal to the probability of eventually staying forever in $M_\lambda$. That is, $Pr^{\MDPtoMC{\Gamma}{\sigma}}(\lozenge \square M) = Pr^{\MDPtoMC{\maxEndCompMDP}{\sigma}}(\lozenge \square M_\lambda)$.
			\end{itemize}
			Moreover, we know that every path (in $\MDPtoMC{\Gamma}{\sigma}$ and in $\MDPtoMC{\maxEndCompMDP}{\sigma}$) will end up, almost surely, in an MEC. Therefore:
			\begin{displaymath}
				\mathbb{E}_{s_{init}}^{\MDPtoMC{\Gamma}{\sigma}}(f^{\sizeMin}_{FixWMP}) = \sum\limits_{M \in MEC(\Gamma)} \underbrace{Pr^{\MDPtoMC{\Gamma}{\sigma}}(\lozenge \square M)}_{= Pr^{\MDPtoMC{\maxEndCompMDP}{\sigma}}(\lozenge \square M_\lambda)} \cdot \underbrace{\mathbb{E}^{\MDPtoMC{M}{\sigma}}(f^{\sizeMin}_{FixWMP})}_{\leq \mathbb{E}^{M_\lambda\sigma}(f_{Mean})} \leq \mathbb{E}_{s_{init}}^{\MDPtoMC{\maxEndCompMDP}{\sigma}}(f_{Mean})
			\end{displaymath}
			This is true for all $\sigma \in strat(\Gamma)$, that is, for all $\sigma \in strat(\Gamma)$, $\mathbb{E}_{s_{init}}^{\MDPtoMC{\Gamma}{\sigma}}(f^{\sizeMin}_{FixWMP}) \leq \mathbb{E}_{s_{init}}^{\MDPtoMC{\maxEndCompMDP}{\sigma}}(f_{Mean})$. Thus, $\mathbb{E}_{s_{init}}^{\Gamma}(f^{\sizeMin}_{FixWMP}) \leq \mathbb{E}_{s_{init}}^{\maxEndCompMDP}(f_{Mean})$.
			\item[(2)] We now prove that $\mathbb{E}_{s_{init}}^{\Gamma}(f^{\sizeMin}_{FixWMP}) \geq \mathbb{E}_{s_{init}}^{\maxEndCompMDP}(f_{Mean})$. We know that, given an MDP $\Gamma$, there exists an optimal memoryless strategy that maximizes the expected mean-payoff in $\Gamma$ (see \cite{Filar12}). Thus, let us consider a memoryless strategy $\sigma \in strat(\maxEndCompMDP)$ that maximizes the expected value of the mean-payoff in $\maxEndCompMDP$. 
			We construct a strategy $\sigma' \in strat(\Gamma)$ such that $\mathbb{E}_{s_{init}}^{\MDPtoMC{\Gamma}{\sigma'}}(f^{\sizeMin}_{FixWMP}) = \mathbb{E}_{s_{init}}^{\MDPtoMC{\maxEndCompMDP}{\sigma}}(f_{Mean})$.
			Let us denote by $M(\sigma)$ the set of MECs of $\Gamma$ that contains at least one BSCC reachable from $s_{init}$ of the Markov chain $\MDPtoMC{\Gamma}{\sigma}$ (there is not necessarily a BSCC in every MEC, if an MEC $M$ is not interesting to end up in, in terms of maximizing the expected mean-payoff. 
			In such case $\sigma$ leaves $M$ and thus no BSCC will appear in $M$). Let $\sigma' \in strat(\Gamma)$ be a strategy that acts like $\sigma$ outside the maximal end components in $M(\sigma)$ and, inside an MEC $M \in M(\sigma)$, acts to get an expected value equal to $\lambda_M^{\sizeMin}$ while staying inside $M$ (this is possible with a finite memory strategy, as we have seen in the proof of Theorem \ref{val_MEC}). 
			In this way, once inside an MEC $M \in M(\sigma)$, the expected value of $f^{\sizeMin}_{FixWMP}$ in $M$ (that is $\lambda^{\sizeMin}_M$, by definition of the strategy $\sigma'$) is the same as the expected value of $f_{Mean}$ in $\MDPtoMC{\maxEndCompMDP}{\sigma}$ (also $\lambda^{\sizeMin}_M$, since every weights appearing inside $M$ are equal to $\lambda^{\sizeMin}_M$). Moreover, since the strategy $\sigma'$ acts like $\sigma$ outside the MECs $M \in M(\sigma)$, the probability of reaching every MEC $M \in M(\sigma)$ is the same in both $\MDPtoMC{\Gamma}{\sigma'}$ and in $\MDPtoMC{\maxEndCompMDP}{\sigma}$. Hence, $\mathbb{E}_{s_{init}}^{\MDPtoMC{\Gamma}{\sigma}}(f^{\sizeMin}_{FixWMP}) = \mathbb{E}_{s_{init}}^{\MDPtoMC{\maxEndCompMDP}{\sigma}}(f_{Mean})$. By choice of $\sigma$, we have $\mathbb{E}_{s_{init}}^{\Gamma^{\sizeMin}_{MEC,\sigma}}(f_{Mean}) = \mathbb{E}_{s_{init}}^{\maxEndCompMDP}(f_{Mean})$. Therefore, $\mathbb{E}_{s_{init}}^{\Gamma}(f^{\sizeMin}_{FixWMP}) = \sup\limits_{\sigma' \in strat(\Gamma)} \mathbb{E}_{s_{init}}^{\MDPtoMC{\Gamma}{\sigma'}}(f^{\sizeMin}_{FixWMP}) \geq \mathbb{E}_{s_{init}}^{\maxEndCompMDP}(f_{Mean})$.
		\end{itemize}
		This concludes the proof.
	\end{proof}
	
	\begin{corollary}
		\label{corollary_1}
		The memory size that is needed to construct an optimal strategy $\sigma'$ that maximizes the expected value of $f^{\sizeMin}_{FixWMP}$ in an MDP $\MDP = \zug{S,E,Act,s_{init},w,\mathbb{P}}$ is bounded above by $|S| \cdot \sizeMin$.
	\end{corollary}
	
	\begin{proof}
		One can construct an optimal strategy $\sigma'$ as follows:
		\begin{itemize}
			\item The first part of $\sigma'$ in $\MDP$ follows the strategy $\sigma$ in $\maxEndCompMDP$ that is used to maximize the expected value of the mean-payoff until it reaches some MEC $M$
			%in which there are BSCCs;
			such that $\sigma$ induces one or more BSCCs in $M$.
			Recall that the corresponding MEC $M$ also exists in $\MDP$;
			\item Then, once inside an MEC $M = (T,A)$ in which it is possible to end up by following $\sigma$ in $\MDP$, we apply a strategy so that an optimal state $v \in T$ maximizing $g(s)$ over all states $s \in T$ is reached almost surely;
			\item Finally, we apply the optimal strategy that maximizes the outcome of the two-player window mean-payoff game from the optimal state $v$.
		\end{itemize}
		The first two steps can be done with a memoryless strategy. The third one requires a memory size bounded above by $|T| \cdot \sizeMin$ for every MEC $M = (T,A)$ in which it is possible to end up by following $\sigma$.
		Hence the memory required is bounded above by $|S| \cdot \sizeMin$, where $S$ is the set of states in $\MDP$.
	\end{proof}
	
	We now examine the algorithms that computes these expected values. In \cite{CDRR15}, Algorithm $DirectFWMP(G,\sizeMin)$ returns the set of winning states for the \emph{direct fixed window mean-payoff} objective in the two-player game $G = (S_1,S_2,s_{init},E,w)$ with threshold 0. This is done in time $O(|S_1 \cup S_2| \cdot |E| \cdot \sizeMin \cdot log(W))$ where $W = \max\limits_{e \in E} w(e)$. 
	
	\begin{algorithm}
		\caption{FixWMP\_MEC($\Gamma,\sizeMin$)}
		\label{fixed_mec}
		\begin{algorithmic}[1]
			\Require{$M = \langle T,E,A,s_{init},w,\mathbb{P} \rangle$ is a weighted Markov decision process that is a maximal end component, $\sizeMin \in \mathbb{N}_0$}
			\Ensure{$\mu$ is equal to $\mathbb{E}^{M}(f^{\sizeMin}_{FixWMP})$}
			\State{$G_{\Gamma} := \langle T_1,T_2,s_{init},E,w' \rangle$ \textit{$\;\;$ :the two-player game obtained from the MDP $M$}} 
			\State{$W = \max\limits_{e \in E} w(e)$}
			\State {$L_B := 0,\; U_B := W+1$}
			\State {$\epsilon := L_B + U_B$}
			\While {$\epsilon > \frac{1}{\sizeMin^{2}}$}
			\State {$\lambda := \frac{L_B + U_B}{2}$}
			\State {$\epsilon := \frac{\epsilon}{2}$}
			\State {$T' := DirectFWMP(\langle T_1,T_2,s_{init},E,w'-\lambda/2 \rangle,2 \cdot \sizeMin)$}
			\If {$(T' \neq \emptyset)$}
			\State {$L_B := \lambda$}
			\Else
			\State {$U_B := \lambda$}
			\EndIf 
			\EndWhile
			\For {$l \in [\sizeMin] $}
			\State {$l_B = l \times L_B; u_B = l \times U_B$}
			\If {$(\lceil l_B \rceil + 1 == \lceil u_B \rceil)$}
			\State {$\mu := \frac{\lfloor u_B \rfloor}{l}$}
			\State break
			\EndIf
			\EndFor
%			\State {$E := \mu$}
			\State
%			\Return {$E$}
			\Return {$\mu$}
		\end{algorithmic}
	\end{algorithm}
	Algorithm~\ref{fixed_mec} computes the expected value of $f^{\sizeMin}_{FixWMP}$ inside an MEC by solving the two-player window mean-payoff game, using the result of Theorem~\ref{val_MEC}. Additionally, a binary search very close to the one used in Algorithm~\ref{exp_val_bscc} is used. This binary search looks for $\lambda_M^{\sizeMin}$ between the two bounds $0$ and $W$.
	Recall that we assume that $\MDP$ has non-negative integer weights.	However, like mentioned before, the value of a path inside the MDP is twice as much as its value in the two-player game. That is why, in line 8 (where this algorithm differs from Algorithm~\ref{exp_val_bscc}), the call to Algorithm $DirectFWMP$ is done with the weights reduced by $\lambda/2$ instead of $\lambda$.
	
	Let us now discuss its complexity. Similar to Algorithm~\ref{exp_val_bscc}, we enter the \texttt{while} loop $O(log(W \cdot \sizeMin^2))$ times. The most expensive operation is the call to algorithm $DirectFWMP$, that runs in time $O((|T_1| + |T_2|) \cdot |E| \cdot \sizeMin \cdot log(W))$ with $|T_1| = |T|$ and $|T_2| = |T| \cdot |A|$. Therefore, the complexity of Algorithm~\ref{fixed_mec} is $O(log(W \cdot \sizeMin^2) \cdot (|T| \cdot |A|) \cdot |E| \cdot \sizeMin \cdot log(W))$.
	\begin{algorithm}
		\caption{Replace\_MEC($\Gamma,\sizeMin$)}
		\label{transform_mec}
		\begin{algorithmic}[1]
			\Require{$\Gamma = \langle S,E,Act,s_{init},w,\mathbb{P} \rangle$ is a weighted Markov decision process and $\sizeMin \in \mathbb{N}_0$}
			\Ensure{$\Gamma'$ is equal to $\maxEndCompMDP$}
			\State{$w' = w$}
			\State {$MEC\_Vect := MEC(\Gamma)$}
			\For{$M = (T,A) \in MEC\_Vect$}
			\State{$\lambda := FixWMP\_MEC(M,\sizeMin)$}
			\For{$(s,a) \in T \times A$}
			\For{$(s,a,s') \in E$}
			\State{$w'((s,a,s')) := \lambda$}
			\EndFor
			\EndFor
			\EndFor
			\State{$\Gamma' = \langle S,E,Act,s_{init},w',\mathbb{P} \rangle$}
			\State
			\Return {$\Gamma'$}
		\end{algorithmic}
	\end{algorithm}
	
	Then, Algorithm~\ref{transform_mec} returns the MDP $\maxEndCompMDP$ when the MDP $\Gamma$ is given as input. The MECs of $\Gamma$, that is $MEC(\Gamma)$ can be computed in time $O(|S| \cdot (|S| + |E|))$ \cite{BK08}. Then, the most expensive operations are the successive calls to Algorithm~\ref{fixed_mec} corresponding to every MEC. Since the sum of the number of states of the MECs of $\Gamma$ is at most the number of states of $\Gamma$, the complexity of this algorithm is in time $O(log(W \cdot \sizeMin^2) \cdot (|S| \cdot |Act|) \cdot (|S| + |E|) \cdot \sizeMin \cdot log(W))$.
	\begin{algorithm}
		\caption{FixWMP\_MDP($\Gamma,s_{init},\sizeMin$)}
		\label{fixed_mdp}
		\begin{algorithmic}[1]
			\Require{$\Gamma = \langle S,E,Act,s_{init},w,\mathbb{P} \rangle$ is a weighted Markov decision process, $s_{init} \in S$ and $\sizeMin \in \mathbb{N}_0$}
			\Ensure{$\mu$ is equal to $\mathbb{E}_{s_{init}}^{\Gamma}(f^{\sizeMin}_{FixWMP})$}
			\State{$\Gamma' = Replace\_MEC(\Gamma,\sizeMin)$}
			\State {$\mu := ExpectedMeanPayoff(\Gamma,s_{init})$}
			\State
			\Return {$\mu$}
		\end{algorithmic}
	\end{algorithm}
	Finally, from an MDP $\Gamma$, Algorithm~\ref{fixed_mdp} constructs the new MDP $\Gamma^{\sizeMin}_{MEC}$. Then, it calls Algorithm $ExpectedMeanPayoff$ on $\Gamma^{\sizeMin}_{MEC}$. Algorithm $ExpectedMeanPayoff$ computes the expected value of the mean-payoff for the optimal strategy of the MDP on which it is called. For an MDP $\Gamma = \langle S,E,Act,s_{init},w,\mathbb{P} \rangle$, it runs in time $O(Avg(|S|,|Act|))$ (the notation $Avg$ was introduced in the preliminaries, in Section~\ref{prelim_mdp}). Therefore, the complexity of Algorithm~\ref{fixed_mdp} is in $O(log(W \cdot \sizeMin^2) \cdot (|S| \cdot |Act|) \cdot (|S| + |E|) \cdot \sizeMin \cdot log(W) + Avg(|S|,|Act|))$.

	Therefore, we have the following theorem:
	\begin{theorem}
		\label{complexity_mdp_fixed}
		The expected value of $f^{\sizeMin}_{FixWMP}$ in an MDP $\Gamma$ can be computed in time $O(p_1(|\Gamma|,\sizeMin))$ where $p_1$ is a polynomial function.
	\end{theorem}

		\subsubsection{Hardness}
	\label{fixed_mdp_hardness}
	We now consider the hardness of the problem. We show that given an MDP $\MDP$ with an initial state $s_{init}$, a window length $\sizeMin$ and a threshold $\lambda$, the \emph{fixed window mean-payoff} problem of checking if $\expect_{s_{init}}^{\Gamma}(f^{\sizeMin}_{FixWMP}) \geq \lambda$ is at least as hard as solving the two-player game for a direct fixed window mean-payoff objective.
	
	\begin{theorem}
		\label{fix_mdp_hardness}
		The fixed window mean-payoff problem for MDP is at least as hard as solving a two-player game for the direct fixed window mean-payoff objective (for polynomial reductions).
	\end{theorem}
	
	Given a two-player game $G = \langle S_1,S_2,s_{init},E,w \rangle$, an initial state $s_{init}$ and a window length $\sizeMin$, we construct an MDP $\Gamma$ such that the expected value of the fixed window mean-payoff in $\Gamma$ is equal to the outcome of the two-player game from $s_{init}$, by using the result of Theorem~\ref{val_MEC}. However, to use that theorem properly, we need the following conditions:
	\begin{itemize}
		\item The window length $\sizeMin$ is even (since, in Theorem~\ref{val_MEC}, the direct fixed window mean-payoff objective considered in the two-player game has a window length equal to $2 \cdot \sizeMin$);
		\item The MDP $\Gamma$ is an MEC, and therefore the two-player game is strongly connected;
		\item Starting the two-player game in $s_{init}$ maximizes the outcome.
	\end{itemize}
	Neither of these three conditions is necessarily ensured by $G$ and $\sizeMin$. Therefore, we construct another two-player game $G^{reset} = \langle S_1',S_2',s_{init},E',w' \rangle$ very similar to $G$ in which some edges are added so that, wherever the game started in $G^{reset}$, Player 2 can do as if the game started in $s_{init}$ by taking one of these edges. However, every new edges has a very high weight (equal to $(W+1) \cdot 2 \cdot \sizeMin$) so that it is not interesting for Player 2 to take more than once a new edge. In this way, the outcome of the game $G$ and $G^{reset}$ are the same from $s_{init}$. Furthermore, playing the game $G^{reset}$ from a state $s$ gives an outcome of at most what can be done from $s_{init}$. Additionally, the two-player game $G^{reset}$ is strongly connected since every vertex is reachable from $s_{init}$ and, with the new edges, $s_{init}$ is reachable from every vertex.
	
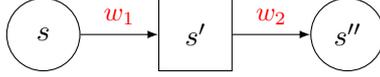
\begin{figure}
		\begin{center}
			\vspace{0pt}
\centering
\scalebox{1}{
	\begin{tikzpicture}
		\node[player1] (s1) at (0,0) {$s$} ;
		\node[player2] (s2) at (2,0) {$s'$} ;
		\node[player1] (s3) at (4,0) {$s''$} ;
		
 \path[-latex] 
	 (s1) edge node[above] {\small \textcolor{red}{$w_1$}} (s2)
	 (s2) edge node[above] {\small \textcolor{red}{$w_2$}} (s3)
		;
	\end{tikzpicture}
}
		\end{center}
		\caption{An example of two consecutive edges in the game $G$. Square states belong to Player 1 and circle states belong to Player 2.}
		\label{original edges}
	\end{figure}

\begin{figure}
		\begin{center}
			\vspace{0pt}
\centering
\scalebox{1}{
	\begin{tikzpicture}
		\node[player1] (s1) at (0,0) {$s$} ;
		\node[player2] (s1') at (2.5,0) {$(s,s',2)$} ;
		\node[player1] (s1'') at (5.5,0) {$(s,s',1)$} ;
		\node[player2] (s2) at (8,0) {$s'$} ;
		\node[player1] (s3) at (10,0) {$s''$} ;
		
 \path[-latex] 
	 (s1) edge node[above] {\small \textcolor{red}{$0$}} (s1')
	 (s1') edge node[above] {\small \textcolor{red}{$w_1$}} (s1'')
	 (s1'') edge node[above] {\small \textcolor{red}{$0$}} (s2)
	 (s2) edge node[above] {\small \textcolor{red}{$w_2$}} (s3)
		;
	\end{tikzpicture}
}
		\end{center}
		\caption{The four consecutive edges in the game $G^{reset}$ that corresponds to the two edges in $G$ of Figure~\ref{original edges}.}
		\label{new edges}
	\end{figure}
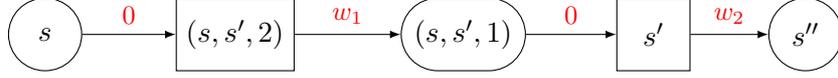

	Moreover, every edge $e$ in $G$ is duplicated to form two consecutive edges $e_1$ and $e_2$ in $G^{reset}$ with $w'(e_1) = 0$ and $w'(e_2) = w(e)$. An example of that duplication can be seen in Figure~\ref{original edges} and Figure~\ref{new edges}.
	
	In this way, the direct fixed window mean-payoff of a path in $G$ for the window length $\sizeMin$ is equal to two times the direct fixed window mean-payoff of the corresponding path in $G^{reset}$ for the window length $2 \cdot \sizeMin$ (this is very similar to the way a game is constructed from an MDP).
	
	Before formally defining $G^{reset}$, we assume that the weights appearing in $G$ are all non-negative integer (w.l.o.g. since, if not, we can use the method for Markov chains described in the preliminaries, in Section~\ref{functions of interest}).
	
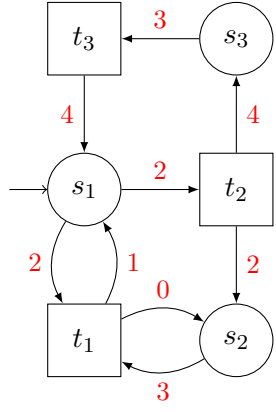
\begin{figure}
		\begin{center}
			\vspace{0pt}
\centering
\scalebox{1}{
	\begin{tikzpicture}
		\node[player1,initial,initial text={}] (s1) at (0,0) {$s_1$} ;
		\node[player1] (s2) at (2,-2) {$s_2$} ;
		\node[player1] (s3) at (2,2) {$s_3$} ;
		
		\node[player2] (t1) at (0,-2) {$t_1$} ;
		\node[player2] (t2) at (2,0) {$t_2$} ;
		\node[player2] (t3) at (0,2) {$t_3$} ;
		
 \path[-latex] 
	 (s1) edge[bend right] node[left] {\small \textcolor{red}{2}} (t1)
	 (s1) edge node[above] {\small \textcolor{red}{2}} (t2)
	 (s2) edge[bend left] node[below] {\small \textcolor{red}{3}} (t1)
	 (s3) edge node[above] {\small \textcolor{red}{3}} (t3)
	 
	 (t1) edge[bend right] node[right] {\small \textcolor{red}{1}} (s1)
	 (t1) edge[bend left] node[above] {\small \textcolor{red}{0}} (s2)
	 (t2) edge node[right] {\small \textcolor{red}{2}} (s2)
	 (t2) edge node[right] {\small \textcolor{red}{4}} (s3)
	 (t3) edge node[left] {\small \textcolor{red}{4}} (s1)
		;
	\end{tikzpicture}
}
		\end{center}
		\caption{An example of a two-player game $G$. The circle states belong to Player 1 and the square states belong to Player 2.} 
		\label{game}
	\end{figure}

\begin{figure}
		\begin{center}
			\vspace{0pt}
\centering
\scalebox{1}{
	\begin{tikzpicture}
	\node[player1,initial,initial text={}] (s1) at (0,0) {$s_1$} ;
	\node[player1] (s2) at (7,-2) {$s_2$} ;
	\node[player1] (s3) at (7,2) {$s_3$} ;

	\node[player2] (t1) at (0,-2) {$t_1$} ;
	\node[player2] (t2) at (7,0) {$t_2$} ;
	\node[player2] (t3) at (0,2) {$t_3$} ;
	
	\node[player2] (s1t12) at (-2,0) {$(s_1,t_1,2)$} ;
	\node[player1] (s1t11) at (-2,-2) {$(s_1,t_1,1)$} ;
	
	\node[player2] (s1t22) at (2,0) {$(s_1,t_2,2)$} ;
	\node[player1] (s1t21) at (5,0) {$(s_1,t_2,1)$} ;
	
	\node[player2] (s2t12) at (5,-2) {$(s_2,t_1,2)$} ;
	\node[player1] (s2t11) at (2,-2) {$(s_2,t_1,1)$} ;
	
	\node[player2] (s3t32) at (5,2) {$(s_3,t_3,2)$} ;
	\node[player1] (s3t31) at (2,2) {$(s_3,t_3,1)$} ;
	
	\path[-latex] 
	(s1) edge[bend right] node[above] {\small \textcolor{red}{0}} (s1t12)
	(s1t12) edge node[left] {\small \textcolor{red}{2}} (s1t11)
	(s1t12) edge[blue,bend right] node[below] {\small \textcolor{blue}{30}} (s1)
	(s1t11) edge node[below] {\small \textcolor{red}{0}} (t1)
	
	(s1) edge node[above] {\small \textcolor{red}{0}} (s1t22)
	(s1t22) edge node[above] {\small \textcolor{red}{2}} (s1t21)
	(s1t22) edge[blue,bend right=45] node[above] {\small \textcolor{blue}{30}} (s1)
	(s1t21) edge node[above] {\small \textcolor{red}{0}} (t2)
	
	(s2) edge node[above] {\small \textcolor{red}{0}} (s2t12)
	(s2t12) edge node[above] {\small \textcolor{red}{3}} (s2t11)
	(s2t12) edge[blue,bend left=5] node[below] {\small \textcolor{blue}{30}} (s1)
	(s2t11) edge node[above] {\small \textcolor{red}{0}} (t1)
	
	(s3) edge node[above] {\small \textcolor{red}{0}} (s3t32)
	(s3t32) edge node[above] {\small \textcolor{red}{3}} (s3t31)
	(s3t32) edge[blue,bend right=10] node[below] {\small \textcolor{blue}{30}} (s1)
	(s3t31) edge node[above] {\small \textcolor{red}{0}} (t3)
	
	(t1) edge node[right] {\small \textcolor{red}{1}} (s1)
	(t1) edge[bend right] node[below] {\small \textcolor{red}{0}} (s2)
	(t2) edge node[right] {\small \textcolor{red}{2}} (s2)
	(t2) edge node[right] {\small \textcolor{red}{4}} (s3)
	(t3) edge node[left] {\small \textcolor{red}{4}} (s1)
	;
	\end{tikzpicture}
}
		\end{center}
		\caption{The two-player game $G^{reset}$ obtained from the game $G$ in Figure~\ref{game} for $\sizeMin = 3$. That is, $(W + 1) \cdot 2 \cdot \sizeMin = 30$. The edges in $E'_5$ appear in blue.} 
		\label{new_game}
	\end{figure}
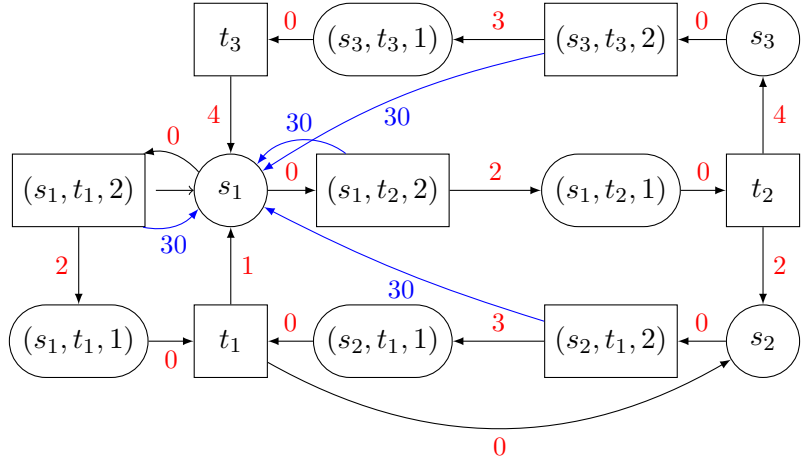

	In the following, $W$ refers to the maximum over all the weights appearing in $G$.
	Formally, the game $G^{reset}$ is defined such that:
	\begin{itemize}
		\item $S_1' = S_1 \cup ((E \cap S_1 \times S_2) \times \lbrace 1 \rbrace)$;
		\item $S_2' = S_2 \cup ((E \cap S_1 \times S_2) \times \lbrace 2 \rbrace)$;
		\item $E' = \underbrace{\lbrace (s,(s,s',2)) \in S_1' \times S_2'\mid (s,s') \in E \cap S_1 \times S_2 \rbrace}_{\text{denoted }E_1'}$ \\ $\cup \underbrace{\lbrace ((s,s',2),(s,s',1)) \in S_2' \times S_1' \mid (s,s') \in E \cap S_1 \times S_2 \rbrace}_{\text{denoted }E_2'}$ \\ $\cup \underbrace{\lbrace ((s,s',1),s') \in S_1' \times S_2' \mid (s,s') \in E \cap S_1 \times S_2 \rbrace}_{\text{denoted }E_3'} \cup \underbrace{E \cap S_2 \times S_1}_{\text{denoted }E_4'}$ \\ $\cup \underbrace{\lbrace ((s,s',2),s_{init}) \in S_2' \times S_1' \rbrace}_{\text{denoted }E_5'}$;
		\item For $e \in E'$, we have:
		\begin{align}
		w'(e)=
		\begin{cases}
		0 & \text{if } e \in E_1';\\
		w(s,s') & \text{if } e = ((s,s',2),(s,s',1)) \in E_2';\\
		0 & \text{if } e \in E_3';\\
		w(s,s') & \text{if } e = (s,s') \in E_4';\\
		(W + 1) \cdot 2 \cdot \sizeMin & \text{if } e \in E_5' \text{ where } W = \max\limits_{e \in E} w(e);
		\end{cases}
		\end{align}
	\end{itemize}
	An example of that construction can be seen in Figure~\ref{new_game}, that is obtained from the two-player game of Figure~\ref{game}.
	
	We first establish the following two lemmas:
	\begin{lemma}
		\label{lemma_G}
		The outcome of the two-player games $G$ and $G^{reset}$ are the same for the direct fixed window mean-payoff objective for some length $\sizeMin \geq 1$. Formally:
		\begin{displaymath}
			\underbrace{\sup\limits_{\sigma_1 \in strat_1(G)} \inf\limits_{\sigma_2 \in strat_2(G)} V_{s_{init}}^{f}(G,\sigma_1,\sigma_2)}_{\text{denoted } g(G,s_{init})} = \underbrace{\sup\limits_{\sigma_1 \in strat_1(G^{reset})} \inf\limits_{\sigma_2 \in strat_2(G^{reset})} 2 \cdot V_{s_{init}}^{f'}(G^{reset},\sigma_1,\sigma_2)}_{\text{denoted }g(G^{reset},s_{init})}
		\end{displaymath}
		where $f = f^{\sizeMin}_{DirFixWMP}$ and $f' = f^{2 \cdot \sizeMin}_{DirFixWMP}$.
	\end{lemma}
	
	\begin{proof}
		By construction of the game $G^{reset}$, if Player 2 has an optimal that never takes an edge in $E_5'$, then the equality holds.
		
		We prove that there exists an optimal strategy for Player 2 (in the game $G^{reset}$) that never takes an edge in $E'_5$. Let $\sigma_1^{reset} \in strat_1(G^{reset})$ and $\sigma_2^{reset} \in strat_2(G^{reset})$ be two optimal strategies for Player 1 and 2 respectively, in the game $G^{reset}$. Let $\pi = \pi_{(G^{reset},s_{init},\sigma_1^{reset},\sigma_2^{reset})} = s_0 \ldots$ and $\lambda = f'(\pi) \leq W$. Then, there exists a sequence of $2 \cdot \sizeMin$ states $\rho = s_n \ldots s_{n + 2 \cdot \sizeMin}$ such that $\wmp(\rho) = \lambda$. Then, since all the weights appearing in $G^{reset}$ are non negative, if an edge in $E_5'$ (that is, an edge of weight $(W+1) \cdot 2 \cdot \sizeMin$) was taken in that sequence, then the mean of the $2 \cdot \sizeMin$ edges of that sequence would have been above $(W+1) \cdot 2 \cdot \sizeMin/2 \cdot \sizeMin = W+1$. Therefore, we can conclude that no edge in $E_5'$ was taken in $\rho$.
		
		Consider the last time an edge $e$ in $E'_5$ is taken in the path $\pi$ before visiting for the first time the sequence of states $\rho$. Then, let us denote by $\rho' = s_{init} \ldots s_n \ldots s_{n+\sizeMin}$ the sequence of states, in $\pi$, that goes from $s_{init}$ (the destination of the edge $e$ since $e \in E'_5$) after seeing $e$. In that case, no edge in $E_5'$ is taken in $\rho'$. Therefore, the sequence of states $\rho$ is reachable from $s_{init}$ without taking any edge in $E_5'$ for Player 2.

		The lemma follows.
	\end{proof}
	
	We use the notations introduced in Lemma~\ref{lemma_G}.
	\begin{lemma}
		\label{lemma_G_s}
		For every state $s \in S'_1$, if we play the game starting from $s$, then the outcome of the game, that is $g(G^{reset},s)$, is at most $g(G^{reset},s_{init})$.
		That is:
		\begin{displaymath}
			\max\limits_{s \in S'_1} g(G^{reset},s) = g(G^{reset},s_{init})
		\end{displaymath}
	\end{lemma}

	\begin{proof}
		Let $\sigma_2 \in strat_2(G^{reset})$ be an optimal strategy from the state $s_{init}$. For every $s \in S'_1$, we prove that $g(G^{reset},s) \leq g(G^{reset},s_{init})$. We define the strategy $\sigma_2^s \in strat_2(G^{reset})$:
		\begin{itemize}
			\item $\forall \ s \ (s,s',2) \in (S'_1 \times S'_2)$, we have $\sigma_2^s(s (s,s',2)) = s_{init} \in S'_1$;
			\item $\forall s_0 \ldots s_n \in (S'_1 \times S'_2)^{k}$, for some $k = \frac{n+1}{2} \geq 2$, we have $\sigma_2^s(s_0 \ldots s_n) = \sigma_2(s_2 \ldots s_n)$.
		\end{itemize}
		By definition of $\sigma_2^s$, an edge in $E'_5$ is taken only once. Let $\sigma_1 \in strat_1(G^{reset})$ and let $\pi = \pi_{(G^{reset},s,\sigma_1,\sigma^s_2)} = s_0 \ldots$. Then, the windows opened in $s_0$ and $s_1$ are closed in $s_2 = s_{init}$ for any threshold $\lambda \leq W$ since $(s_1,s_2) = (s_1,s_{init}) \in E_5'$, that is $w'(s_1,s_2) = W \cdot (2 \cdot \sizeMin)$. Therefore, $f(\pi_{(G^{reset},s,\sigma_1,\sigma^s_2)}) =  f(\pi_{(G^{reset},s_{init},\sigma_1,\sigma_2)}) \leq g(G^{reset},s_{init})$. This is true for every strategy $\sigma_1 \in strat_1(\Gamma{reset})$. Therefore, $g(G^{reset},s) \leq g(G^{reset},s_{init})$. Since this is true for all $s \in S'_1$, we have $\max\limits_{s \in S'_1} g(G^{reset},s) = g(G^{reset},s_{init})$.
	\end{proof}

	We can proceed to the proof of Theorem~\ref{fix_mdp_hardness}.
	\begin{proof}
		Let $G = \langle S_1,S_2,s_{init},E,w \rangle$ be a two-player game and a length $\sizeMin \geq 1$. We denote $G^{reset} = \langle S_1',S_2',s_{init},E',w' \rangle$. We construct an MDP $\Gamma = \langle S,\bar{E},Act,s_{init},\bar{w},\mathbb{P} \rangle$ such that:
		\begin{itemize}
			\item $S = S'_1$;
			\item $\bar{E} \subseteq S \times Act \times S$;
			\item We define an action $\alpha_s$ for every $s \in S'_2$. Then, $Act = \lbrace \alpha_s \mid s \in S'_2 \rbrace \cup \lbrace go \rbrace$;
			\item For $s \in S_1 \cup S_2$, let $Succ(s) = \lbrace s' \in S_1 \cup S_2 \mid (s,s') \in E \rbrace$. Then, for every $s \in S'_1$, we have:
			\begin{align}
			Act(s) =
			\begin{cases}
			\lbrace \alpha_{s'} \mid s' \in Succ(s) \rbrace & \text{if } s \in S_1;\\
			\lbrace go \rbrace & \text{if } s = (a,b,1) \in S_1 \times S_2 \times \lbrace 1 \rbrace\\
			\end{cases}
			\end{align}
			\item For $e = (v,\alpha,v') \in \bar{E}$, we have:
			\begin{align}
			\bar{w}(e) =
			\begin{cases}
			w'((s,s',2),(s,s',1)) & \text{if } v = s \in S, v' = (s,s',1) \in S;\\
			w'(s',s'') & \text{if } v = (s,s',1) \in S, v' = s'' \in S.\\
			(W + 1) \cdot 2 \cdot \sizeMin & \text{if } v = s \in S, v' = s_{init} \in S;\\
			\end{cases}
			\end{align}
			\item Let $v \in S$. 
			\begin{itemize}
				\item If $v = s \in S_1$, then for $\alpha_{s'} \in Act(v)$, there are two transitions: to $(s,s',1)$ and $s_{init}$. Both have the same probability to occur, that is $0.5$. Formally:
				$\mathbb{P}(s,\alpha_{s'},(s,s',1)) = \mathbb{P}(s,\alpha_{s'},s_{init}) = 0.5$;
				\item If $v = (s,s',1) \in S_1 \times S_2 \times \lbrace 1 \rbrace$, then only action $go$ is available. Then, there are $|Succ(s')|$ successor to $v$ for that action. In fact, for every $s'' \in Succ(s')$, we have $\mathbb{P}((s,s',1),go,s'') = \frac{1}{|Succ(s')|}$.
			\end{itemize}
		\end{itemize}
		
		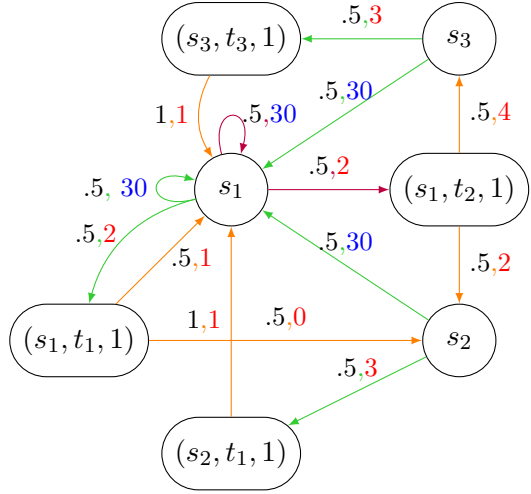
\begin{figure}
			\begin{center}
				\vspace{0pt}
\centering
\scalebox{1}{
	\begin{tikzpicture}
		\node[player1] (s1) at (-1,0) {$s_1$} ;
		\node[player1] (s2) at (2,-2) {$s_2$} ;
		\node[player1] (s3) at (2,2) {$s_3$} ;
		
		\node[player1] (s1t1) at (-3,-2) {$(s_1,t_1,1)$} ;
		\node[player1] (s1t2) at (2,0) {$(s_1,t_2,1)$} ;
		\node[player1] (s2t1) at (-1,-3.5) {$(s_2,t_1,1)$} ;
		\node[player1] (s3t3) at (-1,2) {$(s_3,t_3,1)$} ;
		
 \path[-latex] 
	 (s1) edge[pakistangreen,loop left] node[left] {\small \textcolor{black}{.5}, \textcolor{blue}{30}} (s1)
	 (s1) edge[pakistangreen,bend right] node[left] {\small  \textcolor{black}{.5},\textcolor{red}{2}} (s1t1)
	 
	 (s1) edge[purple,loop above] node[right] {\small  \textcolor{black}{.5},\textcolor{blue}{30}} (s1)
	 (s1) edge[purple] node[above] {\small  \textcolor{black}{.5},\textcolor{red}{2}} (s1t2)
	 
	 (s2) edge[pakistangreen] node[above] {\small  \textcolor{black}{.5},\textcolor{blue}{30}} (s1)
	 (s2) edge[pakistangreen] node[above] {\small  \textcolor{black}{.5},\textcolor{red}{3}} (s2t1)
	 
	 (s3) edge[pakistangreen] node[above] {\small  \textcolor{black}{.5},\textcolor{blue}{30}} (s1)
	 (s3) edge[pakistangreen] node[above] {\small  \textcolor{black}{.5},\textcolor{red}{3}} (s3t3)
	 
	 (s1t1) edge[pseudo] node[right] {\small  \textcolor{black}{.5},\textcolor{red}{1}} (s1)
	 (s1t1) edge[pseudo] node[above] {\small  \textcolor{black}{.5},\textcolor{red}{0}} (s2)
	 
	 (s1t2) edge[pseudo] node[right] {\small  \textcolor{black}{.5},\textcolor{red}{2}} (s2)
	 (s1t2) edge[pseudo] node[right] {\small  \textcolor{black}{.5},\textcolor{red}{4}} (s3)
	 
	 (s2t1) edge[pseudo] node[left] {\small  \textcolor{black}{1},\textcolor{red}{1}} (s1)
	 
	 (s3t3) edge[pseudo,bend right] node[left] {\small  \textcolor{black}{1},\textcolor{red}{1}} (s1)
		;
	\end{tikzpicture}
}
			\end{center}
			\caption{The MDP $\Gamma$ obtained from the two-player game $G$ in Figure~\ref{game}. It ensures that $G_\Gamma = G^{reset}$ modulo renaming of the states where $G^{reset}$ appears in Figure~\ref{new_game} (recall that $G_\Gamma$ is the two-player game obtained from the MDP $\Gamma$). The colors correspond to the available actions. Only in state $s_1$ more than one action is available.}
			\label{mdp_from_game}
		\end{figure}
		
		If the probability $\mathbb{P}'(s,\alpha,s')$ was not defined for two states $s,s' \in S'$, and an action $\alpha \in Act$ then $\mathbb{P}'(s,\alpha,s') = 0$. In Figure~\ref{mdp_from_game}, we can the MDP $\Gamma$ that was constructed from the game in Figure~\ref{game}. 	
		
		Now, by construction, $G_\Gamma$ (the game obtained from the MDP $\Gamma$) and $G^{reset}$ are identical modulo state renaming. Moreover, $\Gamma$ is an MEC since $G^{reset}$ is strongly connected. Then, by Theorem~\ref{val_MEC} and Lemma~\ref{lemma_G_s} we have that:
		\begin{displaymath}
			\mathbb{E}_{s_{init}}^{\Gamma}(f^{\sizeMin}_{FixWMP}) = \max\limits_{s \in S'_1} g(G^{reset},s) = g(G^{reset},s_{init})
		\end{displaymath}
		Moreover, by Lemma~\ref{lemma_G}, we have:
		\begin{displaymath}
			g(G,s_{init}) = g(G^{reset},s_{init})
		\end{displaymath}
		In fact: $ \mathbb{E}_{s_{init}}^{\Gamma}(f^{\sizeMin}_{FixWMP}) = g(G,s_{init})$.
		
		The size of $\Gamma$ is polynomial in the size of $G$. This concludes the proof.
	\end{proof}

		\subsection{\textit{Bounded Window Mean-Payoff}}
	Consider a Markov decision process $\Gamma = \langle S,E,Act,s_{init},w,\mathbb{P} \rangle$. Then, let
	\begin{displaymath}
	\mathbb{E}_{s_{init}}^{\Gamma}(f_{BWMP}) =  \sup\limits_{\sigma \in strat(\Gamma)} \mathbb{E}_{s_{init}}^{\MDPtoMC{\Gamma}{\sigma}}(f_{BWMP})
	\end{displaymath} 
	Similar to the \textit{fixed window mean-payoff}, the \textit{bounded window mean-payoff} is prefix independent. Therefore, we first consider the maximal end components. With the same arguments used in case of fixed window mean-payoff, if we consider an MEC $M = (T,A) \in MEC(\Gamma)$, one can argue that the expected value of the bounded window mean-payoff is the same for every state $s \in T$. This is true in every MEC $M \in MEC(\Gamma)$. Hence, for every $M \in MEC(\Gamma)$, we denote by $\lambda_M$ the (optimal) expected value of $f_{BWMP}$ from every state of $M$, computed among the paths that stay in $M$, that is $\lambda_M = \max\limits_{s \in S} \mathbb{E}_{s}^{M}(f_{BWMP})$. In fact, the value $\lambda_M$ can be expressed using a two-player game. Formally, we have the following theorem:
	\begin{theorem}
		\label{val_MEC_bounded}
		Let $M = (T,A)$ be a MEC of $\Gamma$. Then we have:
		\begin{equation}
		\label{eqn_bounded}
		\lambda_M = \max\limits_{s \in T} \underbrace{2 \cdot \sup\limits_{\sigma_1 \in strat_1(G_M)} \inf\limits_{\sigma_2 \in strat_2(G_M)} \underbrace{V_s^{f_{Mean}}(G_M,\sigma_1,\sigma_2)}_{\text{noted }\bar{f}(s,\sigma_1,\sigma_2)}}_{\text{noted }\bar{g}(s)}
		\end{equation} 
	\end{theorem}
	
	We first make a few observations.
	
	The value $\bar{g}(s)$ denotes two times the outcome of the two-player game $G_M$ resulting from $M$ where the function of interest is the \emph{mean-payoff} function. From \cite{EM79,ZP96}, we know that memoryless strategies are enough to solve the two-player game for the mean-payoff objective\footnote{Similarly to the fixed window mean-payoff objective, since there are finitely many memoryless strategies, the supremum and infimum become maximum and minimum respectively. Also since the two-player game with bounded window objective is determined, the $\sup$ and the $\inf$ can be interchanged.}.
	
	When we construct a two-player game from an MDP, an edge of weight $w$ is split into two edges of weight $0$ and $w$ (which divides the mean-payoff of a path by $2$). Therefore, if the mean-payoff of a path $\pi$ in the MDP equals $2 \cdot \lambda$, then the mean-payoff of its counterpart in the two-player game equals $\lambda$.
	%	is equal to $\lambda$ for a specific length equals to $2 \cdot l$ in the two-player game, then the fixed window mean payoff of the corresponding path in the MDP will be equal to $2 \cdot \lambda$, for a length equal to $l$. 
	Formally, for a strategy $\sigma_1 \in strat_1(M)$, and $\pi \in Paths_s^{\MDPtoMC{M}{\sigma_1}}$ for some state $s \in T$, we have $2 \cdot f_{Mean}(p_v^{\sigma_1}(\pi)) = f_{Mean}(\pi)$. That is why $\bar{g}(s)$ equals $2$ times the outcome of the game.

	Before proving the theorem, we first introduce the following lemma:
	\begin{lemma}
		\label{lemma_fix_bounded}
		Let $\mathcal{M} = \zug{S,E,s_{init},w,\mathbb{P}}$ be a Markov chain and let $\pi = s_0 \ldots \in Paths^\mathcal{M}$. Then, we have:
		\begin{displaymath}
		\forall l \geq 1,\; f_{FixWMP}^l(\pi) \leq f_{mean}(\pi)
		\end{displaymath}
	\end{lemma}
	
	\begin{proof}
		Let $l \geq 1$. We denote $f_{FixWMP}^l(\pi)$ by $\lambda_l$. By definition of $f_{FixWMP}^l$, we have $\pi \in FixWMP(\lambda_l,l)$. Therefore, there exists $n_l \in \nat$ such that for all $i \geq n_l$, we have $WMP(\pi(i \ldots i+l)) \geq \lambda_l$. We prove that $f_{Mean}(\pi(n_\epsilon \ldots)) \geq \lambda_l$. Since $f_{Mean}$ is prefix independent, that would prove that $f_{Mean}(\pi) \geq \lambda_l$. Let $\epsilon > 0$. We show that there exists $n_\epsilon \in \nat$ such that, for all $N \geq n_\epsilon$, $\frac{1}{N - n_\epsilon} \sum\limits_{k = u_0}^{M-1} w(s_k,s_{k+1}) \geq \lambda_l - \epsilon$. 
		
		For all $i \geq n_l$, there exists $p_i \in [l]$ such that $\frac{1}{p_i} \sum\limits_{k=0}^{p_i-1} w(s_k,s_{k+1}) \geq \lambda_l$. Let us denote by $l_i$ the smallest index satisfying this property. Then, analogous to the proof of Theorem~\ref{th_first}, we consider the series $(u_n)_{n \in \nat}$ defined by: 
		\begin{itemize}
			\item $u_0 = n_l$;
			\item For all $n \geq 0$, we have $u_{n+1} = u_n + l_{u_n}$. 
		\end{itemize}
		This series is defined such that:
		\begin{itemize}
			\item For all $n \in \nat$, $\frac{1}{u_{n} - u_0} \sum\limits_{k = u_0}^{u_{n}-1} w(s_k,s_{k+1}) = \frac{1}{u_{n} - u_0} \sum\limits_{m = 0}^{n-1} \underbrace{\sum\limits_{k = u_m}^{u_{m+1}-1} w(s_k,s_{k+1})}_{\substack{\geq \ l_{u_m} \cdot \lambda_l \ = \ (u_{m+1} - u_m) \cdot \lambda_l \\ \text{ by definition of } l_{u_m}}} \geq \lambda_l \cdot \frac{1}{u_{n} - u_0} \underbrace{\sum\limits_{m = 0}^{n-1} u_{m+1} - u_m}_{= u_n - u_0}$. In fact: $\frac{1}{u_{n} - u_0} \sum\limits_{k = u_0}^{u_{n}-1} w(s_k,s_{k+1}) \geq \lambda$\footnote{This does not prove that $f_{Mean}(\pi(u_0 \ldots)) = \liminf\limits_{n \rightarrow \infty} \frac{1}{n-u_0} \sum\limits_{k = u_0}^{n-1} w(s_k,s_{k+1}) \geq \lambda_l$. It only proves that $\limsup\limits_{n \rightarrow \infty} \frac{1}{n-u_0} \sum\limits_{k = u_0}^{n-1} w(s_k,s_{k+1}) \geq \lambda_l$.}
			\item For all $n \geq 0$, $u_{n+1} - u_n = l_{u_n} \geq 1$. Thus, for all $n \geq 0$, $u_{n} - u_0 \geq n$. In fact $\lim\limits_{n \rightarrow \infty} u_n = \infty$.
		\end{itemize}
		
		Let $n_\epsilon = u_{d}$ where $d \in \mathbb{N}$ ensures that: $\frac{d}{l+d} \geq \frac{\lambda_l - \epsilon}{\lambda_l}$ (such an integer $d$ exists since $\lim\limits_{a \rightarrow \infty} \frac{a}{a+l} = 1 > \frac{\lambda_l - \epsilon}{\lambda_l}$). Let $N \geq n_\epsilon$. Let $t \geq d$ be such that $N = u_t + x$ with $x \leq l$. Then:
		\begin{align*}
		\frac{1}{N - u_0} \sum\limits_{k = u_0}^{M-1} w(s_k,s_{k+1}) & = \frac{1}{N - u_0} (\sum\limits_{k = u_0}^{u_t-1} w(s_k,s_{k+1}) + \underbrace{\sum\limits_{k = u_t}^{M-1} w(s_k,s_{k+1})}_{\substack{\geq 0 \\ \text{ since all wieghts are positive }}}) \geq \frac{1}{N - u_0} \underbrace{\sum\limits_{k = u_0}^{u_t-1} w(s_k,s_{k+1})}_{\geq (u_t - u_0) \cdot \lambda_l}\\
		& \geq \frac{u_t - u_0}{N - u_0} \lambda_l = \frac{u_t - u_0}{x + u_t - u_0} \lambda_l \geq \frac{t}{x + t} \lambda_l \geq \frac{d}{x + d} \lambda_l \geq \underbrace{\frac{d}{l + d} \lambda_l}_{\substack{\geq \lambda_l - \epsilon \text{ by } \\ \text{definition of }d}} \geq \lambda_l - \epsilon
		\end{align*} 
		The last inequalities comes the fact that if $a \geq b$, then $\frac{a}{a + x} \geq \frac{b}{b + x}$ since $x \geq 0$. In addition, $u_t - u_0 \geq t \geq d$ and $x \leq l$.
		
		This holds for all $N \geq n_{\epsilon}$. In fact:
		\begin{displaymath}
		\forall \epsilon > 0,\; \exists n_\epsilon \in \nat \text{ such that } \forall N \geq n_\epsilon \text{ we have } \frac{1}{N - u_0} \sum\limits_{k = u_0}^{M-1} w(s_k,s_{k+1}) \geq \lambda_l - \epsilon
		\end{displaymath}
		Therefore, $f_{Mean}(\pi(u_0 \ldots)) = \liminf\limits_{n \rightarrow \infty} \frac{1}{n-u_0} \sum\limits_{k = u_0}^{n-1} w(s_k,s_{k+1}) \geq \lambda_l$. Then $f_{Mean}(\pi) = f_{Mean}(\pi(u_0 \ldots)) \geq \lambda_l = f_{FixWMP}^l(\pi)$. This concludes the proof.
	\end{proof}
	
	We can proceed to the proof of Theorem~\ref{val_MEC_bounded}.	
	\begin{proof}
		Let $v \in T$ be a state such that $\bar{g}(v) = \max\limits_{s \in T} \bar{g}(s)$. We proceed in two steps:
		\begin{itemize}
			\item[(1)] First we show that $\lambda_M \geq \bar{g}(v)$. Let $\sigma_1 \in strat_1(G_M)$ be an optimal memoryless strategy for Player $1$ from a state $v$. Recall (from Section~\ref{subsec_prelim_game}) that any strategy for Player 1 can be seen as a strategy in the MDP $M$. Consider the Markov chain $\MDPtoMC{M}{\sigma_1}$. Then, 
			\begin{displaymath}
				\mathbb{E}_{v}^{\MDPtoMC{M}{\sigma_1}}(f_{BWMP}) = \sum_{\mathcal{B} \in BSCC(\mathcal{M})} Pr_v(\lozenge \mathcal{B}) \cdot c_{\mathcal{B}}
			\end{displaymath}
			where $c_{\mathcal{B}} = \min\limits_{\rho \in ElemCycle(\mathcal{B})} \wmp(\rho)$ (from section~\ref{BWMP_MC}).
			We construct a strategy $\sigma_2 \in strat_2(G_M)$ such that $\bar{f}(v,\sigma_1,\sigma_2) \leq \mathbb{E}_{v}^{\MDPtoMC{M}{\sigma_1}}(f_{BWMP})$ where the function $\bar{f}$ is defined in Equation~\ref{eqn_bounded}.
			
			Let $\mathcal{B}' \in BSCC(\MDPtoMC{M}{\sigma_1})$ be such that $Pr(Reach_{v}(\mathcal{B}')) > 0$ and:
			\begin{displaymath}
				c_{\mathcal{B}'} = \min\limits_{\substack{\mathcal{B} \in BSCC(\MDPtoMC{M}{\sigma_1}) \land \\ Pr_v(\lozenge \mathcal{B}) > 0}} c_{\mathcal{B}}
			\end{displaymath}
			Let $\rho_{\mathcal{B}'} = s_0 \ldots s_{|\rho_{\mathcal{B}'}|} \in ElemCycle^{\mathcal{B}'}$ be such that $\mpay(\rho_{\mathcal{B}'}) = c_{\mathcal{B}'}$ (recall that $\mpay(\rho)$ is equal to mean of the weights appearing in $\rho$). 
			
			Then, let $\sigma_2 \in strat_2(G_M)$ be a strategy such that:
			\begin{itemize}
				\item $\sigma_2$ reaches the state $s_0$ (which is reachable since $s_0 \in \mathcal{B}'$ and $Pr_v(\lozenge \mathcal{B}) > 0$);
				\item Then, $\sigma_2$ takes indefinitely the cycle $\rho_{\mathcal{B}'}$.
			\end{itemize}
			By definition of $\sigma_2$, the path resulting from the two-player game $G_M$ played with strategies $\sigma_1$ for Player 1 and $\sigma_2$ for Player 2 ensures: $\pi_{(G_M,v,\sigma_1,\sigma_2)} = p^{\sigma_1}_v(\rho (\rho_{\mathcal{B}'})^\omega)$ for some $\rho \in FPaths_v^{\MDPtoMC{M}{\sigma_1}}$. Then, we have:
			\begin{displaymath}
				2 \cdot f_{Mean}(\pi_{(G_M,v,\sigma_1,\sigma_2)}) = f_{Mean}(\rho (\rho_{\mathcal{B}'})^\omega) = f_{Mean}((\rho_{\mathcal{B}'})^\omega) = \mpay(\rho_{\mathcal{B}'}) = c_{\mathcal{B}'}
			\end{displaymath}
			Note that $\bar{g}(v) \leq 2 \cdot f_{Mean}(\pi_{(G_M,v,\sigma_1,\sigma_2)})$ since $\bar{g}(v)$ is the value of the outcome of the two-player game when Player 2 opts for an optimal strategy. Moreover, we have:
			\begin{displaymath}
				c_{\mathcal{B}'} \leq \sum_{\mathcal{B} \in BSCC(\mathcal{M})} Pr_v(\lozenge \mathcal{B}) \cdot c_{\mathcal{B}} = \mathbb{E}_{v}^{\MDPtoMC{M}{\sigma_1}}(f_{BWMP})
			\end{displaymath}
			In addition:
			\begin{displaymath}
				\mathbb{E}_{v}^{\MDPtoMC{M}{\sigma_1}}(f_{BWMP}) \leq \sup\limits_{\sigma \in strat(M)} \mathbb{E}_{v}^{\MDPtoMC{M}{\sigma}}(f_{BWMP}) =  \mathbb{E}_{v}^{M}(f_{BWMP}) = \lambda_M
			\end{displaymath}
			
			Therefore, $\lambda_M \geq \mathbb{E}_{v}^{\MDPtoMC{M}{\sigma_1}}(f_{BWMP}) \geq c_{\mathcal{B}'} \geq \bar{g}(v) = \max\limits_ {s \in T} \bar{g}(s)$.
			
			\item[(2)] Now, we show that $\lambda_M \leq \bar{g}(v)$. Let $\sigma_1 \in strat(M)$ be a strategy (not necessarily a memoryless one) in the MEC and let $\pi' \in Paths^{\MDPtoMC{M}{\sigma_1}}_t$ for some $t \in T$. Let $l \geq 1$. For every state $s$ occurring in $\pi'$, there exists a strategy $\sigma_s \in strat_2(G_M)$ such that $2 \cdot f_{Mean}(\pi_{(G_M,s,\sigma_1,\sigma_s)}) \leq \bar{g}(s)$ (since $\sigma$ may not be an optimal strategy for Player 1). Let $\pi \in Paths^{\MDPtoMC{M}{\sigma_1}}_s$ such that $p^{\sigma_1}_s(\pi) = \pi_{(G_M,s,\sigma_1,\sigma_s)}$. Then, with Lemma~\ref{lemma_fix_bounded}, we have $2 \cdot f^{2 \cdot l}_{FixWMP}(\pi_{(G_M,s,\sigma_1,\sigma_s)}) = f^l_{FixWMP}(\pi) \leq f_{mean}(\pi) = 2 \cdot f_{Mean}(\pi_{(G_M,s,\sigma_1,\sigma_s)})$. Therefore: $2 \cdot f^{2 \cdot l}_{FixWMP}(\pi_{(G_M,s,\sigma_1,\sigma_s)}) \leq \bar{g}(s)$. In fact, for every state $s$ occurring in $\pi'$, there exists a strategy $\sigma_s \in strat_2(G_M)$ such that $2 \cdot f^{2 \cdot l}_{FixWMP}(\pi_{(G_M,s,\sigma_1,\sigma_s)}) \leq \bar{g}(s)$. Then, similar to the proof of Theorem~\ref{val_MEC}, we can prove that the fixed window mean-payoff of a path for length $l$ is almost surely at most $\bar{g}(v)$.  This is true for all $l \geq 1$ (since Lemma~\ref{lemma_fix_bounded} holds for all $l \geq 1$).  Moreover, for all $\pi \in Paths^{\MDPtoMC{M}{\sigma_1}}$, we have:
			\begin{displaymath}
				f_{BWMP}(\pi) = \sup\limits_{l \geq 1} f^l_{FixWMP}(\pi)
			\end{displaymath}
			
			Therefore, almost surely, the bounded window mean-payoff of a path is at most $\bar{g}(v)$. Thus, we have $\mathbb{E}_t^{\MDPtoMC{M}{\sigma_1}}(f_{BWMP}) \leq \bar{g}(v)$. This is true for every strategy $\sigma_1 \in strat(M)$ and for all state $t \in T$. That is, $\forall \sigma \in strat(M), \forall t \in T$ we have $\mathbb{E}_t^{\MDPtoMC{M}{\sigma}}(f_{BWMP}) \leq \bar{g}(v)$. Hence $\lambda_M \leq \bar{g}(v) = \max\limits_{s \in T} \bar{g}(s)$.
		\end{itemize}
		The theorem follows. 
	\end{proof}
	
	\begin{corollary}
		\label{corollary_2}
		An optimal strategy for the bounded window mean-payoff objective in MDPs can be found among memoryless strategies.
	\end{corollary}
	
	\begin{proof}
		We construct an optimal strategy in the same way we did for the fixed window mean objective (in the proof of Corollary~\ref{corollary_1}). It differs in the third part. We apply an optimal strategy that maximizes the outcome of the two-player bounded window mean-payoff game from the optimal state. That two-player can be solved with memoryless strategies. Since the first two steps can be done with memoryless strategies, the corollary follows.
	\end{proof}	
	
	Now, to compute $\mathbb{E}^\Gamma_{s_{init}}(f_{BWMP})$, we follow the same steps as we did for the \textit{fixed window mean-payoff objective}, that is:
	\begin{itemize}
		\item We construct a new MDP $\Gamma^{MEC}_B$ from $\Gamma$ where every weight appearing in each MEC $M$ is replaced by $\lambda_M$.
		\item Then, we compute the expected mean-payoff in the new MDP $\Gamma^{MEC}_B$. This is equal to the expected value of the bounded window mean-payoff in $\Gamma$.
	\end{itemize}
	
	The algorithm used here are very similar to Algorithms~\ref{fixed_mec}, ~\ref{transform_mec} and ~\ref{fixed_mdp}. Algorithms~\ref{fixed_mec} is modified into Algorithm $BWMP-MEC$ where, in line 8, the algorithm called is Algorithm $MeanPayoff$ that solves the two-player game for the mean-payoff objective. Moreover, Algorithm $BWMP-MEC$ works as if the $\sizeMin$ given as a parameter (that we do not need anymore, since we are in the bounded case) was equal to $|T|$ the number of states of the $MEC$ on which it is called. That is because the value of $\lambda_M$ will be equal to the mean of an elementary cycle (as we have shown in Section~\ref{BWMP_MC}) whose length is at most $|T|$ by definition. Corresponding to $Replace-MEC$, in Algorithm~\ref{transform_mec}, we have instead $BWMP-Replace-MEC$ where in Line 4 we now call $BWMP-MEC$. Instead of $FixWMP-MDP$, in Algorithm~\ref{fixed_mdp}, we now have $BWMP-MDP$ that calls $BWMP-Replace-MEC$ in Line 1. 
	
	Let us denote by $\mathbb{D}$ the complexity of the $MeanPayoff$ Algorithm (recall that solving the two-player mean-payoff game is in $NP \cap coNP$). In Algorithm~\ref{fixed_mec}, we enter the \texttt{while} loop $O(log(W \cdot |T|^2))$ times. Therefore, the complexity of Algorithm~\ref{fixed_mec} is in $O(log(W \cdot |S|^2) \cdot \mathbb{D})$. Then, in Algorithm~\ref{transform_mec}, Algorithm~\ref{fixed_mec} is called at most $|S|$ times. Therefore, the complexity of Algorithm~\ref{transform_mec} is in $O(log(W \cdot |S|^2) \cdot \mathbb{D} \cdot |S|)$ (since $|T| \leq |S|$ for all MEC $M = (T,A)$).
	
	For an MDP $\Gamma = \langle S,E,Act,s_{init},w,\mathbb{P} \rangle$, Algorithm $ExpectedMeanPayoff$ called on $\Gamma$ runs in time $O(Avg(|S|,|Act|))$. Therefore, the complexity of Algorithm~\ref{fixed_mdp} is in $O(log(W \cdot |S|^2) \cdot \mathbb{D} \cdot |S| + Avg(|S|,|Act|))$.
	
	We have the following theorem:
	\begin{theorem}
		\label{complexity_mdp_bounded}
		Given an MDP and a threshold $\lambda \in \rat$, deciding whether the expected value of $f_{BWMP}$ in the MDP is above $\lambda$ is in \textsc{NP $\cap$ coNP}.
	\end{theorem}

	\subsubsection{Hardness}

We now consider the hardness of the problem. Given an MDP $\MDP$ with an initial state $s_{init}$ and a threshold $\lambda$, we show that  the \emph{bounded window mean-payoff} problem of checking if $\expect_{s_{init}}^{\Gamma}(f_{BWMP}) \geq \lambda$ is at least as hard as solving the two-player game for the mean-payoff objective.

\begin{theorem}
	\label{bounded_mdp_hardness}
	The bounded window mean-payoff problem for MDP is at least as hard as solving a two-player game for the mean-payoff objective (for polynomial reductions).
\end{theorem}

Given a two-player game $G = \langle S_1,S_2,s_{init},E,w \rangle$ (in which we assume that all the weight are non-negative, as we did in Section~\ref{fixed_mdp_hardness}), consider a new two-player game $G^{reset}_B = \langle S'_1,S'_2,s_{init},E',w' \rangle$ 
%very similar to 
that is the same as $G^{reset}$ (see Section~\ref{fixed_mdp_hardness}), except that the edges in $E'_5$ have a weight equal to $(W+1) \cdot |S'_1 \uplus S'_2|$. Considering mean-payoff objective, we have the following lemma (which is analogous to Lemma~\ref{lemma_G}):

\begin{lemma}
	\label{lemma_G_bounded}
	The outcome of the two-player games $G$ and $G^{reset}$ are the same for the mean-payoff objective. Formally:
	\begin{displaymath}
	\underbrace{\sup\limits_{\sigma_1 \in strat_1(G)} \inf\limits_{\sigma_2 \in strat_2(G)} V_{s_{init}}^{f_{Mean}}(G,\sigma_1,\sigma_2)}_{\text{denoted } h(G,s_{init})} = \underbrace{\sup\limits_{\sigma_1 \in strat_1(G^{reset})} \inf\limits_{\sigma_2 \in strat_2(G^{reset})} 2 \cdot V_{s_{init}}^{f_{Mean}}(G^{reset}_B,\sigma_1,\sigma_2)}_{\text{denoted }h(G^{reset},s_{init})}
	\end{displaymath}
\end{lemma}

\begin{proof}
	Similar to the proof of Lemma~\ref{lemma_G}, we prove that there exists an optimal strategy for Player 2 that never takes an edge in $E'_5$. Let $\sigma_1 \in strat_1(G^{reset}_B)$ and $\sigma_2 \in strat_2(G^{reset}_B)$ be two optimal memoryless strategies for Player 1 and Player 2 respectively (see \cite{EM79,ZP96}). Let $\pi = \pi_{(G_B^{reset},s_{init},\sigma_1,\sigma_2)} = s_0 \ldots$. 
	A	ssume towards a contradiction that $\sigma_2$ chooses an edge $e = (s_k,s_{k+1}) \in E'_5$ where $s_{k+1} = s_{init} = s_0$. Then, because both strategies are memoryless, the path $\pi$ is equal to $(s_0 \ldots s_k s_{k+1})^{\omega}$ with $k \leq |S'_1 \uplus S'_2|$. Therefore, since we assume that all the weights appearing in $G$, and therefore in $G^{reset}_B$, are non-negative and since the weight of an edge in $E'_5$ is equal to $(W+1) \cdot |S'_1 \uplus S'_2|$, then we have (recall that the function $MP$ associates to a finite sequence of edges its mean-payoff):
	\begin{displaymath}
		f_{mean}(\pi) = MP(s_0 \ldots s_k s_{k+1}) \geq \frac{1}{k} w'(s_k,s_{k+1}) \geq \frac{1}{|S'_1 \uplus S'_2|} (W+1) \cdot |S'_1 \uplus S'_2| = W+1
	\end{displaymath}
	Any strategy for Player 2 that never takes an edge in $E'_5$ yields an mean-payoff of at most $W$ (by definition of $W$). Hence the contradiction. That is, $\sigma_2$ never takes an edge in $E'_5$. This concludes the proof.
\end{proof}

Then, we have the following lemma (we use the notations introduced in Lemma~\ref{lemma_G_bounded}):
\begin{lemma}
	\label{lemma_G_s_B}
	For every state $s \in S'_1$, if we play the game $G^{reset}_B$ starting from $s$, then the outcome of the game, that is $h(G^{reset}_B,s)$, is at most $h(G^{reset}_B,s_{init})$.
	That is:
	\begin{displaymath}
	\max\limits_{s \in S'_1} h(G^{reset}_B,s) = h(G^{reset}_B,s_{init})
	\end{displaymath}
\end{lemma}
This lemma can be proven with the same proof we used to prove Lemma~\ref{lemma_G_s}. Then, the proof of Theorem~\ref{bounded_mdp_hardness} is identical to the proof of Theorem~\ref{fix_mdp_hardness} by using Lemma~\ref{lemma_G_bounded}, Lemma~\ref{lemma_G_s_B} and Theorem~\ref{val_MEC_bounded}. Specifically, from a two-player game $G$ and a initial state $s_{init}$, we construct an MDP $\Gamma$ such that the two-player games $G_{\Gamma}$ and $G_B^{reset}$ are the same modulo state renaming. Then, the expected value of the bounded window mean-payoff in $\Gamma$ is equal to the maximum over all staring state of the outcome of the two-player game $G_B^{reset}$ for the mean-payoff function (that comes from Theorem~\ref{val_MEC_bounded}). Further, that maximum is achieved by starting in the state $s_{init}$ (from Lemma~\ref{lemma_G_s_B}). Finally, the outcome of the two-player game $G_B^{reset}$ from $s_{init}$ for the mean-payoff function is equal to the outcome of the two-player game $G$ from $s_{init}$ (from Lemma~\ref{lemma_G_bounded}). The reduction is complete.

		\subsection{\textit{Direct Fixed Window Mean-Payoff}}
	Consider an MDP $\Gamma = \langle S,E,Act,s_{init},w,\mathbb{P} \rangle$. Let
	\begin{displaymath}
	\mathbb{E}_{s_{init}}^{\Gamma}(f^{\sizeMin}_{DirFixWMP}) =  \sup\limits_{\sigma \in strat(\Gamma)} \mathbb{E}_{s_{init}}^{\MDPtoMC{\Gamma}{\sigma}}(f^{\sizeMin}_{DirFixWMP})
	\end{displaymath} 
	
	The \textit{direct fixed window mean-payoff} is not prefix independent, and therefore we cannot only look at the MECs. We present an algorithm that is exponential in $\sizeMin$ and then we show the {\sc PSpace-hardness} of our problem via a reduction from the 
	%$k$-$th$ largest subset sum problem (see \cite{HK16}).
	threshold problem for shortest path objectives (see \cite{HK15}).
	
	\subsubsection{Algorithm}	
	Given a weighted Markov decision process $\Gamma = \langle S,E,s_{init},Act,w,\mathbb{P} \rangle$, we construct a new MDP $\Gamma_{\sizeMin} = \langle S',E',s'_{init},Act,w',\mathbb{P}' \rangle$ with:
	\begin{itemize}
		\item $S' = S \times ([W]_0)^{\sizeMin-1} \times \lbrace \frac{p}{q} \mid q \in [\sizeMin],\; p \in [q \cdot W]_0 \rbrace$;
		\item $E \subseteq S' \times Act \times S'$;
		\item $s'_{init} = (s_{init},[W, \ldots, W],W) \in S'$, where $W = \max\limits_{e \in E} w(e)$;
		\item $\forall t = (s,[a_1, \ldots, a_{\sizeMin - 1}],\lambda) \in S'$, we have $Act(t) = Act(s)$;
		\item $\mathbb{P}'(t = (s,[a_1, \ldots, a_{\sizeMin - 1}],\lambda), \alpha, t' = (s',[b_1, \ldots, b_{\sizeMin - 1}],\lambda')) = \mathbb{P}(s,\alpha,s')$ if $t,t' \in S'$, $\alpha \in Act$, and:
		\begin{itemize}
			\item For all $i \in [\sizeMin-2]$, we have $b_i = a_{i+1}$;
			\item $b_{\sizeMin-1} = w(s,\alpha,s')$;
			\item $\lambda' = \min (\lambda,\max\limits_{l \leq \sizeMin} \frac{1}{l} \sum\limits_{k=1}^{l} a_k)$ where we denote by $a_{\sizeMin}$ the number $b_{\sizeMin-1}$.
		\end{itemize}
		\item $w'(t = (s,[a_1, \ldots, a_{\sizeMin - 1}],\lambda),\alpha,t' = (s',[b_1, \ldots, b_{\sizeMin - 1}],\lambda')) = \lambda$ if $t,t' \in S'$ and $\alpha \in Act$.
	\end{itemize}
	Similarly to the previous constructions, if the probability $\mathbb{P}'(s,s')$ is not defined above for two states $s,s' \in S'$, then $\mathbb{P}'(s,s') = 0$. For convenience of notations, for a state $t = (s,[a_1, \ldots, a_{\sizeMin - 1}],\lambda) \in S'$, we have:
	\begin{itemize}
		\item $t^0$ refers to $s$;
		\item for all $k \leq \sizeMin-1$, $t^{1,k}$ refers to $a_k$;
		\item $t^2$ refers to $\lambda$.
	\end{itemize}
	Moreover, for every path $\pi = t_0 t_1 \ldots \in Paths^{\MDPtoMC{\Gamma_{\sizeMin}}{\sigma}}(s'_{init})$ for some $\sigma \in strat(\Gamma_{\sizeMin})$, we denote by $\pi^0 \in S^\omega$ the sequence of states $t_0^0 t_1^0 \ldots$.
	
	Informally, the idea of this construction is the following: consider a state $t \in S'$. This state corresponds to a finite path $\rho$ in $\Gamma$ such that $Last(\rho) = t^0$. Moreover, the last $\sizeMin - 1$ weights encountered in $\rho$ are stored in $t^1$. Finally, $t^2$ keeps track of the minimum window mean-payoff seen so far in $\rho$, which is reflected on the definition of $\lambda'$. Note that the values of $t_0^2 t_1^2 \ldots$ is a non decreasing sequence.
	
	The introduction of the new MDP $\Gamma_{\sizeMin}$ is justified by the following theorem (recall that we denote by $f_{Mean}$ the mean-payoff function): 
	
	\begin{theorem}
		\label{th_mdp_dir}
		For an MDP $\Gamma$, with initial state $s_{init}$, we have:
		\begin{displaymath}
		\mathbb{E}_{s_{init}}^{\Gamma}(f^{\sizeMin}_{DirFixWMP}) = \mathbb{E}_{s'_{init}}^{\Gamma_{\sizeMin}}(f_{Mean})
		\end{displaymath}
		That is, the expected value of the direct fixed window mean-payoff for the window length $\sizeMin$ in $\Gamma$ is equal to the expected value of the mean-payoff in $\Gamma_{\sizeMin}$.
	\end{theorem}
	
	The proof of this theorem is quite elaborate and requires the introduction of several lemmas. We give here the scheme of the proof, the details are provided in the appendix.
	\begin{itemize}
		\item[1.] First, we introduce a function $st_1: strat(\Gamma) \longrightarrow strat(\Gamma_{\sizeMin})$ that maps a strategy in $\Gamma$ to a strategy in $\Gamma_{\sizeMin}$;
		\item[2.] Then, we establish that, for every $\sigma \in strat(\Gamma)$, for every path $\pi \in \MDPtoMC{\Gamma_{\sizeMin}}{st_1(\sigma)}$, the sequence of states $\pi^0$ forms a path in $\MDPtoMC{\Gamma}{\sigma}$ and the mean-payoff of $\pi$ is equal to direct fixed window mean-payoff of $\pi^0$;
		\item[3.] Next, for $\sigma \in strat(\Gamma)$, we introduce a function $s^\sigma: FPaths^{\MDPtoMC{\Gamma}{\sigma}}(s_{init}) \longrightarrow S'$ that maps every finite paths in $\Gamma^\sigma$ into a state in $\Gamma_{\sizeMin}$. We show that the definition of $s^{\sigma}$ is consistent with the definition of $\Gamma_{\sizeMin}$ and $st_1$. That is, for $\sigma \in strat(\Gamma)$, if we consider a path $t_0 t_1 \ldots \in Paths^{\MDPtoMC{\Gamma_{\sizeMin}}{st_1(\sigma)}}(s'_{init})$, we have $s^\sigma(t_0^0 t_1^0 \ldots t_{n-1}^0 t_n^0) = t_n$;
		\item[4.] Afterwards, for all $\sigma \in strat(\Gamma)$, we establish that there exists a bijection between the set of infinite paths of $\MDPtoMC{\MDP}{\sigma}$ and ${\MDPtoMC{\MDP_{\sizeMin}}{st_1(\sigma)}}$ that ensures similar properties ensured by the bijection of Lemma~\ref{bijec_window}, thus proving that $\mathbb{E}_{s_{init}}^{\MDPtoMC{\MDP}{\sigma}}(f^{\sizeMin}_{DirFixWMP}) = \mathbb{E}_{s'_{init}}^{\MDPtoMC{\MDP_{\sizeMin}}{st_1(\sigma)}}(f_{Mean})$. With this result, we are able to prove $\mathbb{E}_{s_{init}}^{\Gamma}(f^{\sizeMin}_{DirFixWMP}) \leq \mathbb{E}_{s'_{init}}^{\Gamma_{\sizeMin}}(f_{Mean})$;
		\item[5.] Finally, we introduce a function $st_2: strat_0(\Gamma_{\sizeMin}) \longrightarrow strat(\Gamma)$ (whose definition is using the function $s^{\sigma}$) that ensures that, for all $\sigma_0 \in strat_0(\Gamma_{\sizeMin})$, we have $\mathbb{E}_{s'_{init}}^{\MDPtoMC{\MDP_{\sizeMin}}{st_1(st_2(\sigma_0))}}(f_{Mean}) = \mathbb{E}_{s'_{init}}^{\MDPtoMC{\MDP_{\sizeMin}}{\sigma_0}}(f_{Mean})$. From that result, it is possible to conclude that $\mathbb{E}_{s_{init}}^{\Gamma}(f^{\sizeMin}_{DirFixWMP}) \geq \mathbb{E}_{s'_{init}}^{\Gamma_{\sizeMin}}(f_{Mean})$. 
	\end{itemize}
	
	Algorithm~\ref{mdp_dir_fix} computes $\mathbb{E}_{s_{init}}^{\Gamma}(f^{\sizeMin}_{DirFixWMP})$ using Theorem~\ref{th_mdp_dir}.
	\begin{algorithm}
		\caption{MDPDirFixWMP($\Gamma,s_{init},\sizeMin$)}
		\label{mdp_dir_fix}
		\begin{algorithmic}[1]
			\Require{$\Gamma = \langle S,E,s_{init},Act,w,\mathbb{P} \rangle$ is a weighted Markov decision process, $s_{init} \in S$ and $\sizeMin \in \mathbb{N}_0$}
			\Ensure{$\lambda$ is equal to $\mathbb{E}_{s_{init}}^{\Gamma}(f^{\sizeMin}_{DirFixWMP})$}
			\State {$\Gamma_{\sizeMin} := \langle S',E',s'_{init},Act',w',\mathbb{P}' \rangle$}
			\State {$\lambda := ExpectedMeanPayoff(\Gamma_{\sizeMin})$}
			\State
			\Return {$\lambda$}
		\end{algorithmic}
	\end{algorithm}
	The size of the MDP $\Gamma_{\sizeMin}$ is in $O(|S| \cdot W^{\sizeMin-1} \cdot W \cdot \sizeMin^2)$. Then, Algorithm $ExpectedMeanPayoffMDP$ runs in time $O(Avg(|S| \cdot W^{\sizeMin} \cdot \sizeMin^2,|Act|))$, when called on $\Gamma_{\sizeMin}$. That is also the complexity of Algorithm~\ref{mdp_dir_fix}.
	
	We have the following theorem:
	\begin{theorem}
		\label{complexity_mdp_direct_fixed}
		The expected value of $f^{\sizeMin}_{DirFixWMP}$ in an MDP $\Gamma$ can be computed in time $O(p_3(|S| \cdot W^{\sizeMin} \cdot \sizeMin^2))$ where $p_3$ is a polynomial function and $W$ is the maximum weight appearing in the MDP $\Gamma$.
	\end{theorem}

	\begin{theorem}
	\label{direct_hardness}
Given an MDP $\MDP$ with an initial state $\initState$, a window length $\sizeMin$ and a $\lambda \in \rat$, deciding whether $\expect_{\initState}^{\Gamma}(\fnDirFixedWindow) \ge \lambda$ is {\sc PSpace-Hard}.
\end{theorem}
\begin{proof}
We show a reduction from the threshold probability problem for shortest path objectives \cite{HK15}. 
An instance of the threshold probability problem is given by an MDP $\Gamma = (S, E, Act, \initState, w, \mathbb{P})$ where w.l.o.g., we have that $w$ assigns positive weights on the edges, $T \subseteq S$ is a set of target states, and for a strategy $\sigma$, the truncated sum $TS^{T} : Paths(\Gamma^{[\sigma]}) \longrightarrow \mathbb{N} \cup {\infty}$ up to $T$ from the initial state $\initState$ is defined as 
$$
TS^{T}(\rho) = \left\{
\begin{array}{ll}
\sum_{i=0}^{n-1} w(e_i) & \mbox{if } \exists n \mbox{ such that } \rho(n) \in T \mbox{ and } \forall i \leq n-1, \mbox{ we have } \rho(i) \not \in T\\
\infty & \mbox{if } \forall i \geq 0, \rho(i) \not \in T,
\end{array}
\right.
$$
where $e_i=(\rho(i),a,\rho(i+1))$, $a \in Act$; for a threshold $L \in \nat$, and a probability threshold $p$, the problem asks to decide if there exists a strategy $\sigma$ such that
$\mathbb{P}_{\Gamma^{[\sigma]},\initState}[\{\rho \in Paths(\Gamma^{[\sigma]}) \mid TS^{T}(\rho) \leq L\}] \geq p$.
The problem is known to be {\sc PSpace-Complete}, even for acyclic MDPs \cite{HK15}. %(also, in \cite{HK15}, they consider only deterministic strategy). 
The target set $T$ is assumed to be made of absorbing states (i.e., with self-loops); the acyclicity is to be interpreted over the rest of the underlying graph.

Let $\Gamma = (S,E,Act,\initState,w,\mathbb{P})$, where $S = T \uplus V$, and $T$ is a set of target vertices.
The acyclicity of that MDP implies that, from the initial state $\initState \not \in T$, it takes at most $|S|-1$ steps to reach a vertex in $T$. %(since we do not consider an MDP where there could be a dead end).
Let $W$ be the maximum weight appearing in $\Gamma$.
We assume that $L \le W \cdot (|S|-1)$, otherwise the problem is trivial.

We construct a new MDP $\Gamma' = (S',E,Act',\initState,w',\mathbb{P}')$ where $S' = S \cup \{s_{\final_1}, s_{\final_2}\}$, $Act' = Act \cup \{\looping,\alpha, \beta\}$.
The set of edges $E'= \{(v,a,s) \:|\: (v,a,s) \in E$, $v \in V$, $s \in S\}$ $\cup$
$\{(t,\alpha,s_{\final_1}) \:|\: t \in T\}$ $\cup$
$\{(t,\beta,s_{\final_2}) \:|\: t \in T\}$ $\cup$
$\{(s_{\final_1},\looping,s_{\final_1})\}$ $\cup$
$\{(s_{\final_2},\looping,s_{\final_2})\}$ $\cup$
$\{(v,\beta,s_{\final_{2}}) \:|\: v \in V$ and there is no outgoing edge from $v$ in $\Gamma\}$.
%The set $E'$ of edges are such that for all $e \in E'$, we have $\mathbb{P}'(e) > 0$, and 
The probability function $\mathbb{P}'$ is defined as:
\begin{itemize}
	\item $\mathbb{P}'(v,a,s) = \mathbb{P}(v,a,s)$ such that $(v,a,s) \in E$, $v \in V$, $s \in S$;
	\item $\mathbb{P}'(t,\alpha,s_{\final_1}) = 1$ for $t \in T$;
	\item $\mathbb{P}'(t,\beta,s_{\final_2}) = 1$ for $t \in T$;
	\item $\mathbb{P}'(s_{\final_j},\looping,s_{\final_j}) = 1$ for $j \in \{1, 2\}$;
	\item $\mathbb{P}'(v,\beta,s_{\final_{2}}) = 1$ for $(v,\beta,s_{\final_{2}}) \in E'$ and $v \in V$;
\end{itemize}
%Furthermore, every edge from a state in $S$ to a state in $T$ (in the original MDP $\Gamma$) is kept with the same probabilities and opposite weight, however it is now going to the state $t$. Otherwise, 
%All the probabilities that are not defined here are equal to 0. 
%The actions $\alpha$ and $\beta$ (playing the same role they did for the proof of the PP-hardness for DirWMP in MDPs) are only available in the states of $T$.
The weight function $w'$ is defined as follows.
\begin{itemize}
	\item $w'(v,a,s) = -w(v,a,s)$ such that $(v,a,s) \in E$, $v \in V$, $s \in S$;
	\item $w'(t, \alpha,s_{\final_1}) = L$, for $t \in T$;
	\item $w'(t, \beta,s_{\final_2}) = W \cdot (|S|-1)$, for $t \in T$;
	\item $w'(s_{\final_1},\looping,s_{\final_1}) = 0$;
	\item $w'(s_{\final_2},\looping,s_{\final_2}) = -\frac{1}{|S|}$, and
	\item $w'(v,\beta,s_{\final_{2}}) = W \cdot (|S|-1)$ for $(v,\beta,s_{\final_{2}}) \in E'$ and $v \in V$.
\end{itemize}
Let $\sizeMin = |S|$.
Starting from $\initState$, since the weights on all the edges on the paths leading to a state in $t \in T$ are negative, the direct fixed window mean-payoff will consider paths until they reach $s_{\final_j}$ for $j \in \{1,2\}$ given that the weights on the edges outgoing from $t$ are positive.

We now call a path to be \emph{good} if $t$ appears in the path for some $t \in T$, and the sum of the edges from $\initState$ to $t$ is at least $-L$, otherwise the path is \emph{bad}.
Note that for a good path, choosing $\alpha$ leads to a direct fixed window mean-payoff of $0$, while choosing $\beta$ leads to direct fixed window mean-payoff of $-\frac{1}{|S|}$.
On the other hand, for a bad path, choosing $\alpha$ gives a direct fixed window mean-payoff of at most $-\frac{1}{|S|}$, while choosing $\beta$ gives a direct fixed window mean-payoff of $-\frac{1}{|S|}$.
Therefore, for an optimal strategy, the direct fixed window mean-payoff for a \emph{good} path is 0, and for a \emph{bad} path, it is $-\frac{1}{|S|}$.

We have $|\Gamma'| = O(\poly(|\Gamma|))$. 
Furthermore, the expected value of the direct fixed window mean-payoff, $\expect_{\initState}^{\Gamma}(\fnDirFixedWindow) \ge p \cdot 0 + (1-p) \cdot -\frac{1}{|S|} = -(1-p) \cdot \frac{1}{|S|}$ iff there is a solution to the threshold probability problem.

Note that since $\sizeMin = |S|$, deciding whether the expected value of the direct fixed window mean-payoff for an MDP is greater than or equal to some threshold is \textsc{PSpace-hard} even when $\sizeMin$ is given in unary. Thus, we cannot expect to have an algorithm that is polynomial in the value of $\sizeMin$ unless \textsc{P=PSpace}\footnote{The reduction does not work for Markov chains since we cannot get a threshold for the window mean-payoff that separates the cases when there is a solution to the threshold probability problem for shortest path objective and when a solution to the problem does not exist. That is, if the sum of path from $\initState'$ to $t$ is below $L$ and the edge corresponding to action $\alpha$ is taken in $t$, we do not know how much below $0$ will the window mean-payoff be. }.
\end{proof}

	\section{Relating Fixed Window Mean-Payoff and Bounded Window Mean-Payoff}
\label{sec:relate}
Let $\Pi$ be the set of strategies in an MDP $M$.
We show the following.
\begin{theorem}
\label{thm-sup_relate}
$\sup\limits_{\sigma \in \Pi}\{\expect^{M[\sigma]} (\fnBoundedWindow)\} = \sup\limits_{\sizeMin} \: \sup\limits_{\sigma \in \Pi}\{\expect^{M[\sigma]} (\fnFixedWindow)\}$.
\end{theorem}
\begin{proof}
First we show in Appendix \ref{app:supsup} that for every pair of sets $A$ and $B$ and for every function $f$, we have that $\sup\limits_{a \in A} \: \sup\limits_{b \in B} f(a,b) = \sup\limits_{b \in B} \: \sup\limits_{a \in A} f(a,b)$.

Thus $\sup\limits_{\sizeMin} \: \sup\limits_{\sigma \in \Pi}\{\expect^{M[\sigma]} (\fnFixedWindow)\} = \sup\limits_{\sigma \in \Pi} \: \sup\limits_{\sizeMin} \{\expect^{M[\sigma]} (\fnFixedWindow)\}$.
Form the definition of $\expect (\fnBoundedWindow)$, in every Markov chain, we have $\expect (\fnBoundedWindow) \geq \sup\limits_{\sizeMin} \{\expect (\fnFixedWindow)\}$.
Thus for every strategy $\sigma \in \Pi$, we have that $\expect^{M[\sigma]}(\fnBoundedWindow) \geq \sup\limits_{\sizeMin} \{\expect^{M[\sigma]} (\fnFixedWindow)\}$.
Hence $\sup\limits_{\sigma \in \Pi}\{\expect^{M[\sigma]} (\fnBoundedWindow)\} \geq \sup\limits_{\sigma \in \Pi} \: \sup\limits_{\sizeMin} \{\expect^{M[\sigma]} (\fnFixedWindow)\}$.

For the other direction, we note that in an MDP, there exists a memoryless strategy $\sigma_0$ such that $\sup\limits_{\sigma \in \Pi}\{\expect^{M[\sigma]} (\fnBoundedWindow)\} = \expect^{M[\sigma_0]} (\fnBoundedWindow)$.
Now, from the proof of Theorem \ref{thm:MC_BWMP}, we have $\sup\limits_{\sizeMin} \{\expect^{M[\sigma_0]} (\fnFixedWindow)\} = \expect^{M[\sigma_0]} (\fnBoundedWindow)$.
Hence $\sup\limits_{\sigma \in \Pi} \: \sup\limits_{\sizeMin} \{\expect^{M[\sigma]} (\fnFixedWindow)\} \geq \sup\limits_{\sigma \in \Pi} \{\expect^{M[\sigma_0]} (\fnFixedWindow)\} = \sup\limits_{\sigma \in \Pi}\{\expect^{M[\sigma]} (\fnBoundedWindow)\}$.
\end{proof}

As a direct implication of the above, we also have the following proposition.
\begin{proposition} \label{prop:relate}
In an MDP $M$, for every $\lambda \in \reals$, the following holds:
$\exists \sigma \in \Pi \: : \: \expect^{M[\sigma]} (\fnBoundedWindow) > \lambda \Longleftrightarrow \exists \sizeMin \exists \sigma' \in \Pi \: : \: \expect^{M[\sigma'] (\fnFixedWindow)} > \lambda$
\end{proposition}
\begin{proof}
For every $\lambda$, if there exists a strategy $\sigma \in \Pi$ such that 
%$\lambda < \exists \displaystyle{\sigma_{\sigma \in \Pi}}\: \expect^{M[\sigma]} (\fnBoundedWindow)$ 
$\lambda < \expect^{M[\sigma]} (\fnBoundedWindow)$, then this is equivalent to $\lambda < \sup\limits_{\sigma \in \Pi}\{\expect^{M[\sigma]} (\fnBoundedWindow)\}$.
Similarly, for every $\lambda$ if there exist an $\sizeMin$ and a ${\sigma \in \Pi}$, such that
%with $\lambda < \exists \sizeMin \exists \displaystyle{\sigma' _{\sigma' \in \Pi}}\: \in \Pi
$\lambda < \expect^{M[\sigma] (\fnFixedWindow)}$, then this is equivalent to $\lambda < \sup\limits_{\sizeMin} \: \sup\limits_{\sigma \in \Pi}\{\expect^{M[\sigma]} (\fnFixedWindow)\}$.

Now for every $\lambda$ with
$\lambda < \sup\limits_{\sigma \in \Pi}\{\expect^{M[\sigma]} (\fnBoundedWindow)\} \Longleftrightarrow \lambda < \sup\limits_{\sizeMin} \: \sup\limits_{\sigma \in \Pi}\{\expect^{M[\sigma]} (\fnFixedWindow)\}$ is equivalent to $\sup\limits_{\sigma \in \Pi}\{\expect^{M[\sigma]} (\fnBoundedWindow)\} = \sup\limits_{\sizeMin} \: \sup\limits_{\sigma \in \Pi}\{\expect^{M[\sigma]} (\fnFixedWindow)\}$ which we know is true by Theorem \ref{thm-sup_relate}, and we are done.
\end{proof}

\begin{remark}
We note that the following however is not true:
For every $\lambda$, $\exists \sigma \in \Pi \: : \: \expect^{M[\sigma]} (\fnBoundedWindow) \geq \lambda \Longleftrightarrow \exists \sizeMin \exists \sigma' \in \Pi \: : \: \expect^{M[\sigma'] (\fnFixedWindow)} \geq \lambda$.
\end{remark}
	\begin{figure}
		\begin{center}
			\vspace{-15pt}
\centering
\scalebox{1}{
	\begin{tikzpicture}
		\node[player,initial,initial text={}] (s2) at (0,0) {$s_0$} ;
		\node[player] (s3) at (4,0) {$s_1$} ;
		\node[player] (s4) at (8,0) {$s_2$} ;
		
 \path[-latex]  
				(s2) edge[bend left=15] node[above] {\small 1, \textcolor{red}{-1}} (s3)
				(s3) edge[bend left=15] node[below] {\small 0.5, \textcolor{red}{2}} (s2)				
				
				(s3) edge[bend left=15] node[above] {\small 0.5, \textcolor{red}{0}} (s4)				
				(s4) edge[bend left=15] node[below] {\small 1, \textcolor{red}{0}} (s3)
		;
	\end{tikzpicture}
}
		\end{center}
		\caption{$\fnBoundedWindow$ gives $0$ while $\fnFixedWindow$ is less than $0$ for all $\sizeMin$} 
		\label{fig:Fig_relate}
	\end{figure}
	
Consider the Markov chain in Figure \ref{fig:Fig_relate}.
Note that this an MDP $M$ with a single strategy, say $\sigma$.
We see that $\expect^{M[\sigma]} (\fnBoundedWindow) = 0$ (equal to the minimum mean cycle) while for every path $\pi$ starting from $s_0$, for every $\sizeMin$, with probability $1$, we have $\fnFixedWindow (\pi) < 0$ and hence $\expect^{M[\sigma]} (\fnFixedWindow) < 0$.
Thus the statement above is not true for $\lambda=0$.

Also consider $\lambda \leq \sup\limits_{\sizeMin} \: \sup\limits_{\sigma \in \Pi}\{\expect^{M[\sigma]} (\fnFixedWindow)\}$.
As in the proof of Proposition \ref{prop:relate}, this is \emph{not} equivalent to $\lambda \leq \exists \sizeMin \exists \sigma' \in \Pi \: : \:\expect^{M[\sigma']} (\fnFixedWindow)$.
In general, for some set $A$, it may be the case that $\sup\limits_{a \in A} f(a) > f(a)$ for every $a \in A$.

	\section{Window Mean-Cost} \label{sec:WMC}
Window mean-cost is similar to window mean-payoff with the difference that here the objective is to \emph{minimize} the cost over a window length or to ensure that the mean-cost over a window length is less than a threshold.
In particular, given an $\sizeMin \in \nat$ and a threshold $\lambda \in \reals$ and a position in a path over a Markov chain, we check if the window mean-cost is less than or equal to $\lambda$ over some window of length $l$ starting from that position, where $1 \le l \le \sizeMin$.
	Consider a length $l \in \mathbb{N}$ and a sequence of $l+1$ states (that is $l$ edges) $\sigma = s_0 \ldots s_l$. We first define the function $\wtc$ (for \textit{Window Total cost}) such that: 
	\begin{displaymath}
		\wtc(\sigma) = \min_{k \in [l]} \sum_{i = 0}^{k-1} w(s_i,s_{i+1})
	\end{displaymath} 
	The value $WTC(\sigma)$ is the minimum total cost one can ensure over a window of length $k \in [l]$ starting from $s_0$.
	Similarly, we define $WMC$ (for \textit{Window mean-cost}) such that:
	\begin{displaymath}
		\wmc(\sigma) = \min_{k \in [l]} \frac{1}{k} \sum_{i = 0}^{k-1} w(s_i,s_{i+1})
	\end{displaymath} 
	Basically, $\wmc(\sigma)$ is the minimum mean-cost one can ensure over a window of length $k \in [l]$ starting from $s_0$. For a given path $\pi \in Paths^{\Lambda}$, a threshold $\lambda \in \rat$, a position $i \in \nat$ and $l \leq \sizeMin$, we say that the window $\pi(i \ldots (i+l))$ is \emph{closed} if $\wmc(\pi(i \ldots (i+l))) \leq \lambda$. Otherwise, the window is \emph{open}.
	We note that the inductive property of windows also holds for the above definition of window mean-cost.
 	
	Let $\Lambda$ be an MC or an MDP.
	Given an initial state $s_{init}$ and a rational threshold $\lambda \in \rat$, we define the following objectives.
	\begin{itemize}
		\item Given $\sizeMin \in \nat$, the \textit{good window} cost objective
		
		\begin{align}	
		\goodWindowMCObj = \Big\lbrace\play \in Paths^\Lambda \;\vert\; \wmc(\pi(0 \ldots \sizeMin)) \leq \lambda\Big\rbrace\label{eq:goodWindowObj},
		\end{align}
		
		requires that there exists a window starting in the first position of $\pi$ and of size at most $\sizeMin$ over which the mean-cost is bounded above by the threshold $\lambda$.
		
		\item Given $\sizeMin \in \nat$, the \textit{direct fixed window mean-cost} objective
		
		\begin{align}	
		\directFixedWindowMCObj =  \Big\lbrace\play \in Paths^\Lambda \;\vert\; \forall\, j \geq 0,\; \playSuffix{j} \in \goodWindowMCObj \Big\rbrace\label{eq:directFixedWindowObj}
		\end{align}
		requires that good windows of size at most $\sizeMin$ exist in all positions along the play.
		
		\item The \textit{direct bounded window mean-cost} objective
		
		\begin{align}	
		\directBoundedWindowMCObj =  \Big\lbrace\play \in Paths^\Lambda \;\vert\; \exists \sizeMin>0,\; \play \in \directFixedWindowMCObj \Big\rbrace\label{eq:directBoundeddWindowObj}
		\end{align}
		requires that there exists a bound $\sizeMin$ such that the play satisfies the direct fixed objective.
		
		\item Given $\sizeMin \in \nat$, the \textit{fixed window mean-cost}  objective
		\begin{align}
		\fixedWindowMCObj =  \Big\lbrace\play \in Paths^\Lambda \;\vert\; \exists\, i \in \mathbb{N},\; \playSuffix{i} \in \directFixedWindowMCObj \Big\rbrace\label{eq:fixedWindowObj}
		\end{align}
		is the \textit{prefix-independent} version of the direct fixed window objective: it requires for the existence of a suffix of the play satisfying it.
		
		\item The \textit{bounded window mean-cost} objective
		
		\begin{align}	
		\boundedWindowMCObj =  \Big\lbrace\play \in Paths^\Lambda \;\vert\; \exists \sizeMin>0,\; \play \in \fixedWindowMCObj \Big\rbrace\label{eq:boundedWindowObj}
		\end{align}
		is the prefix-independent version of the direct bounded window objective.
	\end{itemize}

%	\subsection{Functions of interest}
%	\label{functions of interest}
	For each of these decision problems, similar to window mean-payoff, we associate a value to every infinite path. We define the following functions, respectively for the \textit{fixed}, \textit{direct fixed}, \textit{bounded} and \textit{direct bounded} \textit{window mean-cost} problem:
	\begin{center}
		$
		\begin{array}{l c l}
		f^{\sizeMin}_{FixWMC}(\pi) & = & \inf \{\lambda \in \rat \mid \pi \in FixWMC(\lambda,\sizeMin)\}\\
		f^{\sizeMin}_{DirFixWMC}(\pi) & = & \inf \{\lambda \in \rat \mid \pi \in DirFixWMC(\lambda,\sizeMin)\}\\
		f_{DirBWMC}(\pi) & = & \inf \{\lambda \in \rat \mid \pi \in DirBWMC(\lambda)\}\\
		f_{BWMC}(\pi) & = & \inf \{\lambda \in \rat \mid \pi \in BWMC(\lambda)\}\\
		\end{array}
		$
	\end{center}

The window mean-cost can be reduced to window mean-payoff by negating all the weights on the edges, and thus for every problem considered above for window mean-payoff, a solution to the problem and taking its negation solves the corresponding problem for window mean-cost.
\section{Conclusion} \label{sec:conclusion}
In this paper, we have studied in the context of MDPs the window mean-payoff objectives~\cite{CDRR15} that were originally introduced for two-player games. Those window objectives guarantee stability of the mean-payoff along outcomes contrary to the classical mean-payoff objectives.
We have provided algorithms to compute the expected window mean-payoff value in MDPs for three variants: the fixed prefix independent variant which fixes the length of the window and asks for the window property to eventually hold forever, the bounded version which does not bound the window length a priori but asks for a bound to exist on each outcome, and the direct fixed window that fixes the length and asks for the property to hold from the very beginning of the outcome. The complexity of our algorithms are listed in the first column of Tab.~\ref{sumMDP}. For the {\sf FixWMP}  problem, the complexity of our algorithm is fully polynomial when the window length is bounded polynomially in the size of the MDP, a natural assumption if we want to obtain stability of the mean-payoff over reasonable time periods.
	\begin{table}
		\begin{tabular}{| l | l | l |}
			\hline
			Problem & Complexity & Hardness \\ \hline 
			FixWMP & $\mathsf{Poly}(|\Gamma|,\sizeMin)$ & Two-player game DirFixWMP \\ [.3cm]
			BWMP & \textsc{NP $\cap$ coNP} & Two-player game MeanPayoff \\ [.3cm]
			DirFixWMP & $\mathsf{Poly}(|\Gamma| \cdot W^{\sizeMin} \cdot \sizeMin)$ & {\sc PSpace}\\ 
			\hline
		\end{tabular}
		\caption{\label{sumMDP}A summary of the complexity and hardness of solving different problems in Markov decision processes.}
	\end{table}

For each of the three problems, we have also provided hardness results. The {\sf FixWMP} for MDP is shown to be as hard as two player-games while {\sf BWMP} is shown to be as hard as classical two-player mean-payoff games, a problem for which we do not have any polynomial time algorithm so far. Finally, we have shown that surprisingly requesting the window property to hold from the beginning of the outcome seems harder: for the {\sf DirFixWMP}, we were able to prove {\sc PSpace-Hardness} even when the length $l_{\it max}$ is given in unary (while this problem has a 
%fully 
polynomial time solution in the non direct case).  

The complexity of the algorithms for Markov chains in summeraized in Tab.~\ref{tab:complexityMCs}.

\begin{table}
	\begin{tabular}{| l | l | l |}
		\hline
		Problem & Complexity & Pseudopolynomial? \\ \hline 
		FixWMP & $\mathsf{Poly}(|\mathcal{M}|,\sizeMin)$ & $\sizeMin$ \\ [.3cm]
		BWMP & $\mathsf{Poly}(|\mathcal{M}|)$ & - \\ [.3cm]
		DirFixWMP & $\mathsf{Poly}(|\mathcal{M}|,\sizeMin,W)$ & $\sizeMin,W$ \\ \hline
	\end{tabular}
	\caption{\label{tab:complexityMCs}A summary of the complexity of solving different problems in Markov chains.}
\end{table}

% ^
%\section{Discussion}

%We end our paper by discussing our definitions of window mean-payoff.
Our definitions allow us to reduce our problems to weighted reachability problems: once an MEC is reached, the optimal value there is the value of the $2$-player game from the best starting state in the MEC (see e.g. Lemma~\ref{val_MEC}), or we need to reach another (better) MEC.
Alternatively, there can be other definitions of window mean-payoff in which the probabilistic behavior inside the MECs also influences the optimal value.
%For example, 
Here is a natural alternative.
%One can consider the property that given a state $s$ in an MDP, a window length $\sizeMin$, a value threshold $\lambda$, and a probability threshold $p$, whether there exists a strategy such that starting from $s$, for all reachable positions, the probability of the window mean-payoff over window length $\sizeMin$ is at least $p$.
Given a state $s$ in an MC $\mathcal{M} = \zug{S,E,\initState,w,\mathbb{P}}$, a window length $\sizeMin$, a value threshold $\lambda$, and a probability threshold $p$, we ask whether all reachable states from $s$ (noted using CTL like syntax $\forall \Box$) ensure that, with probability at least $p$, the window mean-payoff for the next $\sizeMin$ steps is at least $\lambda$. Formally, a state $s$  ensures this alternative good window property, noted $s \models \varphi_{GW}(p, \sizeMin, \lambda)$, if and only if:
\begin{align*}
	\big [\displaystyle{\sum_{\pi \in \fpaths_{\sizeMin}^{\mathcal{M}}(s)} Pr(\pi)} \text{ such that } \wmp^{\sizeMin}(\pi(0, \infty)) \geq \lambda \big ] \ge p
\end{align*}
\noindent
Then, the Markov chain $\mathcal{M} \models \forall \Box \varphi_{GW}(p, \sizeMin, \lambda)$ if $\initState \models \forall \Box \varphi_{GW}(p, \sizeMin, \lambda)$. In the case of an MDP $\Gamma$, we ask whether there exists a strategy $\sigma$ such that $\MDP^{[\sigma]} \models \forall \Box \varphi_{GW}(p, \sizeMin, \lambda)$.

Unfortunately, such a definition leads to intractability even for MCs (contrary to the definitions that we have choosen).
%We can show that this problem is already {\sc PP-Hard} for MCs.
\begin{theorem} \label{thm:alternative}
	Given an MC $\mathcal{M} = \zug{S,E,\initState,w,\mathbb{P}}$, a window length  $\sizeMin$, and a probability threshold $p$, the problem of deciding whether $\mathcal{M} \models \forall \Box \varphi_{GW}(p, \sizeMin, \lambda)$ is {\sf PP-Hard}.
\end{theorem}
\begin{figure}[h]
	\begin{center}
		%\documentclass[preview]{standalone}
%
%\usepackage{fullpage}%
%\usepackage[T1]{fontenc}%
%\usepackage[utf8]{inputenc}%
%\usepackage[main=francais,english]{babel}%
%
%\usepackage{graphicx}%
%\usepackage{url}%
%\usepackage{abstract}%
%
%\usepackage{mathpazo}%
%
%\usepackage{listings}%
%
%\usepackage{times}
%\usepackage{amssymb}
%\usepackage{mathtools}
%
%\usepackage{array}
%\usepackage{multirow}
%
%\usepackage{xcolor}
%\newcommand{\length}{1.75cm}
%
%\usepackage{pifont}
%\newcommand{\cmark}{\ding{51}}
%\newcommand{\xmark}{\ding{55}}
%\definecolor{pakistangreen}{rgb}{0.2, 0.8, 0.2}
%
%\newcommand{\Y}{\textcolor{white}{aaaaa}\textcolor{pakistangreen}{\cmark}}
%\newcommand{\N}{\textcolor{white}{aaaaa}\textcolor{red}{\xmark}}
%\newcommand{\I}{\textcolor{white}{aaaaa}\textcolor{blue}{\textbf{?}}}
%
%\usepackage{footnote}
%\makesavenoteenv{tabular}
%
%\usepackage{tikz}
%\usetikzlibrary{arrows,shapes,snakes,automata,backgrounds,positioning}
%
%\tikzstyle{player}=[state,draw,rounded rectangle,align=center]
%\tikzstyle{widget}=[draw=red,rectangle, rounded rectangle=10pt,dashed,minimum size=6mm,fill=yellow]
%\tikzset{every loop/.style={looseness=7}}
%
%\tikzstyle{player1}=[state,draw,rounded rectangle,align=center]
%\tikzstyle{player2}=[state,draw,rectangle,align=center]
%\begin{document}
%	\vspace{-15pt}
%	\hspace{-15pt}
%	\centering
	\scalebox{1}{
		\begin{tikzpicture}
		\node[player,initial,initial text={}] (s0) at (-3,0) {$\initState$} ;
		\node[player] (sinit) at (0,0) {$s_0$} ;
		\node[player] (s1) at (2,0) {$s_1$} ;
		\node (s7) at (3,0) {\LARGE $\cdots$} ;
		\node[player] (s3) at (4,0) {$s_{n-1}$} ;
		\node[player] (s4) at (6,0) {$s_n$} ;
		
		\path[-latex]  (sinit) edge[bend left=20] node[above] {\small 0.5, \textcolor{red}{$a_1$}} (s1)
		(sinit) edge[bend right=20] node[below] {\small 0.5, \textcolor{red}{0}} (s1)
		
		(s3) edge[bend left=20] node[above] {\small 0.5, \textcolor{red}{$a_{n}$}} (s4)
		(s3) edge[bend right=20] node[below] {\small 0.5, \textcolor{red}{0}} (s4)
		
		(s4) edge[loop right] node[right] {\small 1, \textcolor{red}{$0$}} (s4)
		
		(s0) edge node[above] {\small 1, \textcolor{red}{$- (L+1)$}} (sinit)
		;
		\end{tikzpicture}
	}
	
%\end{document}
%pdflatex file
%convert -density 300 file.pdf -quality 90 file.png
	\end{center}
	\caption{PP-hardness for Markov chain for alternative definition of window mean-payoff} 
	\label{fig-PPhardnessalternative}
\end{figure}
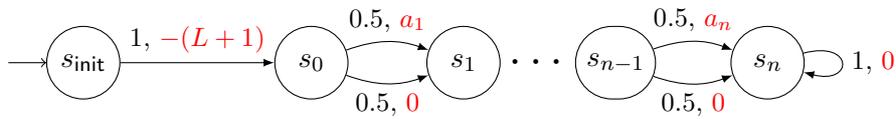 
\begin{proof}
We show a reduction from $k$-th largest subset which has recently been shown to be PP-complete \cite{HK16}.
The $k$-th largest subset problem is stated as given a finite set $A$, a size function $h: A \rightarrow \nat$ assigning strictly positive integer values to elements of $A$, and two naturals $K, L \in \nat$, decide if there exist $n_B \geq K$ distinct subsets $S_j \subseteq A$, where $1 \le j \le n_B$, such that $\sum_{o \in S_j} h(o) \le L$ for all these $K$ or more subsets.

The Markov chain that is constructed from an instance ($A=\{o_1 \ldots o_n\}$, $h$,$K$,$L$) of the $K$-th largest subset problem is shown in Figure~\ref{fig-PPhardnessalternative}.
In this construction, the window length $\sizeMin = n+1$, the probability $p = \frac{2^n-K+1}{2^n}$ and $a_i=h(o_i)$ for $1 \le i \le n$. We claim that there exists at least $K$ subsets of sum lower than or equal to $L$ if and only if $\initState \not \models \forall \Box \varphi_{GW}(p, \ell,0)$.

Note that $s_{i} \models \forall \Box \varphi_{GW}(p, \ell,0)$ for all $i \in [n]_0$ since every payoff appearing on the edges on the paths starting from $s_0$ is non-negative.
This implies that $\initState \models \forall \Box \varphi_{GW}(p, \ell,0) \Leftrightarrow s_{init} \models \varphi_{GW}(p, \ell,0)$. 
Let $n_B$ be the number of subsets of $A$ of sum lower than or equal to $L$ (which is also equal to the number of paths (out of $2^n$) from $s_0$ to $s_n$ of sum lower than or equal to $L$).
Then, if we denote by $p_B$ the probability that the window mean payoff from $\initState$ is not below $0$, %it can be expressed as the number of paths of sum higher then $L$ on the total number of paths from $s_0$ to $s_n$ (there are $2^n$ of them), t
we have that $p_B = \frac{2^n-n_B}{2^n}$. It follows that: $n_B \geq K \Leftrightarrow p_B \leq \frac{2^n-K}{2^n} \Leftrightarrow p_B < p \Leftrightarrow \initState \not \models \varphi_{GW}(p, \ell,0)$.

Hence, deciding if $\initState \models \forall \Box \varphi_{GW}(p, \ell,0)$ holds in a Markov chain is coPP-hard (even when $\sizeMin$ is given in unary). 
The class PP being closed under complementation, it is also {\sc PP-Hard}.
\end{proof}

\noindent
Also, in MDPs, randomized strategies are necessary.
% to satisfy the property $\forall \Box \varphi_{GW}(p, \sizeMin, \lambda)$ 
%(see Appendix \ref{app:randomized} for a proof.).
\begin{theorem} \label{thm:randomized}
	There exist an MDP $\MDP$, a probability threshold $p$, and a window length $\sizeMin$ such that for all deterministic strategies $\sigma$, we have $\MDP^{[\sigma]} \not \models \forall \Box \varphi_{GW}(p, \sizeMin, \lambda)$, while there exists a randomized strategy $\sigma'$ such that $\MDP^{[\sigma']} \models \forall \Box \varphi_{GW}(p, \sizeMin, \lambda)$.
\end{theorem}
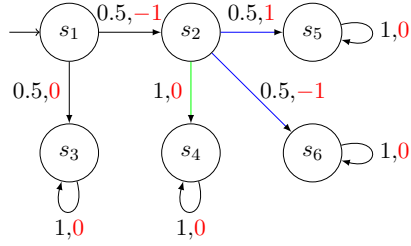
\begin{figure}
	\centering
	%\documentclass[a4paper,11pt]{standalone}
%
%\usepackage{tikz}
%\usetikzlibrary{arrows,shapes,snakes,automata,backgrounds,positioning}
%
%\tikzstyle{player1}=[state,draw,rectangle,align=center]
\tikzstyle{player2}=[state,draw,rounded rectangle,align=center]
%\tikzstyle{widget}=[draw=red,rectangle, rounded rectangle=10pt,dashed,minimum size=6mm,fill=yellow]
%\tikzset{every loop/.style={looseness=7}}
%
%\begin{document}
%		\vspace{-15pt}
%		\centering
		\scalebox{.8}{
			\begin{tikzpicture}
				\node[player2,initial,initial text={}] (sinit) at (0,0) {$s_1$} ;
				\node[player2] (s2) at (2,0) {$s_2$} ;
				\node[player2] (s3) at (0,-2) {$s_3$} ;
				\node[player2] (s4) at (2,-2) {$s_4$} ;
				\node[player2] (s5) at (4,0) {$s_5$} ;
				\node[player2] (s6) at (4,-2) {$s_6$} ;
				
				 \path[-latex]  
						(sinit) edge node[above] {0.5,\textcolor{red}{$-1$}} (s2)
						(sinit) edge node[left] {0.5,\textcolor{red}{$0$}} (s3)
						
						(s2) edge[draw=blue] node[above] {0.5,\textcolor{red}{$1$}} (s5)
						(s2) edge[draw=blue] node[right] {0.5,\textcolor{red}{$-1$}} (s6)
						(s2) edge[draw=green] node[left] {1,\textcolor{red}{$0$}} (s4)
						
						(s3) edge[loop below] node[below] {1,\textcolor{red}{$0$}} (s3)
						(s4) edge[loop below] node[below] {1,\textcolor{red}{$0$}} (s4)
						(s5) edge[loop right] node[right] {1,\textcolor{red}{$0$}} (s5)
						(s6) edge[loop right] node[right] {1,\textcolor{red}{$0$}} (s6)
				;
			\end{tikzpicture}
		}
		%\includegraphics[width=1\textwidth]{example_new.pdf}
		%\caption{\label{fig-example} An excerpt of the MDP for Example~\ref{exm-prelims}. Tasks in \textbf{bold} are active, tasks in \textit{italics} have missed a deadline.}
%\end{document}
	\caption{An example of an MDP where probabilities are in black and weights in red. We consider $\ell = 2$ and $p = 0.6$.}
	\label{fig-exampleMDP-random}
\end{figure}
\begin{proof}
Consider the example in Figure \ref{fig-exampleMDP-random}.
Let $\sizeMin = 2$.
Note that the states $s_3, s_4, s_5$ and $s_6$ satisfy the property $\forall \Box \varphi_{GW}(1, 2, 0)$.
When the blue action is chosen from $s_2$, then in the corresponding MC we have that $s_1 \models \varphi_{GW}(0.75, 2, 0)$ corresponding to the finite paths $s_1s_3s_3$ and $s_1s_2 s_5$, and $s_2 \models \varphi_{GW}(0.5, 2, 0)$ for the path $s_2s_5s_5$.
We call the deterministic strategy {\sf blue} when the blue action is chosen from $s_2$.
Thus we have $\MDP^{[\sf blue]} \models \forall \Box \varphi_{GW}(0.5, 2, 0)$.

Now if the green action is chosen from $s_2$, then in the resulting MC we have $s_1 \models \varphi_{GW}(0.5, 2, 0)$ and $s_2 \models \varphi_{GW}(1, 2, 0)$.
We call the deterministic strategy {\sf green} when the green action is chosen from $s_2$.
Thus we have $\MDP^{[\sf green]} \models \forall \Box \varphi_{GW}(0.5, 2, 0)$.

Now for every randomized strategy $\sigma'$, we have $\MDP^{[\sigma']} \models \forall \Box \varphi_{GW}(p, 2, 0)$ for $p > 0.5$.
In particular, let $\sigma_1$ be a randomized strategy in which the blue action is chosen with probability $0.6$, and the green action with probability $0.4$.
Then we have $p = \min(0.75 \cdot 0.6 + 0.5 \cdot 0.4, 0.5 \cdot 0.6+1 \cdot 0.4) = \min(0.65, 0.7) = 0.65$.
Thus $\MDP^{[\sigma_1]} \models \forall \Box \varphi_{GW}(0.65, 2, 0)$ while none of the deterministic strategies satisfies the property for $p=0.65$.
Moreover, we can show that $\max(\{p: \MDP^{[\sigma']} \models \forall \Box \varphi_{GW}(p, 2, 0) \text{, and } \sigma' \text{ is a strategy in \MDP }\}) = \frac{2}{3}$ when the blue action is chosen from $s_2$ with probability $\frac{2}{3}$, and the green action is chosen with probability $\frac{1}{3}$.
\end{proof}

	\bibliographystyle{plain}
	\bibliography{papers}

\begin{thebibliography}{10}

\bibitem{BK08}
C.~Baier and J{-}P. Katoen.
\newblock {\em Principles of model checking}.
\newblock {MIT} Press, 2008.

\bibitem{BBCFK14}
Tom{\'{a}}s Br{\'{a}}zdil, V{\'{a}}clav Brozek, Krishnendu Chatterjee, Vojtech
  Forejt, and Anton{\'{\i}}n Kucera.
\newblock Two views on multiple mean-payoff objectives in markov decision
  processes.
\newblock {\em Logical Methods in Computer Science}, 10(1), 2014.

\bibitem{BCFK17}
Tom{\'{a}}s Br{\'{a}}zdil, Krishnendu Chatterjee, Vojtech Forejt, and
  Anton{\'{\i}}n Kucera.
\newblock Trading performance for stability in markov decision processes.
\newblock {\em J. Comput. Syst. Sci.}, 84:144--170, 2017.

\bibitem{BFKN16}
Tom{\'{a}}s Br{\'{a}}zdil, Vojtech Forejt, Anton{\'{\i}}n Kucera, and Petr
  Novotn{\'{y}}.
\newblock Stability in graphs and games.
\newblock In {\em 27th International Conference on Concurrency Theory, {CONCUR}
  2016, August 23-26, 2016, Qu{\'{e}}bec City, Canada}, volume~59 of {\em
  LIPIcs}, pages 10:1--10:14. Schloss Dagstuhl - Leibniz-Zentrum fuer
  Informatik, 2016.

\bibitem{BHR16}
V{\'{e}}ronique Bruy{\`{e}}re, Quentin Hautem, and Jean{-}Fran{\c{c}}ois
  Raskin.
\newblock On the complexity of heterogeneous multidimensional games.
\newblock In {\em 27th International Conference on Concurrency Theory, {CONCUR}
  2016, August 23-26, 2016, Qu{\'{e}}bec City, Canada}, volume~59 of {\em
  LIPIcs}, pages 11:1--11:15. Schloss Dagstuhl - Leibniz-Zentrum fuer
  Informatik, 2016.

\bibitem{CDRR13}
Krishnendu Chatterjee, Laurent Doyen, Mickael Randour, and
  Jean{-}Fran{\c{c}}ois Raskin.
\newblock Looking at mean-payoff and total-payoff through windows.
\newblock In {\em Automated Technology for Verification and Analysis - 11th
  International Symposium, {ATVA} 2013, Hanoi, Vietnam, October 15-18, 2013.
  Proceedings}, volume 8172 of {\em Lecture Notes in Computer Science}, pages
  118--132. Springer, 2013.

\bibitem{CDRR15}
Krishnendu Chatterjee, Laurent Doyen, Mickael Randour, and
  Jean{-}Fran{\c{c}}ois Raskin.
\newblock Looking at mean-payoff and total-payoff through windows.
\newblock {\em Inf. Comput.}, 242:25--52, 2015.

\bibitem{EM79}
A.~Ehrenfeucht and J.~Mycielski.
\newblock Positional strategies for mean payoff games.
\newblock {\em International Journal of Game Theory}, 8(2):109--113, Jun 1979.

\bibitem{Filar12}
Jerzy Filar and Koos Vrieze.
\newblock {\em Competitive Markov decision processes}.
\newblock Springer Science \& Business Media, 2012.

\bibitem{HK15}
Christoph Haase and Stefan Kiefer.
\newblock The odds of staying on budget.
\newblock In {\em Automata, Languages, and Programming - 42nd International
  Colloquium, {ICALP} 2015, Kyoto, Japan, July 6-10, 2015, Proceedings, Part
  {II}}, pages 234--246, 2015.

\bibitem{HK16}
Christoph Haase and Stefan Kiefer.
\newblock The complexity of the kth largest subset problem and related
  problems.
\newblock {\em Inf. Process. Lett.}, 116(2):111--115, 2016.

\bibitem{HPR18}
Paul Hunter, Guillermo~A. P{\'{e}}rez, and Jean{-}Fran{\c{c}}ois Raskin.
\newblock Looking at mean payoff through foggy windows.
\newblock {\em Acta Inf.}, 55(8):627--647, 2018.

\bibitem{Karp78}
Richard~M. Karp.
\newblock A characterization of the minimum cycle mean in a digraph.
\newblock {\em Discrete Mathematics}, 23:309--311, 1978.

\bibitem{Put94}
Martin~L. Puterman.
\newblock {\em Markov Decision Processes: Discrete Stochastic Dynamic
  Programming}.
\newblock John Wiley \& Sons, Inc., New York, NY, USA, 1st edition, 1994.

\bibitem{Tarjan72}
Robert Tarjan.
\newblock Depth-first search and linear graph algorithms.
\newblock {\em SIAM journal on computing}, 1(2):146--160, 1972.

\bibitem{ZP96}
Uri Zwick and Mike Paterson.
\newblock The complexity of mean payoff games on graphs.
\newblock {\em Theoretical Computer Science}, 158(1-2):343--359, 1996.

\end{thebibliography}
	
	\newpage
	\begin{center}
{\LARGE \textbf{Appendix}}
\end{center}		
\appendix
\section{Proof of $\sup\limits_{a \in A} \sup\limits_{b \in B} f(a,b) = \sup\limits_{b \in B} \sup\limits_{a \in A} f(a,b)$}
\label{app:supsup}
For every $a \in A$, $b \in B$ and function $f$
\begin{lemma}
$\sup\limits_{a \in A} \sup\limits_{b \in B} f(a,b) = \sup\limits_{b \in B} \sup\limits_{a \in A} f(a,b)$
\end{lemma}
\begin{proof}
%Let $\lambda_1 = sup_{a \in A} \sup\limits_{b \in B} f(a,b)$ and $\lambda_2 = \sup\limits_{b \in B} \sup\limits_{a \in A} f(a,b)$.
Consider $\lambda < \sup\limits_{b \in B} \sup\limits_{a \in A} f(a,b)$.
Then there exists $b_0 \in B$ such that $\lambda \le \sup\limits_{a \in A} f(a, b_0) \le \sup\limits_{a \in A} \sup\limits_{b \in B} bf(a, b)$.
Thus for every $\lambda < \sup\limits_{b \in B} \sup\limits_{a \in A} f(a,b)$, we have $\lambda \leq \sup\limits_{a \in A} \sup\limits_{b \in B} f(a,b)$.
Hence $\sup\limits_{b \in B} \sup\limits_{a \in A} f(a,b) \leq \sup\limits_{a \in A} \sup\limits_{b \in B} f(a,b)$.

We can analogously show that $\sup\limits_{a \in A} \sup\limits_{b \in B} f(a,b) \leq \sup\limits_{b \in B} \sup\limits_{a \in A} f(a,b)$ and we are done.
\end{proof}

\section{Proof of Lemma~\ref{th_first}} \label{lem:dirBWMPequalsBWMP}
\begin{proof}
	By definition, $f_{DirBWMP}(\pi) \leq f_{BWMP}(\pi)$.
	
	We now prove that $f_{BWMP}(\pi) \leq f_{DirBWMP}(\pi)$. Let $\lambda \in \mathbb{R}$ such that $\lambda < f_{BWMP}(\pi)$. We prove that $\lambda \leq f_{DirBWMP}(\pi)$. Let $\nu \in \mathbb{R}$ such that $\lambda < \nu < f_{BWMP}(\pi)$. Since $\nu < f_{BWMP}(\pi)$, we have $\pi \in BWMP(\nu)$. Therefore, there exists $n_\nu \in \nat$, $l_\nu > 0$ such that, for all $j \geq n_\nu$, we have $\wmp(\pi(j \ldots j+l_\nu)) = \max\limits_{k \in [l_\nu]} \frac{1}{k} \sum\limits_{i = 0}^{k-1} w(s_{j+i},s_{j+i+1}) \geq \nu$. Let $d = \min\limits_{i \in [n_\nu-1]_0} \sum\limits_{k = i}^{n_\nu -1} (w(s_k,s_{k+1}) - \lambda)$ be the minimum over all index $i \in [n_\nu-1]_0$ of the sum of the weights, reduced by $\lambda$, from $s_i$ to $s_\nu$. Then, let $h = \lceil \frac{\mid d \mid}{\nu - \lambda}\rceil$ and $l_\lambda = n_\nu + l_\nu \cdot (1 + h)$. 
	
	We prove that $\pi \in DirFixWMP(\lambda,l_\lambda)$:
	\begin{itemize}
		\item For all $j \geq n_\nu$, $\pi(j,\infty) \in GW(\lambda,l_\lambda)$ since $\pi(j,\infty) \in GW(\nu,l_\nu)$, $\lambda < \nu$ and $l_\lambda \geq l_\nu$.
		\item Let $j < n_\nu$. We prove that there exists $l_j \in \nat$ such that:
		\begin{itemize}
			\item The length $l_j$ considered is at most $l_\lambda$: $l_j \leq l_\lambda$;
			\item The window mean-payoff from $s_j$ to $s_{j+l_j}$ is at least $\lambda$: $\wmp(\pi(j \ldots j+l_j)) \geq \lambda$.
		\end{itemize}
		
		For all $k \geq n_\nu$, there exists $p_k \in [l_\nu]$ such that $\frac{1}{p_k} \sum\limits_{i = 0}^{p_k-1} w(s_{k+i},s_{k+i+1}) \geq \nu$. Let us denote by $l_k \in [l_\nu]$ the smallest index satisfying that property.
		
		Consider the series $(u_n)_{n \in \nat}$ defined by: 
		\begin{itemize}
			\item $u_0 = n_\nu$;
			\item For all $n \geq 0$, we have $u_{n+1} = u_n + l_{u_n}$. 
		\end{itemize}
		Now given some $n \in \nat$, we partition $\pi(u_0 \dots u_n)$ into $\pi(u_0 \dots u_1), \dots, \pi(u_{n-1} \dots u_n)$, such that the mean weight over each of $\pi(u_r \dots u_{r+1})$ for every $r \in [n-1]_0$ is at least $\nu$.
		%				The series $(u_n)_{n \in \nat}$ is defined such that, for all $n \in \nat$, the sum of the weights from $u_0 = n_\nu$ to $u_n$ can be partitioned into several sums from $u_k$ to $u_{k+1}$, for $k \in [n-1]_0$, so that each sum has an average payoff above $\nu$. Formally, for all $n \in \nat$:
		Formally,
		\begin{displaymath}
		\sum\limits_{k = u_0}^{u_{n} - 1} w(s_k,s_{k+1}) = \sum\limits_{i = 0}^{n - 1} \underbrace{\sum\limits_{k = u_i}^{u_{i+1} - 1} w(s_k,s_{k+1})}_{\substack{\geq \ l_{u_i} \cdot \nu \ = \ (u_{i+1} - u_i) \cdot \nu \\ \text{ by definition of }l_{u_i}}} \geq \nu \cdot \sum\limits_{i = 0}^{n - 1} (u_{i+1} - u_i) = \nu \cdot (u_n - u_0)
		\end{displaymath}
		
		Furthermore, $u_{h} - u_0 = \sum\limits_{n=0}^{h-1} \underbrace{u_{n+1} - u_n}_{=l_{u_n}} = \sum\limits_{n=0}^{h-1} l_{u_n}$. Moreover, for all $n \geq 0$, $1 \leq l_{u_n} \leq l_\nu$. That is: $h \leq u_{h} - u_0 \leq h \cdot l_\nu$. 
		
		Let $l_j = u_h - j$. We have:
		\begin{itemize}
			\item $l_j = u_h - j \leq u_h = u_0 + u_h - u_0 \leq \underbrace{u_0}_{=n_\nu} + h \cdot l_\nu \leq l_\lambda$ (recall that $l_\lambda = n_\nu + l_\nu \cdot (1 + h)$);
			\item We prove that $\wmp(\pi(j \ldots j+l_j)) = \wmp(\pi(j \ldots u_h)) \geq \lambda$. We have:			
			\begin{align*}
			\sum\limits_{k = j}^{u_{h} - 1} w(s_k,s_{k+1}) & = \sum\limits_{k = j}^{u_0 - 1} w(s_k,s_{k+1}) + \sum\limits_{k = u_0}^{u_{h} - 1} w(s_k,s_{k+1})\\
			\end{align*}
			Moreover, we have:
			\begin{enumerate}[(i)]
				\item $\sum\limits_{k = j}^{u_0 - 1} w(s_k,s_{k+1}) = \sum\limits_{k = j}^{u_0 - 1} (w(s_k,s_{k+1}) - \lambda + \lambda) = \underbrace{\sum\limits_{k = j}^{u_0 - 1} (w(s_k,s_{k+1}) - \lambda)}_{\substack{\geq d \\ \text{ by definition of } d}} + (u_0 - j) \cdot \lambda$. Therefore: $\sum\limits_{k = j}^{u_0 - 1} w(s_k,s_{k+1}) \geq d + (u_0 - j) \cdot \lambda$;
				\item $\sum\limits_{k = u_0}^{u_{h} - 1} w(s_k,s_{k+1}) \geq \nu \cdot (u_h - u_0)$, by definition of $(u_n)_{n \in \nat}$.
			\end{enumerate}
			Therefore:
			\begin{align*}
			\sum\limits_{k = j}^{u_{h} - 1} w(s_k,s_{k+1}) & \geq d + (u_0 - j) \cdot \lambda + \nu \cdot (u_h - u_0) \underbrace{- \lambda \cdot (u_h - u_0) + \lambda \cdot (u_h - u_0)}_{=0}\\
			& = d + \underbrace{((u_h - u_0) + (u_0 - j))}_{= u_h - j} \cdot \lambda + (\nu - \lambda) \cdot \underbrace{(u_h - u_0)}_{\geq h}\\
			& = \lambda \cdot (u_h - j) + d + (\nu - \lambda) \cdot \underbrace{h}_{=\lceil \frac{\mid d \mid}{\nu - \lambda}\rceil \geq \frac{\mid d \mid}{\nu - \lambda}} \geq \lambda \cdot (u_h - j) + \underbrace{d + \mid d \mid}_{\geq 0}\\
			& \geq \lambda \cdot (u_h - j)
			\end{align*}
			In fact:
			\begin{displaymath}
			\frac{1}{u_h - j} \sum\limits_{k = j}^{u_{h} - 1} w(s_k,s_{k+1}) \geq \lambda
			\end{displaymath}
			Therefore, $\wmp(\pi(j \ldots j+l_j)) = \wmp(\pi(j \ldots u_h)) \geq \lambda$. 
		\end{itemize}
	\end{itemize}
	Since this is true for all $j \geq 0$, we have $\pi \in DirFixWMP(\lambda,l_\lambda)$ and therefore $\pi \in DirBWMP(\lambda)$, that is $\lambda \leq f_{DirBWMP}(\pi)$. Since this is true for all $\lambda < f_{BWMP}(\pi)$, we have in fact $f_{DirBWMP}(\pi) = f_{BWMP}(\pi)$. 
	
	We note that this proof holds even if all the weights in the Markov chain are equal to any realch number, not only non negative integers.
\end{proof}

\section{Second Algorithm for the Direct Fixed Case in MCs}
\label{dirfix_2}
Given weighted Markov chain $\mathcal{M} = \langle S,E,s_{init},w,\mathbb{P} \rangle$ we consider a structure similar to a weighted Markov chain $\mathcal{M}_{l_{max}} = \langle S',\mathbb{P}',s_{init},L \rangle$ where the weights are on the states instead of the edges, and are defined using the function $L: S' \longrightarrow \rat$. We have:
\begin{itemize}
	\item $S' = s_{init} \cup FPaths^{\mathcal{M}}_{l_{max}}$;
	\item $\mathbb{P}'(\pi,\pi') = \mathbb{P}(s,s')$ if $\pi = s_0 s_1\ldots s_{n} s$, $\pi'= s_1\ldots s_{n} s s'$ and $\pi,\pi' \; \in \; FPaths^{\mathcal{M}}_{l_{max}}$;
	\item $\mathbb{P}'(s_{init},\pi) \; = \; \mathbb{P}(\pi)$ if $\pi = s_0 s_1 \ldots s_n \; \in \; FPaths^{\mathcal{M}}_{l_{max}}$, where $s_0 = s_{init}$;
	\item $L(s_{init}) = 0$;
	\item $L(\pi) = \wmp(\pi(0 \ldots l_{max})) \text{, if } \pi \in S' \setminus \{s_{init}\}$.
\end{itemize}
Similarly to the previous construction, if the probability $\mathbb{P}'(s,s')$ was not defined for two states $s,s' \in S'$, then $\mathbb{P}'(s,s') = 0$.

\begin{figure}
	\begin{center}
		\vspace{-15pt}
\scalebox{1}{
	\begin{tikzpicture}
		\node[player,initial,initial text={}] (sinit) at (0,0) {$s_0$} ;
		\node[player] (s011) at (-3,-1.5) {$s_0,s_1,s_1$ \\ \textcolor{red}{1.5}};
		\node[player] (s023) at (3,-1.5) {$s_0,s_2,s_3$ \\ \textcolor{red}{2}} ;
		
		\node[player] (s111) at (-3,-3.5) {$s_1,s_1,s_1$ \\ \textcolor{red}{2}} ;
		
		\node[player] (s233) at (5,-3.5) {$s_2,s_3,s_3$ \\ \textcolor{red}{2}} ;
		\node[player] (s234) at (1,-3.5) {$s_2,s_3,s_4$ \\ \textcolor{red}{1.5}} ;
		
		\node[player] (s333) at (9,-7) {$s_3,s_3,s_3$ \\ \textcolor{red}{3}} ;
		\node[player] (s334) at (7,-5.5) {$s_3,s_3,s_4$ \\ \textcolor{red}{3}} ;
		\node[player] (s343) at (5,-7) {$s_3,s_4,s_3$ \\ \textcolor{red}{2}} ;
		\node[player] (s344) at (-1,-5.5) {$s_3,s_4,s_4$ \\ \textcolor{red}{2}} ;
		
		\node[player] (s433) at (7,-8.5) {$s_4,s_3,s_3$ \\ \textcolor{red}{1.5}} ;
		\node[player] (s434) at (1,-7) {$s_4,s_3,s_4$ \\ \textcolor{red}{1}} ;
		\node[player] (s443) at (-1,-8.5) {$s_4,s_4,s_3$ \\ \textcolor{red}{1}} ;
		\node[player] (s444) at (-3,-7) {$s_4,s_4,s_4$ \\ \textcolor{red}{1}} ;
		
		\path[-latex]	(sinit) edge node[above] {} (s011)
				(sinit) edge node[above] {} (s023)
	
				(s011)  edge node[left] {$1$} (s111)

				(s111)  edge[loop below] node {$1$} (s111)
			
				(s023)  edge node[left] {} (s233)
				(s023)	edge node[right] {} (s234)

				(s233)  edge[bend left=25] node[left] {} (s333)
				(s233)	edge node[left] {} (s334)

				(s234)  edge node[left] {} (s343)
				(s234)	edge node[right] {} (s344)

				(s333)  edge[loop below] node {} (s333)
				(s333)	edge node[left] {} (s334)

				(s334)  edge node[right] {} (s343)
				(s334)	edge node[right] {} (s344)

				(s343)  edge node[right] {} (s433)
				(s343)	edge[bend left=5] node[left] {} (s434)

				(s344)  edge node[right] {} (s443)
				(s344)	edge node[left] {} (s444)

				(s433)  edge node[right] {} (s333)
				(s433)	edge node[left] {} (s334)

				(s434)  edge[bend left=5] node[left] {} (s343)
				(s434)	edge node[right] {} (s344)

				(s443)  edge node[left] {} (s433)
				(s443)	edge node[left] {} (s434)

				(s444)  edge node[right] {} (s443)
				(s444)	edge[loop above] node {} (s444)
		;
	\end{tikzpicture}
}
	\end{center}
	\caption{The \textit{path Markov chain} obtained from the weighted Markov chain of Figure~\ref{eq_weighted_markov}, with $l_{max} = 2$. The label of each state is indicated in red inside of them. The probabilities are not indicated on the edges because every single one has the same: $0.5$, at the exception of the two edges going inside state $s_1,s_1,s_1$, which have probability 1.} 
	\label{window_markov}
\end{figure}
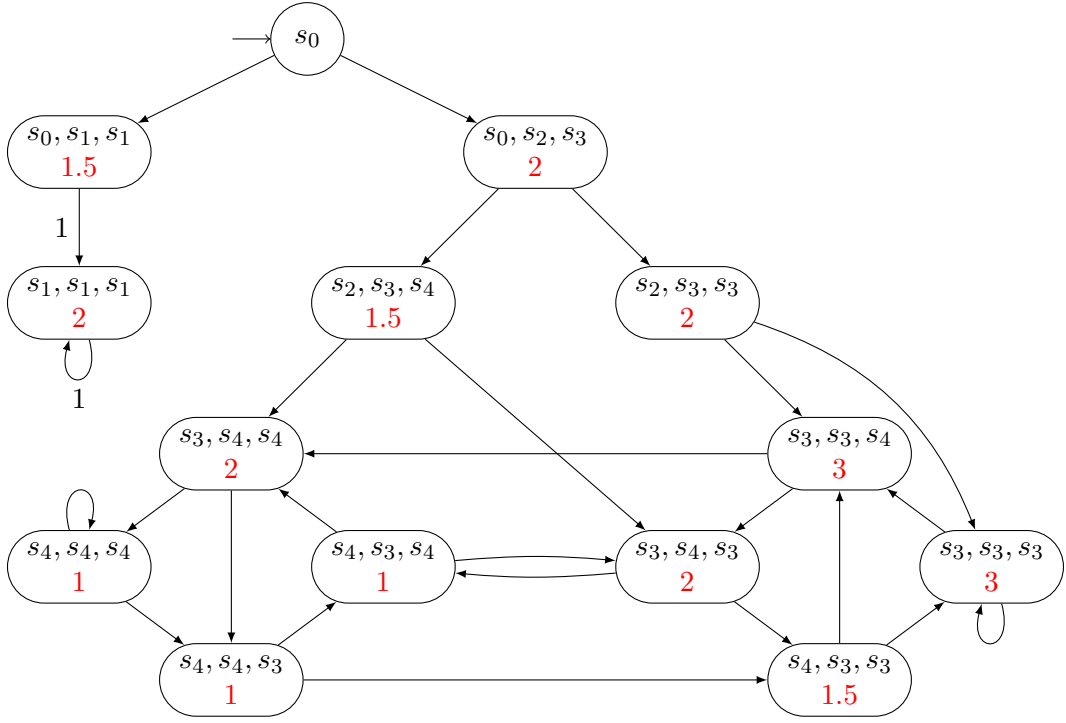
An illustration of this construction can be seen in Figure~\ref{window_markov}. In this example, the new Markov chain $\mathcal{M}_{\lambda}$ unfolds every possible paths of length $l_{max} = 2$ of the weighted Markov chain.

Let us now define $g^{l_{max}}_{DirFixWMP}: Paths^{\mathcal{M}_{l_{max}}}(s_{init}) \rightarrow \rat$ such that:
\begin{displaymath}
\forall \pi \in Paths^{\mathcal{M}_{l_{max}}}(s_{init}), \; g^{l_{max}}_{DirFixWMP}(\pi) = \sup_{\lambda} \{\lambda \mid \forall i \geq 1,\; \lambda \leq L(\pi(i))\}
\end{displaymath}

The expected value in the new Markov chain $\mathcal{M}_{l{max}}$ can be defined in a similar way that we did in $\mathcal{M}$:
\begin{displaymath}
\mathbb{E}_{s_{init}}^{\mathcal{M}_{l_{max}}}(g^{l_{max}}_{DirFixWMP}) = \sum_{r \in \mathbb{Q}} r \cdot Pr((g^{l_{max}}_{DirFixWMP})^{-1}(\mathcal{M}_{l_{max}},s_{init},r))
\end{displaymath} 

We claim that:
\begin{theorem}
	\label{path_th}
	For all weighted Markov chain $\mathcal{M}$, we have:
	\begin{displaymath}
	\mathbb{E}_{s_{init}}^{\mathcal{M}}(f^{l_{max}}_{DirFixWMP}) = \mathbb{E}_{s_{init}}^{\mathcal{M}_{l_{max}}}(g^{l_{max}}_{DirFixWMP})
	\end{displaymath} 
\end{theorem}

We first prove the following lemma.
\begin{lemma}
	There exists a bijection $h: Paths^{\mathcal{M}}(s_{init}) \rightarrow Paths^{\mathcal{M}_{l_{max}}}(s_{init})$ such that:
	\begin{itemize}
		\item $\forall \pi \in Paths^{\mathcal{M}}(s_{init}),\; f^{l_{max}}_{DirFixWMP}(\pi) = g^{l_{max}}_{DirFixWMP}(h(\pi))$, that is the value of a path $\pi$ in $\mathcal{M}$ is the same as the value of $h(\pi)$ in $\mathcal{M}_{l_{max}}$;  
		\item $\forall j \geq l_{max},\; \mathbb{P}(\pi(0 \ldots j)) = \mathbb{P}'(h(\pi)(0 
		\ldots (j-l_{max}+1)))$, that is any finite fragment of the path $\pi$ has the same probability to occur in $\mathcal{M}$ as the corresponding path fragment of $h(\pi)$ in $\mathcal{M}_{l_{max}}$, while considering that the $l_{max}$ first states of $\pi$ are merged into one state in $\mathcal{M}_{l_{max}}$.
	\end{itemize}
	\label{lemma}
\end{lemma}

\begin{proof}
	Let $h: Paths^{\mathcal{M}}(s_{init}) \rightarrow Paths^{\mathcal{M}_{l_{max}}}(s_{init})$ be such that for $\pi = s_0 s_1\ldots  \in Paths^{\mathcal{M}}(s_{init})$, where $s_0 = s_{init}$, we have $h(\pi) = t_0 t_1\ldots  \in Paths^{\mathcal{M}_{l_{max}}}(s_{init})$ with 
	$t_0 = s_{init} \in S'$ and $\forall i \geq 0,\; t_{i+1} = (s_i\ldots s_{i + l_{max}}) \in S'$.

	\begin{itemize}
		\item Let us first argue that $h$ is well defined, that is for any $\pi \in Paths^{\mathcal{M}}(s_{init})$, $h(\pi) \in Paths^{\mathcal{M}_{l_{max}}}(s_{init})$. We have that $\mathbb{P}'(t_0,t_1) > 0$ because $t_1 = s_0\ldots s_{l_{max}} \in FPaths^{\mathcal{M}}_{l_{max}}(s_{init})$, since $t_1$ is a prefix of $\pi \in Paths^{\mathcal{M}}(s_{init})$. Moreover, $\forall i \geq 1$, $\mathbb{P}'(t_i,t_{i+1}) = \mathbb{P}(s_{i + l_{max} - 1},s_{i + l_{max}}) > 0$. Therefore, $h(\pi) \in Paths^{\mathcal{M}_{l_{max}}}(s_{init})$;
		
		\item Now we prove that if, for $\pi,\pi' \in Paths^{\mathcal{M}}(s_{init})$ we have $h(\pi) = h(\pi')$, then $\pi = \pi'$. Let $\pi = s_0 s_1\ldots,\; \pi' = s'_0 s'_1\ldots,\; h(\pi) = t_0 t_1\ldots $ and $h(\pi') = t'_0 t'_1\ldots$. Let $i \in \mathbb{N}$. Since $h(\pi) = h(\pi')$, we have $(s_i\ldots s_{i + l_{max}}) = t_{i+1} = t'_{i+1} = (s'_i\ldots s'_{i + l_{max}})$ and therefore $s_i = s'_i$. In fact, $\forall i \in \mathbb{N}, s_i = s'_i$. That is, $\pi = \pi'$. Thus, we can conclude that $h$ is injective;
		
		\item Let $\pi' = t_0 t_1 \ldots \in Paths^{\mathcal{M}_{l_{max}}}(s_{init})$. We construct a path $\pi \in Paths^{\mathcal{M}}(s_{init})$ such that $h(\pi) = \pi'$ and $\forall j \geq l_{max},\; \mathbb{P}(\pi(0 \ldots j)) = \mathbb{P}'(\pi'(0 
		\ldots (j-l_{max}+1)))$. For all $i \geq 1$, $t_i \neq s_{init}$ because there is no edge going to the state $s_{init}$ in $\mathcal{M}^{l_{max}}$, by definition of $\mathcal{M}^{l_{max}}$. Therefore, for all $i \geq 1$, let $t_i = (s_{k_{i-1}} \ldots s_{k_{i + l_{max} - 1}}) \in FPaths^{\mathcal{M}}_{l_{max}}$. We can now define $\pi = s_{k_0} s_{k_1} \ldots$ (where $s_{k_{0}} = s_{init}$ since $\pi' \in Paths^{\mathcal{M}_{l_{max}}}(s_{init})$). Then:
		\begin{itemize}
			\item  $\mathbb{P}(s_{k_0} \ldots s_{k_{l_{max}}}) = \mathbb{P}'(t_0,t_1) > 0$
			\item $\forall i \geq l_{max}$, $\mathbb{P}(s_{k_i},s_{k_{i+1}}) = \mathbb{P}'(t_{i - l_{max} + 1},t_{i - l_{max} + 2}) > 0$
		\end{itemize}
		This implies that $\pi \in Paths^{\mathcal{M}}(s_{init})$, since $\forall i \in \nat,\; \mathbb{P}(s_{k_i} s_{k_{i+1}}) > 0$. By construction, we have $h(\pi) = \pi'$. Moreover, for $j \geq l_{max}$, we have
		\begin{align*}
		\mathbb{P}(\pi(0 \ldots j)) = \mathbb{P}(s_{k_0} \ldots s_{k_{l_{j}}}) & = \mathbb{P}(s_{k_0} \ldots s_{k_{l_{max}}}) \cdot \prod \limits_{l = l_{max}}^{j-1} \mathbb{P}(s_{k_{l}},s_{k_{l+1}})\\
		& = \mathbb{P}'(t_0,t_1) \cdot \prod \limits_{l = l_{max}}^{j-1} \mathbb{P}'(t_{l - l_{max} + 1},t_{l - l_{max} + 2}) \\
		& = \prod \limits_{l = 0}^{j-l_{max}} \mathbb{P}'(t_{l},t_{l+1})\\
		& = \mathbb{P}'(\pi'(0 \ldots (j-l_{max}+1)))
		\end{align*}
		
		Thus, we can conclude that $h$ is surjective, and therefore bijective. Moreover, $h$ ensures that $\forall j \geq l_{max},\; \mathbb{P}(\pi(0 \ldots j)) = \mathbb{P}'(h(\pi)(0 
		\ldots (j-l_{max}+1)))$;
		
		\item 	Finally, we prove that $h$ maintains the value of each path.
		Let $\pi = s_0 s_1 \ldots \in Paths^{\mathcal{M}}(s_{init})$ such that $h(\pi) = t_0 t_1 \ldots \in Paths^{\mathcal{M}_{l_{max}}}(s_{init})$. We prove that $f^{l_{max}}_{DirFixWMP}(\pi) = g^{l_{max}}_{DirFixWMP}(h(\pi))$. We have that $f^{l_{max}}_{DirFixWMP}(\pi) = sup \{\lambda \mid \pi \in FixWMP(\lambda,l_{max})\}$. Since $\mathcal{M}$ is a finite Markov chain, there is a finite number of paths of length $l_{max}$ in $\mathcal{M}$. We can deduce that:
		\begin{itemize}
			\item $\exists i \in \mathbb{N}, f^{l_{max}}_{DirFixWMP}(\pi) = \wmp(s_i \ldots s_{i+l_{max}}) = L(t_{i+1})$;
			\item $\forall n \in \mathbb{N}, f^{l_{max}}_{DirFixWMP}(\pi) \leq \wmp(s_n \ldots s_{n+l_{max}}) = L(t_{n+1})$.
		\end{itemize}
		
		Since $g^{l_{max}}_{DirFixWMP}(h(\pi)) = \sup \limits_{\lambda} \{\lambda \mid \forall i \in \nat,\; \lambda \leq L(t_{i+1}) \}$, then $g^{l_{max}}_{DirFixWMP}(h(\pi)) = f^{l_{max}}_{DirFixWMP}(\pi)$. In fact, $\forall \pi \in Paths^{\mathcal{M}}(s_{init}),\; f^{l_{max}}_{DirFixWMP}(\pi) = g^{l_{max}}_{DirFixWMP}(h(\pi))$.
	\end{itemize}
	Therefore, the function $h$ ensures the desired properties.
\end{proof}

We can now proceed to the proof of Theorem~\ref{path_th}.
\begin{proof}
	Consider the function $h$ satisfying the conditions of Lemma~\ref{lemma}.
	For $\lambda,\mu \in \rat$, let us note
	\begin{align*}
		FP^{\mathcal{M}}_\lambda = \{& \pi \in FPaths^{\mathcal{M}}(s_{init}) \mid |\pi| \geq l_{max} \ \land \\
		& \wmp(\pi(|\pi|- l_{max} \ldots |\pi|)) = \lambda \ \land \\
		& \forall i < |\pi|- l_{max},\; \wmp(\pi(i \ldots i+l_{max})) \neq \lambda \}
	\end{align*}
	and
	\begin{align*}
		FP^{\mathcal{M}}_{\lambda,\mu} = \{& \pi \in FP^{\mathcal{M}}_\lambda \mid \exists j < |\pi|- l_{max},\; \wmp(\pi(|\pi|- l_{max} \ldots |\pi|)) = \mu \} 
		\\
		\cup \ \{& \pi \in FP^{\mathcal{M}}_{\mu} \mid \exists j < |\pi|- l_{max},\; \wmp(\pi(|\pi|- l_{max} \ldots |\pi|)) = \lambda\} 
	\end{align*}
	
	Let $r \in \rat$. Then we have:
	\begin{center}
		$ 
		\begin{array}{rcl}
		(f^{l_{max}}_{DirFixWMP})^{-1}(\mathcal{M},s_{init},r) & = & \{ \pi \in Paths^{\mathcal{M}}(s_{init}) \mid \exists i \in \nat,\; \wmp(\pi(i \ldots (i+l_{max}))) = r \land \\[2pt]
		& & \ \ \forall j \in \nat,\; \wmp(\pi(j \ldots (j+l_{max})) \geq r)\} \\[2pt]
		& = & \{ \pi \in Paths^{\mathcal{M}}(s_{init}) \mid \exists i \in \nat,\; \wmp(\pi(i \ldots (i+l_{max})) = r)\} \\[2pt]
		& & \setminus \{\pi \in Paths^{\mathcal{M}}(s_{init}) \mid \exists i \in \nat,\; \wmp(\pi(i \ldots (i+l_{max}))) = r \land \\[2pt]
		& & \ \ \ \exists j \in \nat,\; \wmp(\pi(j \ldots (j+l_{max})) < r)\}\\[2pt]
		& = & \biguplus \limits_{\pi \in FP^{\mathcal{M}}_r} Cyl(\pi) \setminus (\biguplus \limits_{p < r} \biguplus \limits_{\pi \in FP^{\mathcal{M}}_{p,r}} Cyl(\pi))\\[2pt]
		\end{array}
		$
	\end{center}
	
 We extend the definition of $h$ to sets of infinite paths: for $P \subseteq Paths^{\mathcal{M}}(s_{init})$, let $h(P) = \lbrace \pi \in Paths^{\mathcal{M}_{l_{max}}}(s_{init}) \mid \exists \pi' \in P,\; h(\pi') = \pi \rbrace$. Then, by construction of $h$, we have:
 \begin{displaymath}
	 	(g^{l_{max}}_{DirFixWMP})^{-1}(\mathcal{M}_{l_{max}},s_{init},r) = h((f^{l_{max}}_{DirFixWMP})^{-1}(\mathcal{M},s_{init},r))
 \end{displaymath} 
 Since all unions are disjoint in the previous series of equalities and because $h$ is a bijection, we have:
 \begin{displaymath}
	 	(g^{l_{max}}_{DirFixWMP})^{-1}(\mathcal{M}_{l_{max}},s_{init},r) = \biguplus \limits_{\pi \in FP^{\mathcal{M}}_r} h(Cyl(\pi)) \setminus (\biguplus \limits_{p < r} \biguplus \limits_{\pi \in FP^{\mathcal{M}}_{p,r}} h(Cyl(\pi)))
 \end{displaymath} 
 Let $\pi \in FP^{\mathcal{M}}_r$ and $\bar{\pi} \in Cyl(\pi)$. Then:
	\begin{displaymath}
		h(Cyl(\pi)) = Cyl(\ h(\bar{\pi}(0 \ldots |\pi| - l_{max} + 1)) \ )
	\end{displaymath}
	with $|\pi| \geq l_{max}$ since $\pi \in FP^{\mathcal{M}}_r$. Therefore:
	\begin{center}
		$
		\begin{array}{c c l}
		Pr(h(Cyl(\pi))) & = & Pr(Cyl(h(\bar{\pi}(0 \ldots (|\pi| - l_{max}+1)))))\\
		& = & \mathbb{P}'(h(\bar{\pi}(0 \ldots (|\pi| - l_{max}+1)))) \\
		& = & \mathbb{P}(\bar{\pi}(0 \ldots |\pi|))\\
		& = & \mathbb{P}(\pi)\\
		& = & Pr(Cyl(\pi))
		\end{array}
		$
	\end{center}
	
	Finally, we have:
	\begin{center}
		$
		\begin{array}{c c l}
		& & Pr((g^{l_{max}}_{DirFixWMP})^{-1}(\mathcal{M}_{l_{max}},s_{init},r))\\
		& = & Pr(\biguplus \limits_{\pi \in FP^{\mathcal{M}}_r} h(Cyl(\pi)) \setminus (\biguplus \limits_{p < r} \biguplus \limits_{\pi \in FP^{\mathcal{M}}_{p,r}} h(Cyl(\pi))))\\
		& = & \sum \limits_{\pi \in FP^{\mathcal{M}}_r} Pr(h(Cyl(\pi))) - \sum \limits_{p < r} \sum \limits_{\pi \in FP^{\mathcal{M}}_{p,r}} Pr(h(Cyl(\pi)))\\
		& = & \sum \limits_{\pi \in FP^{\mathcal{M}}_r} Pr(Cyl(\pi)) - \sum \limits_{p < r} \sum \limits_{\pi \in FP^{\mathcal{M}}_{p,r}} Pr(Cyl(\pi))\\
		& = & Pr(\biguplus \limits_{\pi \in FP^{\mathcal{M}}_r} Cyl(\pi) \setminus (\biguplus \limits_{p < r} \biguplus \limits_{\pi \in FP^{\mathcal{M}}_{p,r}} Cyl(\pi)))\\
		& = & Pr((f^{l_{max}}_{DirFixWMP})^{-1}(\mathcal{M},s_{init},r))
		\end{array}
		$
	\end{center}
	In fact $\forall r \in \rat$:
	\begin{displaymath}
		Pr((g^{l_{max}}_{DirFixWMP})^{-1}(\mathcal{M}_{l_{max}},s_{init},r)) = Pr((f^{l_{max}}_{DirFixWMP})^{-1}(\mathcal{M},s_{init},r))
	\end{displaymath}
	
	It follows that:
	\begin{displaymath}
		\mathbb{E}_{s_{init}}^{\mathcal{M}}(f^{l_{max}}_{DirFixWMP}) = \mathbb{E}_{s_{init}}^{\mathcal{M}_{l_{max}}}(g^{l_{max}}_{DirFixWMP})
	\end{displaymath} 
\end{proof}

Let us now focus on the computation of the expected value of $g^{l_{max}}_{DirFixWMP}$ on a \textit{path Markov chain} $\mathcal{M} = \langle S,\mathbb{P},s_{init},AP,L \rangle$. That is the purpose of Algorithm~\ref{val_reg}. The idea is that the value of a paths $\pi$, that is $g^{l_{max}}_{DirFixWMP}(\pi)$, depends on the value of the states visited by $\pi$. More specifically, let us consider the set of values $Val = \{ m \in \rat \mid \exists s \in S,\; L(s) = m \}$ such that $Val = \{ m_1,m_2, \ldots ,m_{|Val|}\}$ with $m_1 < m_2 < \ldots < m_{|Val|}$. Let $\pi \in Paths^{\mathcal{M}_{l_{max}}}(s_{init})$. Then:
\begin{itemize}
	\item $g^{l_{max}}_{DirFixWMP}(\pi) = m_1 \Leftrightarrow \pi \models \lozenge m_1$
	\item $g^{l_{max}}_{DirFixWMP}(\pi) = m_2 \Leftrightarrow \pi \models \lozenge m_2 \land \lnot \lozenge m_1 \Leftrightarrow \pi \models (\lnot m_1) U (m_2 \land \lnot \lozenge m_1)$
	\item $g^{l_{max}}_{DirFixWMP}(\pi) = m_3 \Leftrightarrow \pi \models \lozenge m_3 \land \lnot \lozenge (m_1 \vee m_2) \Leftrightarrow \pi \models (\lnot (m_1 \vee m_2)) U (m_3 \land \lnot \lozenge (m_1 \vee m_2))$
\end{itemize}
More generally, we have:
\begin{displaymath}
\forall i \leq |Val|,\; g^{l_{max}}_{DirFixWMP}(\pi) = m_i \Leftrightarrow \pi \models (\lnot \underset{j < i}{\lor} m_j) U (m_i \land  \lnot \lozenge \underset{j < i}{\lor} m_j)
\end{displaymath}
We use the above in Algorithm~\ref{val_reg}. In the first $\mathsf{for}$ loop, the set $Val$ is constructed, while associating, for each $m \in Val$, the set of states $S_m = \lbrace s \in S \mid L(s) = m \rbrace$. If $W$ be the maximal weight that appears in the Markov chain, then that \textsf{for} loop takes time $O(|S| \cdot \log(W))$. In the following, for each $m \in Val$, the probability that $g^{l_{max}}_{DirFixWMP}(\pi) = m$ is computed. This computation takes time $O(Mat(S) \cdot \log(W))$. In fact, the complexity of this algorithm is in $O(|S| \cdot Mat(S) \cdot \log(W))$ (since $|Val| \leq |S|$).
	
\begin{algorithm}
	\caption{ValWindowMarkovChain($\mathcal{M},s_{init},l_{max}$)}
	\label{val_reg}
	\begin{algorithmic}[1]
		\Require{$\mathcal{M}=\langle S,\mathbb{P},s_{init},AP,L \rangle$ is a regular Markov Chain, $s_{init} \in S$ and $l_{max} \in \mathbb{N}_0$}
		\Ensure{$E$ is equal to $\mathbb{E}_{s_{init}}^{\mathcal{M}_{l_{max}}}(g^{l_{max}}_{DirFixWMP})$}
		\State {$Val := AssocTable(\textsf{float},\textsf{state list}$) \textit{$\;\;$ :float are sorted in ascending order}}
		\For {$s \in S$}
		\If {$s \neq s_{init}$}
		\If {$L(s) \not \in Val$}
		\State {$Val.add(L(s),[ ])$}
		\Else
		\State {$Val.(L(s)).append(s)$}
		\EndIf
		\EndIf
		\EndFor
		\State {$GreaterVal := [ ]$}
		\State {$Current := [ ]$}
		\State {$E := 0$}
		\For {$m \in Val$}
		\State {$Current := Val(m)$}
		\State {$Prob_m := 0$}
		\For {$s \in Current$}
		\State {$C = GreaterVal \setminus Current; B = {s}$}
		\State {$Prob_m \mathrel{+}= Pr_{s_{init}}(C \ U \ B) \cdot Pr_s(\square GreaterVal)$}
		\EndFor 
		\State {$GreaterVal := GeraterVal \setminus Current$}
		\State {$E \mathrel{+}= Prob_m \cdot m$}
		\EndFor
		\Return {$E$}
	\end{algorithmic}
\end{algorithm}

Then, we can use the result of the previous theorem to compute the expected value of $f^{l_{max}}_{DirFixWMP}$ in a weighted Markov chain $\mathcal{M} = \langle S,E,\mathbb{P},w \rangle$. That is done in algorithm~\ref{val_weigh}. The size of the regular Markov chain that is constructed is in $O(|S|^{l_{max}}$). Therefore, the complexity of this algorithm is in $O(|S|^{l_{max}} \cdot Mat(|S|^{l_{mas}}) \cdot \log(W))$ where $W$ is still the maximal weight that appears in the Markov chain. 

Although that complexity seems way worst that the complexity of the previous algorithm (since it is exponential in $l_{max}$), the main asset of this algorithm is that it is logarithmic in $W$, whereas the previous algorithm is polynomial in $W$. 
\begin{algorithm}
	\caption{DirFixWMP2($\mathcal{M},s_{init},l_{max}$)}
	\label{val_weigh}
	\begin{algorithmic}[1]
		\Require{$\mathcal{M}=\langle S,E,s_{init},\mathbb{P},w \rangle$ is a weighted Markov Chain, $s_{init} \in S$ and $l_{max} \in \mathbb{N}_0$}
		\Ensure{$E$ is equal to $\mathbb{E}_{s_{init}}^{\mathcal{M}}(f^{l_{max}}_{DirFixWMP})$}
		\State {$\mathcal{M}_{l_{max}} := \langle S',\mathbb{P}',s_{init},AP,L \rangle$}
		\State {$E := ValWindowMarkovChain(\mathcal{M}_{l_{max}})$}
		\State
		\Return {$E$}
	\end{algorithmic}
\end{algorithm}

\begin{remark}
	A possibly useful observation is that the value of path that has rheached a given state can not exceed the maximum of the value of the BSCCs reachable from that state. Therefore, with some computation on the Markov chain $\mathcal{M}_{l_{max}}$, we could avoid some further computations in the \textit{path Markov chain}.
\end{remark}

\section{Detailed Proof of Theorem~\ref{th_mdp_dir}}
\label{dirfix_mdp}
	We follow the five steps:
	\begin{itemize}
		\item[1.] We introduce the function $st_1: strat(\Gamma) \longrightarrow strat(\Gamma_{l_{max}})$ such that, $\forall \sigma \in strat(\Gamma)$:
		\begin{displaymath}
		\forall (t_0 \ldots t_n) \in (S')^{+}, st_1(\sigma)(t_0 \ldots t_n) = \sigma(t_0^0 \ldots t_n^0)
		\end{displaymath}
		In the following, for $\sigma \in strat(\Gamma)$, the strategy $st_1(\sigma) \in strat(\Gamma_{l_{max}})$ will be denoted $\sigma_{l_{max}}$.
		
		\begin{proposition}
			\label{proposition_pi0}
			For all $\sigma \in strat(\Gamma)$, if $\pi \in Paths^{\MDPtoMC{\Gamma_{\sizeMin}}{\sigma_{l_{max}}}}(s'_{init})$, then $\pi^0 \in Paths^{\MDPtoMC{\Gamma}{\sigma}}(s_{init})$.
		\end{proposition}
		
		\begin{proof}
			This comes from the definition of $st_1$ and because, for all $t \in S'$, $Act(t) = Act(t^0)$.
		\end{proof}

		\item[2.] We establish Lemma~\ref{lemma_average_WMP} that links the mean-payoff over a path $\pi$ in $\MDPtoMC{\Gamma_{\sizeMin}}{\sigma_{l_{max}}}$ and the direct fixed window mean-payoff over $\pi^0$ in $\MDPtoMC{\Gamma}{\sigma}$.
		\begin{lemma}
			Let $\sigma \in strat(\Gamma)$ and let $\pi = t_0 \ldots \in Paths^{\MDPtoMC{\Gamma_{\sizeMin}}{\sigma_{l_{max}}}}(s'_{init})$. Then $\pi^0 \in 
			Paths^{\MDPtoMC{\Gamma}{\sigma}}(s_{init})$ and
			\begin{displaymath}
			f^{l_{max}}_{DirFixWMP}(\pi^0) = f_{Mean}(\pi)
			\end{displaymath}
			\label{lemma_average_WMP}
		\end{lemma}
		
		\begin{proof}
			We have that $\pi^0 \in 
			Paths^{\MDPtoMC{\Gamma}{\sigma}}(s_{init})$ by Proposition~\ref{proposition_pi0}. Therefore, $f^{l_{max}}_{DirFixWMP}(\pi^0)$ is defined.
			
			By definition of $\Gamma_{\sizeMin}$, we have: $\forall i \geq 0,\; t_i^2 \geq t_{i+1}^2$ since $t_{i+1}^2 = \min (t_{i}^2,x)$ where $x$ can be expressed as a function of the weights of the MDP. Moreover, for all $i \geq 0$, $w'(t_{i-1}, \sigma_{l_{max}}(t_0 \ldots t_{i-1}), t_{i}) = t_i^2$. Since $\Gamma_{l_{max}}$ is finite, the series $w'(t_{i-1},\sigma_{l_{max}}(t_0 \ldots t_{i-1}),t_{i})$ is a non-increasing series of rational numbers that are included in the finite set $\lbrace \frac{p}{q} \mid q \in [l_{max}],\; p \in [q \cdot W]_0 \rbrace$,  therefore there exists $j \in \nat$, and $\nu \in \rat$ such that:
			\begin{displaymath}
			\forall i \geq j,\; w'(t_i,\sigma_{l_{max}}(t_0 \ldots t_i),t_{i+1}) = \nu = t_i^2
			\end{displaymath}
			It follows that $f_{mean}(\pi) = \nu$. 
			
			Moreover, for all $i \geq 0$, $t_i^{1,l_{max}-1} = w(t_{i-1}^0,\sigma_{l_{max}}(t_0 \ldots t_i),t_{i}^0) = w(t_{i-1}^0,\sigma(t_0^0 \ldots t_i^0),t_{i}^0)$. Hence, by definition of $\Gamma_{l_{max}}$ (specifically, the condition on $\lambda'$ so that $\mathbb{P}' > 0$) we have $t_{l_{max}}^2 = \max\limits_{l \leq l_{max}} \frac{1}{l} \sum\limits_{k=0}^{l-1}
			w(t_k^0,\sigma(t_0^0 \ldots t_k^0),t_{k+1}^0)$. More generally:
			\begin{displaymath}
			\forall n \geq 0,\; t_{n+l_{max}}^2 = \min\limits_{k \leq n} \max\limits_{l \leq l_{max}} \frac{1}{l} \sum\limits_{i=k}^{k+l-1} w(t_i^0,\sigma(t_0^0 \ldots t_i^0),t_{i+1}^0)
			\end{displaymath}
			Then, by definition of $f_{DirFixWMP}^{l_{max}}$, we have:
			\begin{displaymath}
				f^{l_{max}}_{DirFixWMP} (\pi^0) = \inf\limits_{n \geq 0} t_{n+l_{max}}^2 = t^2_j = \nu = f_{mean}(\pi)
			\end{displaymath}
		\end{proof}

		\item[3.] Let $\sigma \in strat(\Gamma)$. We define by induction the function $s^{\sigma}: S^{+} \longrightarrow S'$ such that:
		\begin{itemize}
			\item $\forall s_0 \in S^0$, we have $s^\sigma(s_0) = (s_0,[W, \ldots, W],W)$;
			\item $\forall s_0 s_1 \ldots s_{n+1} \in S^{n+1}$, let $t_{n} = s^\sigma(s_0 s_1 \ldots s_{n-1} s_n)$ and $t_{n+1} = s^\sigma(s_0 s_1 \ldots s_{n} s_{n+1})$. Then, we have:
			\begin{itemize}
				\item[$\square$] $t_{n+1}^0 = s_{n+1}$;
				\item[$\square$] for $k \leq l_{max}-2$, we have $t_{n+1}^{1,k} = t_{n}^{1,k+1}$;
				\item[$\square$] $t_{n+1}^{1,l_{max}-1} = w(s_n,\sigma(s_0 \ldots s_n),s_{n+1})$;
				\item[$\square$] $t_{n+1}^2 = \min (t_n^2,\max\limits_{l \leq l_{max}} \frac{1}{l} \sum\limits_{k=1}^{l} t_{n}^{k})$ where $t_{n}^{l_{max}} = t_{n+1}^{l_{max}-1}$.
			\end{itemize}
		\end{itemize}
		Let $s_0 s_1 \ldots s_{n} s_{n+1} \in FPaths^{\MDPtoMC{\Gamma}{\sigma}}(s_{init})$. For all $i \leq n$, we denote $s^\sigma(s_0 s_1 \ldots s_{i})$ by $t_i$. Then, by construction of $s^\sigma$, for all $i \leq n$, we have:
		\begin{equation}
		\label{eq_s}
		\mathbb{P}'(t_i,\sigma(s_0 \ldots s_i),t_{i+1}) = \mathbb{P}(s_i,\sigma(s_0 \ldots s_i),s_{i+1}) > 0
		\end{equation}
		
		Note that, for a sequence of states $s_0 \ldots s_{n+1} \in S^{+}$, if a partial function $\iota: S^{+} \longrightarrow Act$ is defined on all sequence $s_0 \ldots s_i \in S^{+}$ for $0 \leq i \leq n$, then $s^{\iota}(s_0 \ldots s_n)$ is also defined (in the same way that $s^{\sigma}(s_0 \ldots s_n)$ is defined if $\sigma \in strat(\Gamma)$).	
		
		Moreover, $s^\sigma$ ensures the following property (recall that $\sigma_{l_{max}}$ refers to $st_1(\sigma)$):
		\begin{lemma}
			\label{lemma_s_property}
			Let $t_0 t_1 \ldots  \in Paths^{\MDPtoMC{\Gamma_{\sizeMin}}{\sigma_{l_{max}}}}(s'_{init})$. Then:
			\begin{displaymath}
			s^\sigma(t_0^0 t_1^0 \ldots t_{n-1}^0 t_n^0) = t_n
			\end{displaymath}
		\end{lemma}
		
		This property follows directly from the definition of $s^{\sigma}$ and $\Gamma_{l_{max}}$. Still, we give a complete proof of that property.
		\begin{proof}
			We prove the following inductive property, defined for $n \geq 0$:
			\begin{center}
				$H(n): \forall t_0 t_1 \ldots  \in Paths^{\MDPtoMC{\Gamma_{\sizeMin}}{\sigma_{l_{max}}}}(s'_{init})$, $s^{\sigma}(t_0^0 t_1^0 \ldots t_{n-1}^0 t_n^0) = t_n$
			\end{center}
			
			\begin{itemize}
				\item $H(0)$ is true since, for all $t_0 t_1 \ldots  \in Paths^{\MDPtoMC{\Gamma_{\sizeMin}}{\sigma_{l_{max}}}}(s'_{init})$, we have $t_0 = s'_{init}$. Therefore $t_0^0 = s_{init}$ and $s^\sigma(t_0^0) = s^\sigma(s_{init}) = s'_{init} = t_0$. 
				\item Assume that $H(n)$ holds for some $n \geq 0$. Let $t_0 t_1 \ldots \in Paths^{\MDPtoMC{\Gamma_{\sizeMin}}{\sigma_{l_{max}}}}(s'_{init})$. We denote $t'_{n+1} = s^\sigma(t_0^0 t_1^0 \ldots t_{n}^0 t_{n+1}^0)$. We prove that $t'_{n+1} = t_{n+1}$. According to $H(n)$, $s^\sigma(t_0^0 \ldots t_{n}^0) = t_{n}$. By definition of $s$ and $\mathbb{P}'$ (recall that, since $t_0 t_1 \ldots \in Paths^{\MDPtoMC{\Gamma_{\sizeMin}}{\sigma_{l_{max}}}}(s'_{init})$, we have $\mathbb{P}'(t_n,\sigma_{l_{max}}(t_0 \ldots t_n),t_{n+1}) > 0$):
				\begin{itemize}
					\item[$\square$] $t^{' 0}_{n+1} = t_{n+1}^0$;
					\item[$\square$] for $k \leq l_{max} - 1$, we have $t^{' 1,k}_{n+1} = t^{1,k+1}_{n} = t^{1,k}_{n+1}$;
					\item[$\square$] $t^{' 1,l_{max}-1}_{n+1} = w(t_n^0,\sigma_{l_{max}}(t_0 \ldots t_n),t^0_{n+1}) = t^{1,l_{max}-1}_{n+1}$;
					\item[$\square$] $t^{' 2}_{n+1} = \min (t_n^2,\max\limits_{l \leq l_{max}} \frac{1}{l} \sum\limits_{k=1}^{l} t_{n}^{k}) = t^{2}_{n+1}$.
				\end{itemize}
				Therefore $t'_n = t_n$. Since this is true for all $t_0 t_1 \ldots \in Paths^{\MDPtoMC{\Gamma_{\sizeMin}}{\sigma_{l_{max}}}}(s'_{init})$, $H(n+1)$ holds. 
			\end{itemize}
			The lemma follows.
		\end{proof}
		
		\item[4.] Now, we can establish a property that the function $st_1$ ensures:
		\begin{lemma}
			Let $\sigma \in strat(\Gamma)$. There exists a bijection $h^{\sigma}: Paths^{\MDPtoMC{\Gamma}{\sigma}}(s_{init}) \rightarrow Paths^{\MDPtoMC{\Gamma_{\sizeMin}}{\sigma_{l_{max}}}}(s'_{init})$ such that, $\forall \pi \in Paths^{\MDPtoMC{\Gamma}{\sigma}}(s_{init})$:
			\begin{itemize}
				\item $f^{l_{max}}_{DirFixWMP}(\pi) = f_{Mean}(h^\sigma(\pi))$, that is the direct fixed window mean-payoff of $\pi$ in $\MDPtoMC{\Gamma}{\sigma}$ is equal to the mean-payoff of $h^\sigma(\pi)$ in $\MDPtoMC{\Gamma_{\sizeMin}}{\sigma_{l_{max}}}$;  
				\item $\forall j \geq 0,\; \mathbb{P}(\pi(0 \ldots j)) = \mathbb{P}'(h^\sigma(\pi)(0 
				\ldots j))$, that is any finite fragment of the path $\pi$ has the same probability to occur in $\MDPtoMC{\Gamma}{\sigma}$ as the corresponding path fragment of $h^\sigma(\pi)$ in $\MDPtoMC{\Gamma_{\sizeMin}}{\sigma_{l_{max}}}$.
			\end{itemize}
			\label{lemma_direct_mdp}
		\end{lemma}
		
		\begin{proof}
			Let $\sigma \in strat(\Gamma)$.	Let $h^\sigma: Paths^{\MDPtoMC{\Gamma}{\sigma}}(s_{init}) \rightarrow Paths^{\MDPtoMC{\Gamma_{\sizeMin}}{\sigma_{l_{max}}}}(s'_{init})$ be such that for $\pi = s_0 s_1 \ldots \in Paths^{\MDPtoMC{\Gamma}{\sigma}}(s_{init})$, where $s_0 = s_{init}$, we have $h^{\sigma}(\pi) = t_0 t_1\ldots \in Paths^{\MDPtoMC{\Gamma_{\sizeMin}}{\sigma_{l_{max}}}}(s'_{init})$ with 
			\begin{displaymath}
			\forall i \geq 0,\; t_i = s^\sigma(s_0 \ldots s_i) \in S'
			\end{displaymath}
			We have, for all $i \geq 0$, $t_i^0 = s_i$. Therefore, by definition of $st_1$, for all $i \geq 0$, we have $\sigma_{l_{max}}(t_0 \ldots t_i) = \sigma(s_0 \ldots s_i)$.
			
			Then:
			\begin{itemize}
				\item Let $\pi \in Paths^{\MDPtoMC{\Gamma}{\sigma}}(s_{init})$. By Equation~\ref{eq_s}, $\forall i \geq 0$, we have: $\mathbb{P}'(t_i,\sigma_{l_{max}}(t_0 \ldots t_i),t_{i+1}) = \mathbb{P}'(t_i,\sigma(s_0 \ldots s_i),t_{i+1}) = \mathbb{P}(s_i,\sigma(s_0 \ldots s_i),s_{i+1}) > 0$. Therefore $h^{\sigma}(\pi) \in Paths^{\MDPtoMC{\Gamma_{\sizeMin}}{\sigma_{l_{max}}}}(s'_{init})$. This holds for all $\pi \in Paths^{\MDPtoMC{\Gamma}{\sigma}}(s_{init})$. That is, $h^\sigma$ is well defined and $\forall j \geq 0$ we have $\mathbb{P}(\pi(0 \ldots j)) = \mathbb{P}'(h^\sigma(\pi)(0 \ldots j))$.
				
				\item In the definition of $h^\sigma$, for all $i \geq 0$, we have $t_i^0 = s_i$. 
				It is easy to see that, for all $\pi,\pi' \in Paths^{\MDPtoMC{\Gamma}{\sigma}}(s_{init})$ such that $\pi \neq \pi'$, we have $h^{\sigma}(\pi) \neq h^{\sigma}(\pi')$. That is, $h^\sigma$ is injective.
				
				\item Let $\pi_{l_{max}} = t_0 t_1 \ldots \in Paths^{\MDPtoMC{\Gamma_{\sizeMin}}{\sigma_{l_{max}}}}(s'_{init})$. We construct a path $\pi \in Paths^{\MDPtoMC{\Gamma}{\sigma}}(s_{init})$ such that $h^\sigma(\pi) = \pi_{l_{max}}$. Let $\pi = \pi_{l_{max}}^0 \in Paths^{\MDPtoMC{\Gamma}{\sigma}}(s_{init})$ (by Proposition~\ref{proposition_pi0}). Moreover, $h^\sigma(\pi) = t'_0 t'_1 \ldots$ where, for all $n \geq 0$, 
				\begin{align*}
				t'_n = s^\sigma(t_0^0 \ldots t_{n}^0) = t_n
				\end{align*}
				according to Lemma~\ref{lemma_s_property}.	That is, $h^\sigma(\pi) = \pi_{l_{max}}$. This proves that $h^\sigma$ is surjective, and therefore bijective. 
				\item Let $\pi = s_0 s_1 \ldots \in Paths^{\MDPtoMC{\Gamma}{\sigma}}(s_{init})$ and let $h^{\sigma}(\pi) = t_0 t_1 \ldots \in Paths^{\MDPtoMC{\Gamma_{\sizeMin}}{\sigma_{l_{max}}}}(s'_{init})$. Then $\pi = (h^{\sigma}(\pi))^0$. Then, Lemma~\ref{lemma_average_WMP} gives us that $f^{l_{max}}_{DirFixWMP}(\pi) = f_{Mean}(h^\sigma(\pi))$.
			\end{itemize}
			In fact, the function $h^\sigma$ ensures the properties specified in Lemma~\ref{lemma_direct_mdp}.
		\end{proof}

		\item[5.] We now define the function $st_2: strat_0(\Gamma_{l_{max}}) \longrightarrow strat(\Gamma)$ that maps a memoryless strategy in $\Gamma_{l_{max}}$ into a strategy in $\Gamma$. The function $st_2$ is defined by induction such that for all memoryless strategy $\tilde{\sigma}^{l_{max}} \in strat_0(\Gamma_{l_{max}})$:
		\begin{itemize}
			\item Let $s_0 \in (S)^{1}$. Then $st_2(\tilde{\sigma}^{l_{max}})(s_0) = \tilde{\sigma}^{l_{max}}((s_0,[W, \ldots, W],W))$;
			\item Let $s_0 \ldots s_n \in (S)^{n+1}$. Then $st_2(\tilde{\sigma}^{l_{max}})(s_0 \ldots s_n) = \tilde{\sigma}^{l_{max}}(s^{st_2(\tilde{\sigma}^{l_{max}})}(s_0 s_1 \ldots s_n))$. It is well defined since $st_2(\tilde{\sigma}^{l_{max}})$ is defined by induction and therefore $st_2(\tilde{\sigma}^{l_{max}})(s_0 s_1 \ldots s_i)$ is already defined for $i \leq n-1$.
		\end{itemize}
		We define this function only on the set of memoryless strategies of $\Gamma_{l_{max}}$ since we know that there exists a memoryless strategy that maximizes the expected value of the mean-payoff in $\Gamma_{l_{max}}$. 
		
		The function $st_2$ is defined such that, for all $\tilde{\sigma}^{l_{max}} \in strat_0(\Gamma_{l_{max}})$, the strategies $\tilde{\sigma}^{l_{max}}$ and  $st_1(st_2(\tilde{\sigma}^{l_{max}}))$ coincide on the valid paths of $\MDPtoMC{\MDP_{\sizeMin}}{\tilde{\sigma}^{l_{max}}}$ and $\MDPtoMC{\MDP_{\sizeMin}}{st_1(st_2(\tilde{\sigma}^{l_{max}}))}$. This is stated in the following lemma:
		\begin{lemma}
			\label{lemma_st2}
			Let $\tilde{\sigma}^{l_{max}} \in strat_0(\Gamma_{l_{max}})$ and $t_0 t_1 \ldots \in Paths^{\MDPtoMC{\MDP_{\sizeMin}}{\tilde{\sigma}^{l_{max}}}}(s'_{init}) \cup \\ Paths^{\MDPtoMC{\MDP_{\sizeMin}}{st_1(st_2(\tilde{\sigma}^{l_{max}}))}}(s'_{init})$. Then, for all $n \geq 0$, $st_1(st_2(\tilde{\sigma}^{l_{max}}))(t_0 \ldots t_n) = \tilde{\sigma}^{l_{max}}(t_0 \ldots t_n)$.
		\end{lemma}
		
		\begin{proof}
			Let $\tilde{\sigma}^{l_{max}} \in strat_0(\Gamma_{l_{max}})$. Let $\pi = t_0 t_1 \ldots \in Paths^{\MDPtoMC{\MDP_{\sizeMin}}{\tilde{\sigma}^{l_{max}}}}(s'_{init}) \cup \\ Paths^{\MDPtoMC{\MDP_{\sizeMin}}{st_1(st_2(\tilde{\sigma}^{l_{max}}))}}(s'_{init})$. We prove the following inductive property, defined for $n \geq 0$:
			\begin{center}
				$H^{\pi}(n):$ we have $t_0 \ldots t_n \in FPaths^{\MDPtoMC{\MDP_{\sizeMin}}{st_1(st_2(\tilde{\sigma}^{l_{max}}))}}(s_{init})$ and $\tilde{\sigma}_{l_{max}}(t_0 \ldots t_n) = \tilde{\sigma}^{l_{max}}(t_0 \ldots t_n)$.
			\end{center}
			
			\begin{itemize}
				\item We have $t_0 = s'_{init}$. Hence:
				\begin{itemize}
					\item[$\square$] $t_0 \in FPaths^{\MDPtoMC{\MDP_{\sizeMin}}{st_1(st_2(\tilde{\sigma}^{l_{max}}))}}(s_{init})$;
					\item[$\square$] $st_1(st_2(\tilde{\sigma}^{l_{max}}))(t_0) = st_2(\tilde{\sigma}^{l_{max}})(t_0^0) = st_2(\tilde{\sigma}^{l_{max}})(s_{init}) = \tilde{\sigma}^{l_{max}}(s'_{init}) = \tilde{\sigma}^{l_{max}}(t_0)$.
				\end{itemize}
				Thus, $H^\pi(0)$ holds.
				
				\item Assume that $H^\pi(n-1)$ holds for some $n \geq 1$. Suppose that $\pi \in Paths^{\MDPtoMC{\MDP_{\sizeMin}}{\tilde{\sigma}^{l_{max}}}}(s'_{init})$. In that case, by definition of $\mathbb{P}'$, we have:
				\begin{displaymath}
				\mathbb{P}'(t_{n-1},\tilde{\sigma}^{l_{max}}(t_0 \ldots t_{n-1}),t_{n}) = \mathbb{P}(t^0_{n-1},\tilde{\sigma}^{l_{max}}(t_0 \ldots t_{n-1}),t^0_{n}) > 0
				\end{displaymath}
				Moreover, by $H^\pi(n-1)$, we have that:
				\begin{itemize}
					\item[$\square$] $t_0 \ldots t_{n-1} \in FPaths^{\MDPtoMC{\MDP_{\sizeMin}}{st_1(st_2(\tilde{\sigma}^{l_{max}}))}}(s_{init})$;
					\item[$\square$] $\tilde{\sigma}^{l_{max}}(t_0 \ldots t_{n-1}) = st_1(st_2(\tilde{\sigma}^{l_{max}}))(t_0 \ldots t_{n-1})$.
				\end{itemize}
				In fact:
				\begin{displaymath}
				\mathbb{P}'(t_{n-1},st_1(st_2(\tilde{\sigma}^{l_{max}}))(t_0 \ldots t_{n-1}),t_{n}) = \mathbb{P}(t_{n-1},\tilde{\sigma}^{l_{max}}(t_0 \ldots t_{n-1}),t_{n}) > 0
				\end{displaymath}
				Thus, $t_0 \ldots t_n \in FPaths^{\MDPtoMC{\MDP_{\sizeMin}}{st_1(st_2(\tilde{\sigma}^{l_{max}}))}}(s_{init})$. This is also the case if $\pi \in \\ Paths^{\MDPtoMC{\MDP_{\sizeMin}}{st_1(st_2(\tilde{\sigma}^{l_{max}}))}}(s'_{init})$. By Lemma~\ref{lemma_s_property}, we have:
				\begin{displaymath}
				s^{st_2(\tilde{\sigma}^{l_{max}})}(t_0^0 \ldots t_n^0) = t_n
				\end{displaymath}
				Since $\tilde{\sigma}^{l_{max}}$ is memoryless, we have $\tilde{\sigma}^{l_{max}}(t_n) = \tilde{\sigma}^{l_{max}}(t_0 \ldots t_n)$. In fact:
				\begin{align*}
				st_1(st_2(\tilde{\sigma}^{l_{max}}))(t_0 \ldots t_n) & = st_2(\tilde{\sigma}^{l_{max}})(t_0^0 \ldots t_n^0) = \tilde{\sigma}^{l_{max}}(s^{st_2(\tilde{\sigma}^{l_{max}})}(t_0^0 \ldots t_n^0)) \\ 
				& = \tilde{\sigma}^{l_{max}}(t_n) = \tilde{\sigma}^{l_{max}}(t_0 \ldots t_n)
				\end{align*}
				Hence, $H^\pi(n+1)$ holds.
			\end{itemize}
			Since this holds for every $\pi = t_0 t_1 \ldots \in Paths^{\MDPtoMC{\MDP_{\sizeMin}}{\tilde{\sigma}^{l_{max}}}}(s'_{init}) \cup Paths^{\MDPtoMC{\MDP_{\sizeMin}}{st_1(st_2(\tilde{\sigma}^{l_{max}}))}}(s'_{init})$, the lemma follows.
		\end{proof}
	\end{itemize}

	We can now proceed to the proof of the Theorem~\ref{th_mdp_dir}.
	\begin{proof}
		We proceed in two steps:
		\begin{itemize}
			\item First we prove that $\mathbb{E}_{s_{init}}^{\Gamma}(f^{l_{max}}_{DirFixWMP}) \leq \mathbb{E}_{s'_{init}}^{\Gamma_{l_{max}}}(f_{Mean})$. Let $\sigma \in strat(\Gamma)$. Then, using Lemma~\ref{lemma_direct_mdp}, we can prove that $\mathbb{E}_{s_{init}}^{\MDPtoMC{\MDP}{\sigma}}(f^{l_{max}}_{DirFixWMP}) = \mathbb{E}_{s'_{init}}^{\MDPtoMC{\MDP_{\sizeMin}}{\sigma_{l_{max}}}}(f_{Mean})$ with the same arguments used in the proof of Theorem~\ref{path_th}. This is true for all $\sigma \in strat(\Gamma)$. This implies that, for all $\sigma \in strat(\Gamma)$, there exists $\sigma^{l_{max}} \in strat(\Gamma_{l_{max}})$ such that $\mathbb{E}_{s_{init}}^{\MDPtoMC{\MDP}{\sigma}}(f^{l_{max}}_{DirFixWMP}) = \mathbb{E}_{s'_{init}}^{\MDPtoMC{\MDP_{\sizeMin}}{\sigma^{l_{max}}}}(f_{Mean})$. Therefore:
			\begin{align*}
				\mathbb{E}_{s_{init}}^{\Gamma}(f^{l_{max}}_{DirFixWMP}) & = \sup\limits_{\sigma \in strat(\Gamma)} \mathbb{E}_{s_{init}}^{\MDPtoMC{\MDP}{\sigma}}(f^{l_{max}}_{DirFixWMP}) \\ & \leq \sup\limits_{\sigma^{l_{max}} \in strat(\Gamma_{l_{max}})} \mathbb{E}_{s'_{init}}^{\MDPtoMC{\MDP_{\sizeMin}}{\sigma^{l_{max}}}}(f^{l_{max}}_{DirFixWMP}) \\ & = \mathbb{E}_{s'_{init}}^{\Gamma_{l_{max}}}(f_{Mean})
			\end{align*}
			\item Now we prove that $\mathbb{E}_{s_{init}}^{\Gamma}(f^{l_{max}}_{DirFixWMP}) \geq \mathbb{E}_{s'_{init}}^{\Gamma_{l_{max}}}(f_{Mean})$. There exists $\sigma_0^{l_{max}} \in strat_0(\Gamma_{l_{max}})$ such that
			\begin{displaymath}
				\mathbb{E}_{s'_{init}}^{\MDPtoMC{\MDP_{\sizeMin}}{\sigma_0^{l_{max}}}}(f_{Mean}) = \mathbb{E}_{s'_{init}}^{\Gamma_{l_{max}}}(f_{Mean})
			\end{displaymath} 
			From Lemma~\ref{lemma_st2}, we can conclude that the two Markov chains $\MDPtoMC{\MDP_{\sizeMin}}{st_1(st_2(\sigma_0^{l_{max}}))}$ and $\MDPtoMC{\MDP_{\sizeMin}}{\sigma_0^{l_{max}}}$ are identical if we do not consider the states that are not reachable from $s'_{init}$. Therefore:
			\begin{displaymath}
				\mathbb{E}_{s'_{init}}^{\MDPtoMC{\MDP_{\sizeMin}}{st_1(st_2(\sigma_0^{l_{max}}))}}(f_{Mean}) = \mathbb{E}_{s'_{init}}^{\MDPtoMC{\MDP_{\sizeMin}}{\sigma_0^{l_{max}}}}(f_{Mean})
			\end{displaymath}
			Then, since $st_2(\sigma_0^{l_{max}}) \in strat(\MDP)$, with the same arguments we used in the previous item, we can justify that:
			\begin{displaymath}
				\mathbb{E}_{s_{init}}^{\MDPtoMC{\MDP}{st_2(\sigma_0^{l_{max}})}}(f^{l_{max}}_{DirFixWMP}) = \mathbb{E}_{s'_{init}}^{\MDPtoMC{\MDP_{\sizeMin}}{st_1(st_2(\sigma_0^{l_{max}}))}}(f_{Mean})
			\end{displaymath}
			 Therefore, 
			 \begin{align*}
				 \mathbb{E}_{s_{init}}^{\Gamma}(f^{l_{max}}_{DirFixWMP}) & = \sup\limits_{\sigma \in strat(\Gamma)} \mathbb{E}_{s_{init}}^{\MDPtoMC{\MDP}{\sigma}}(f^{l_{max}}_{DirFixWMP}) \geq \mathbb{E}_{s_{init}}^{\MDPtoMC{\MDP_{\sizeMin}}{st_2(\sigma_0^{l_{max}})}}(f^{l_{max}}_{DirFixWMP}) \\
				 & =  \mathbb{E}_{s'_{init}}^{\MDPtoMC{\MDP_{\sizeMin}}{st_1(st_2(\sigma_0^{l_{max}}))}}(f_{Mean}) = \mathbb{E}_{s'_{init}}^{\MDPtoMC{\MDP_{\sizeMin}}{\sigma_0^{l_{max}}}}(f_{Mean})\\
				 & = \mathbb{E}_{s'_{init}}^{\Gamma_{l_{max}}}(f_{Mean})
			 \end{align*}
			 That is, $\mathbb{E}_{s_{init}}^{\Gamma}(f^{l_{max}}_{DirFixWMP}) \geq \mathbb{E}_{s'_{init}}^{\Gamma_{l_{max}}}(f_{Mean})$.
		\end{itemize}
		In fact:
		\begin{displaymath}
			\mathbb{E}_{s_{init}}^{\Gamma}(f^{l_{max}}_{DirFixWMP}) = \mathbb{E}_{s'_{init}}^{\Gamma_{l_{max}}}(f_{Mean})
		\end{displaymath} 
		This concludes the proof.		
	\end{proof}

\end{document}